\documentclass[fleqn, useAMS,usenatbib]{mnras}
\usepackage{amssymb, graphicx, verbatim}
\usepackage{amsmath}

\usepackage{ulem}
\usepackage[T1]{fontenc}
\usepackage{siunitx}

\newcommand{\e}[1]{\times 10^{#1}}

\newcommand{\msun}{M$_\odot$}

\def\ti {$^{44}\mbox{Ti}$}

\def\ni {$^{56}$Ni}

\def\co {$^{56}$Co}

\def\kms {km~s$^{-1}$}

\setcitestyle{round,aysep={},yysep={,}}

\newcommand{\arj}[1]{\textcolor{black}{{#1}}}

\title[Properties of gamma-ray decay lines in 3D core-collapse supernova models]{Properties of gamma-ray decay lines in 3D core-collapse supernova models, with application to SN 1987A and Cas A}
\author[Anders Jerkstrand]{A. Jerkstrand$^{1,2}$\thanks{E-mail:anders@mpa-garching.mpg.de}, 
A. Wongwathanarat$^{1}$, H.-T. Janka$^{1}$, M. Gabler$^{1,9}$, D. Alp$^5$, R. Diehl$^3$,
\newauthor K. Maeda$^4$, J. Larsson$^5$, C. Fransson$^2$, A. Menon$^6$, A. Heger$^{7,8}$\\
$^{1}$Max-Planck Institut f\"ur Astrophysik, Karl-Schwarzschild Str. 1, D-85748 Garching, Germany\\
$^2$Department of Astronomy, Stockholm University, The Oskar Klein Centre, AlbaNova, SE-106 91 Stockholm, Sweden\\
$^{3}$Max-Planck-Institut f\"ur extraterrestrische Physik, Giessenbachstr. 1, 85741 Garching, Germany\\
$^{4}$Department of Astronomy, Kyoto University, Kitashirakawa-Oiwake-cho, Sakyo-ku, Kyoto, 606-8502, Japan\\
$^{5}$Department of Physics, KTH Royal Institute of Technology, The Oskar Klein Centre, AlbaNova, SE-106 91 Stockholm, Sweden\\
$^6$Anton Pannekoek Institute for Astronomy, University of Amsterdam, 1090 GE Amsterdam, The Netherlands\\
$^7$Monash Centre for Astrophysics, School of Physics and Astronomy, Monash University, VIC 3800, Australia\\
$^8$Tsung-Dao Lee Institute, Shanghai 200240, People's Republic of China\\
$^9$LATO/DCET, Universidade Estadual de Santa Cruz, Rod. Jorge Amado, km 16, Ilheus, BA, CEP 45662-900, Brazil\\
}
\begin{document}

\date{}

\pagerange{\pageref{firstpage}--\pageref{lastpage}} \pubyear{2017}

\maketitle

\label{firstpage}

\begin{abstract}
Comparison of theoretical line profiles to observations provides important tests for supernova explosion models. We study the shapes of radioactive decay lines predicted by current 3D core-collapse explosion simulations, and compare these to observations of SN 1987A and Cas A.  Both the widths and shifts of decay lines vary by several thousand kilometers per second depending on viewing angle. The line profiles can be complex with multiple peaks. By combining observational constraints from \co~decay lines, \ti~decay lines, and Fe IR lines, we delineate a picture of the morphology of the explosive burning ashes in SN 1987A. For $M_{\rm ZAMS}=15-20$ \msun~progenitors exploding with $\sim1.5 \times 10^{51}$ erg, ejecta structures suitable to reproduce the observations involve a bulk asymmetry of the \ni~of at least $\sim$400 \kms~and a bulk velocity of at least 1500 \kms. By adding constraints to reproduce the UVOIR bolometric light curve of SN 1987A up to 600d, an ejecta mass around 14 \msun~is favoured. We also investigate whether observed decay lines can constrain the neutron star (NS) kick velocity. The model grid provides a constraint $V_{\rm NS} > V_{\rm redshift}$, and applying this to SN 1987A gives a NS kick of at least 500 \kms. For Cas A, our single model provides a satisfactory fit to the NuSTAR observations and reinforces the result that current neutrino-driven core-collapse SN models achieve enough bulk asymmetry in the explosive burning material. Finally, we investigate the internal gamma-ray field and energy deposition, and compare the 3D models to 1D approximations. 
\end{abstract}

\begin{keywords}
supernovae: general - supernovae: individual: SN 1987A, Cas A - stars: evolution 
\end{keywords}

\section{Introduction} 
From the fractured and heterogenous appearance of supernova (SN) remnants, it is clear that significant deviations from spherical symmetry is the norm for core-collapse supernovae \citep[e.g.,][]{Hughes2000, Fesen2006, Larsson2013}. The remnant morphologies tell us that SNe neither retain a stratified shell structure, nor become homogeneously stirred on small scales, but rather have distinct compositional regions mixed on macroscopic scales with plumes, clumps, and filaments of various sizes and shapes. Whereas the view of remnants gives a picture biased towards shock-interaction regions, multiple lines of spectroscopic evidence from unresolved radioactivity-powered SNe paint a similar picture \citep[e.g.,][]{Meikle1989,Fillipenko1989,Maeda2008}. Polarization measurements typically indicate significant asymmetry that increases with time, suggesting strongest deformation in the core \citep{Hoflich1991, Leonard2006, Wang2008}. Asymmetric explosions are also evidenced from the high observed kick velocities of neutron stars \citep{Arzoumanian2002,Hobbs2005}.

On the theory side, an important first step in understanding how such structures come about was the discovery of Rayleigh-Taylor instabilities behind the reverse shock created at the He/H interface \citep{Chevalier1976}. Later, both other types of instabilities and other important composition interfaces have been identified \citep[e.g.,][]{Mueller1991,Herant1994,Burrows1995,Janka1996,Buras2006, Kifonidis2006}. These play important roles in understanding why also Type Ib/c (and Type IIb) supernovae can achieve large degrees of mixing \citep{Shigeyama1990,Hachisu1991,Nomoto1995,Wongwat2017}.

The spatial distribution of different elements in the SN depends on a combination of progenitor properties, explosion dynamics, and long-term mixing processes. In the explosion phase, hydrodynamic instabilities in the neutrino-heated layer behind the shock lead to the growth of low-mode plumes and significant shock deformation \citep[see][for a recent review]{Janka2016}. These deformations also serve as seeds to trigger the later Rayleigh-Taylor mixing \citep{Kifonidis2003}. Whereas the later mixing and \ni~heating modify the initially emerging structures, much of their basic properties, set in the first second of explosion, can be retained to various degrees  \citep{Wongwat2015}. Therefore, observed or inferred morphologies of explosive burning ashes provide a direct link to the explosion mechanism. At the same time, detailed analyses and quantitative results require modelling the long-term evolution in 3D, and testing these models with 3D radiative transfer codes.

Among the many types of observations of SNe, those of radioactive decay lines at gamma-ray and X-ray energies play a rather unique role. They probe directly the innermost distribution of explosive burning ashes, with no model-dependence on the emissivity as for UVOIR emission. They are therefore direct, powerful probes of the explosion dynamics. The transfer of radiation is straightforward (mainly Compton scattering), and lines are stand-alone with no risk of blending. The limitation of their use lies in the challenge of measuring them, requiring space-born detectors that allow only observations of the most nearby events. In addition, for the case of \co~decay lines in H-rich SNe, only a small fraction of the decay line emission escapes without being degraded to gamma-ray and X-ray continuum over the life-time of the \co~(few hundred days). One therefore does not obtain a clean view of the \co~distribution, but only what emerges after a significant degree of Compton interaction. For \ti~lines, observations can however be carried out in the optically thin phase due to its longer decay time of 85y.

For these reasons combined, there is currently only three SNe with gamma-ray decay line observations; SN 1987A \citep{Teegarden1991,Kurfess1992,Grebenev2012, Boggs2015}, Cas A \citep{Iyudin1994,Vink2001,Renaud2006, Siegert2015, Grefenstette2017} and SN 2014J \citep{Churazov2014,Diehl2015,Isern2016}. Nevertheless, these observations have played an important role in constraining the inner ejecta morphologies in SNe, and to a significant extent still remain to be fully interpreted by the development and analysis of realistic 3D models. Planned new instruments such as e-ASTROGAM \citep{deAngelis2018} and AMEGO \citep{McEnery2019} will further aid in the use of radioactive decay lines as an important supernova diagnostic, motivating efforts in this field. Sensitivities better by about two orders of magnitude can be achieved with foreseeable technology compared to currently operating instruments such as INTEGRAL and NuSTAR.

With a growing library of realistic 3D-models evolved to the homologous phase \citep[days/weeks, ][]{Wongwat2013, Wongwat2015, Wongwat2017, Muller2018}, some considering also the heating/hydrodynamic effect of the \co~decay (Gabler et al., in prep.), it becomes imperative to develop computational tools to simulate their emission across the electromagnetic spectrum. Further, not only the emergent radiation is of interest, but also the internal conditions which can be used both for improving 1D models and for providing the physical foundation for understanding the SN emission and establishing analytic treatments useful for big sample analyses. It is well recognized that significant differences arise between 2D and 3D \citep[e.g.,][]{Kane2000,Joggerst2010, Hammer2010, Ellinger2012}. For example were significantly faster \ni~bullets found in 3D compared to 2D by \citet{Hammer2010}. It is therefore likely of higher scientific return to pursue 3D modelling even if many more 2D models can be computed and analysed for the same cost.

In this paper we present the first part of a new framework to calculate spectral emission of SNe in 3D by extension of the SUMO platform \citep{Jerkstrand2011,Jerkstrand2012}. Here we cover the new photon transport algorithm on a 3D spherical coordinate grid. We then apply this to study the transport and deposition of gamma-ray decay line radiation in 3D core-collapse models, and properties of the emergent decay line profiles. In section \ref{sec:models} we present the 3D models used for the simulations. After a brief review of the radioactivities in section \ref{sec:decays}, in section \ref{sec:transportcode} we describe the transport algorithm. In section \ref{sec:analysis} we analyze the output and generic properties of the simulated decay lines. In section \ref{sec:87Acomp} we make comparisons with SN 1987A (\co~and \ti~decay lines), and in section \ref{sec:CasAcomp} we make comparisons with Cas A (\ti~decay lines). In section \ref{sec:thecorrelation} we present a correlation between NS kick and line shifts uncovered by the model grid. In section \ref{sec:gammafield} we analyze the internal gamma-ray field and deposition from \co, and compare this to 1D models. Section \ref{sec:infrared} ties in the observations of infrared iron lines for the conclusions of the explosive ashes morphology in SN 1987A. Section \ref{sec:discussion} provides a discussion of our results, which are finally reviewed in a summary and conclusion section \ref{sec:conclusions}. We would also like to draw to the attention of the reader the paper by \citet{Alp2019}, a companion paper to this one where X-ray and gamma-ray light curves and SEDs are studied. These two papers provide a continuation of high-energy radiation modelling of CCSNe using self-consistent 3D models, building upon models with parameterized mixing in 1D \citep{Chan1987,Pinto1988,Kumagai1989,Bussard1989,The1990,Leising1990} and 3D \citep{Burrows1995-MC}.

\section{Ejecta models} 
\label{sec:models}
\arj{We use six hydrogen-rich models and one hydrogen-poor (Type IIb) from \citet{Wongwat2015}, \citet{Wongwat2017}, Gabler et al. (in prep.) and Utrobin et al. (in prep), summarised in Table \ref{table:ejecta}. The progenitors are in most cases $M_{\rm ZAMS}=15$ \msun~single stars, from \citet{Woosley1988} (model B15, BSG progenitor), \citet{Woosley1995} (W15 series, RSG progenitor), and \citet{Limongi2000} (L15 series, RSG progenitor). Model M15-7b had a BSG progenitor resulting from a 15 \msun+7 \msun~merger \citep{Menon2017}. All progenitors except the \citet{Menon2017} one was evolved without wind mass loss but this is expected to be small at these main-sequence masses and LMC metallicities. The explosion simulations used the \texttt{PROMETHEUS} code \citep{Fryxell1991,Mueller1991}, with inner boundary neutrino luminosity functions and ray-by-ray gray neutrino transport \citep[see][for details]{Wongwat2013}. An important distinction for the two-model sets (W15 and L15, two explosion simulations run for the same progenitor) is the following: L15-1 and L15-2 had significantly different neutrino boundary luminosities, giving two significantly different explosion energies (1.7 and 2.8 $\times 10^{51}$ erg). Model W15-1 and W15-2, on the other hand, had the same neutrino luminosities and only differ in the random perturbations imposed}. 

\begin{table*}
\centering
\begin{tabular}{ccccccccccc}
\hline
Model & Time & $E$  & $M_{\rm ejecta}$   & $M$(\ni)    & $M$(X)     &   $M$(\ti)   & \ni~shift & \ti~shift & X shift &  NS kick $V_{\rm NS}$ \\
         &    (d)        &         ($10^{51}$ erg)          & (\msun) & (\msun) & (\msun) & (\msun) & (\kms)              & (\kms)  & (\kms)         & (\kms) \\
        
\hline
 \arj{B15-1L}  &  \arj{152} & \arj{1.4} & \arj{14.1} & \arj{0.030} & \arj{0.060} & \arj{$5.5\e{-4}$} & \arj{145} & \arj{216}  & \arj{288} &  \arj{104} \\ 
   
 \arj{L15-1L}     &     \arj{150}  &  \arj{1.7} &  \arj{13.6} &  \arj{0.031}    &  \arj{0.12}    &  \arj{$3.8\e{-3}$}    & \arj{398} & \arj{414} & \arj{388} & \arj{304}  \\ 
       L15-2        &   1.16 &          2.8          &       13.5   &  0.035 & 0.16 & $4.5\e{-3}$ &  81 & 128  & 115 & 117 \\ 
      W15-1  &            1.25  &           1.5 & 13.9  &    0.05 & 0.13 & $2.9\e{-3}$ & 430 & 510  & 482 & 672 \\ 

  \arj{W15-2L}   &    \arj{150} & \arj{1.5}  & \arj{13.9} & \arj{0.053}     & \arj{0.083}   &   \arj{$2.9\e{-3}$}     &    \arj{517} & \arj{521} & \arj{584} & \arj{719}  \\ 
  
 \arj{M15-7b1}  &    \arj{1.00}  & \arj{1.4}  & \arj{19.4} &   \arj{0.047} & \arj{0.10} &  \arj{$9.1\e{-4}$}  &  \arj{473}  & \arj{633}  & \arj{787} & \arj{688} \\
\hline
\arj{W15-IIb} &    \arj{0.44}  &   \arj{1.5}   & \arj{3.5} & \arj{0.053} & \arj{0.083} & \arj{$2.9\e{-3}$} & \arj{1380}  & \arj{1310}  &  \arj{1424} & \arj{719} \\
\hline
\end{tabular}
\caption{\arj{Ejecta models used. An ``L'' suffix means it is a long-term simulation including effects of \ni/\co~decay and reflection shocks (Gabler et al, in prep). The element shift velocities (columns 8-10, also referred to as ``bulk asymmetry'') are the magnitude of the momentum vector divided by the total element mass. The NS kicks are from Wongwathanarat et al. in prep.}}
\label{table:ejecta}
\end{table*}

\arj{The ejecta are represented on a spherical coordinate grid of size $N_r\times N_\theta \times N_\phi$, where $N_\theta=90$ and $N_\phi=180$ for all models (uniform 2$^\circ$ spacing in each angular direction). The number of radial grid zones $N_r$ varies between 871 and 1730 for the H-rich models, and equals 466 for the IIb model. The total number of grid points is between $\left(1.4-2.0\right)\e{7}$.
The Wongwathanarat simulations (L15-2, W15-1, M15-7b1, W15-IIb) have evolved the ejecta to around 1d after explosion, at which point accelerations have abated and the ejecta are close to coasting (having reached the final velocity distribution if one ignores \ni~heating). The Gabler simulations (B15-1L, L15-1L, W15-2L) have evolved the ejecta for longer (several months), and have considered the effects of radioactive decays and inner boundary reverse shock reflection. As the Eulerian simulations used a fixed outer grid radius, the maximum velocity tracked at time $t_{\rm last}$ is $v_{\rm max} \sim R_{\rm max}/t_{\rm last}$. The maximum grid velocities are between 9,000-17,000 \kms~for the H-rich models, and 26,000 \kms~for the IIb model}. 

The nucleosynthesis calculations of the input models used a 13-element $\alpha$-network 
($^4$He -- $^{56}$Ni) 
for models B15 and M15, and a 9-element network (excluding $^{32}$S, $^{36}$Ar, $^{48}$Cr, and $^{52}$Fe) for model series W15 and L15. 
\arj{\ni~is the main nucleus produced in explosive silicon burning, and these $\alpha$-networks would be expected to reproduce the \ni~production and morphology reasonably well. A significant amount of mass ($\sim$0.1 \msun) experiences, however, strong enough neutrino processing to change the electron fraction $Y_{\rm e}$, and the detailed composition here becomes uncertain due to the difficulty of simulating this in detail. If $Y_{\rm e} < 0.49$ in a cell at a time when nuclear reactions occur, reactions formally producing \ni~increase element ``X'' (a catch-all for iron-group nuclei of unspecified detail) rather than \ni. 
  For the considered explosion models here, the fraction of the X material that needs to be \ni~(in addition to that formed in non-X regions) to match SN 1987A (0.07 \msun) ranges from 30-70\%. When simulating \ni~lines, we combined the \ni~with such a fraction of the X material to have a total mass of 0.07 \msun.}

\arj{For \ti~the situation is yet more complex. In the W15 and L15 models, the amount of $^{44}$Ti computed by the $\alpha$ network is overproduced by more than a factor 30 compared to the directly observed value in SN 1987A \citep[$1.5\e{-4}$ \msun,][]{Boggs2015} as well as what is obtained in the detailed nucleosynthesis calculations\footnote{\arj{More detailed nucleosynthesis yields were calculated in post-processing \citep{Wongwat2017}. This brought the \ti~mass down from $2.9\e{-3}$ \msun~using the $\alpha$-network to $1.6\e{-5}-1.6\e{-4}$ \msun~using the large network, where the range corresponds to different treatments of the neutron excess which is not robustly inferred from the current simulations. Crucially, if one considers the $1.6\e{-4}$ \msun~result, some 95\% of the \ti~was made in neutrino-processed material, a physically distinct region to the shock nucleosynthesis region. In relation, \ni~is made to 50\% in the neutrino-processed region and to 50\% in the shock region at $Y_e=0.5$. The shock region production is the same as in the $\alpha$-network (0.05 \msun), but another 0.05 \msun~is added from the neutrino-processed region. 
The reason for the overproductio of \ti~in the hydro runs may lie in the restriction of the $\alpha$-network to 9 species in these runs, as the missing $^{32}$S, $^{36}$Ar, $^{48}$Cr, $^{52}$Fe elements can partly go into $^{44}$Ti. The mass of these four elements is too small to affect yields of dominant species like \ni, but the error in a minor species like \ti~becomes large. For line profiles, a problem arises if $^{32}$S, $^{36}$Ar, $^{48}$Cr, $^{52}$Fe are predominantly made in different regions than the \ti.}}}. \arj{In addition, the detailed nucleosynthesis runs of W15-IIb \citep{Wongwat2017} show that the bulk of \ti~is in fact made in the neutrino-processed regions. It is thus plausible that the X distribution represents the \ti~morphology better than the \ti~category itself. We therefore used the X distribution when simulating the \ti~lines.}

\arj{In addition to these issues, if one takes the spatial distribution of \ti~from the tracer particle simulations for W15-IIb, the morphology is somewhat different than obtained using the $\alpha$-network for either \ti~or X.
The mean velocity shift of the \ti~and X distributions are around 1300 \kms~using the $\alpha$-network, but 2000 \kms~using the tracers, a 50\% increase. For \ni~the corresponding values are 1380 and 1700 \kms, a 30\% increase. Although the tracer simulations have more accurate nucleosynthesis, they involve numerical errors for the tracer positions arising from the path integrations at later evolutionary times, i.e. after the nucleosynthesis is completed. We will therefore use the $\alpha$-network element distributions in this paper. A further improvement would require to carry out large network calculations coupled with the hydrodynamics}.

\subsection{Homologous structure}
\label{sec:homology}
\arj{The Monte Carlo transport algorithm developed in this paper relies on an assumption of homology for the ejecta, i.e. $V(r,t)=r(t)/t$.
The long-term simulations ($\sim$150d), and the W15-IIb model, have reached this state to good approximation. This is true also for the BSG models evolved to $\sim$1d (although the \ni~heating and reflected shock effects can still affect the structure somewhat at later times). For the RSG models at $\sim$1d this state is, however, not yet reached. The reverse shock has still not traversed the innermost ejecta, and many ejecta layers are not at a position given by the homology relation, typically being at a radius larger than $Vt$ due to the large initial radius of the RSG. For these models, we first make an assumption that the velocities on the grid are the final ones. We then use the homology relation to approximate their radius at time $t$, so $r(t) = V(r,t_{\rm final})/t$ (rather than $r(t) = r(t_{\rm final}) + V(r,t_{\rm final})/t$) which will be accurate for $t \gg t_{\rm final}$. This involves a remapping where material from different cells will become mixed.  Doing so, issues with microscopic mixing arise, and these models may not be suitable for spectral modelling sensitive to this. For gamma decay lines there is, however, no sensitivity to microscopic mixing so the procedure should be satisfactory for the purposes of this paper. Also, we will mainly focus our analysis on the models that do not rely on this remapping step (B15-1L, L15-1L, W15-2L, M15-7b1, and W15-IIb), and limit the use of the 1-day RSG models (L15-2, W15-1) to section \ref{sec:thecorrelation}.}

\section{Radioactive decays} 
\label{sec:decays}
\arj{For \co$\rightarrow$ $^{56}$Fe decays (111d decay time), we calculate emergent line profiles in the 846.77 and 1238.28 keV lines, and for  \ti$\rightarrow$ $^{44}$Sc decays (85y decay time) in the 67.87 and 78.34 keV lines. The daughter nucleus of \ti, $^{44}$Sc, decays at 1157 keV within a few hours (no other significant radiation), and we calculated also this. Other \co~decay lines, such as 1037.84 keV, 1360.22 keV, 1771.35 keV and 2598.46 keV lines have branching fractions 14\%, 4.3\%, 16\%, and 17\% relative to the 847 keV line, and would be generally harder to observe. If needed, their line profiles would correspond to good approximation to the 847 keV/1238 keV line profiles at a somewhat shifted time. We do however include such other channels when calculating the internal gamma field and deposition (Section \ref{sec:gammafield}).}

\section{Transport code} 
\label{sec:transportcode}
\arj{A new version of the SUMO spectral formation code \citep{Jerkstrand2011,Jerkstrand2012} operating in 3D is under construction. The full description of this code, and results for calculations of UVOIR line profiles and spectra will be presented in a forthcoming paper. Here we focus on the methodology for the 3D photon packet transfer in the code, and results for the gamma-ray field which can be used to study the simplest case of line formation, namely that of radioactive decay lines. We will also address the properties of the internal gamma-ray field and deposition which governs the powering of the thermal gas and gives the first indications of UVOIR  line formation}. 

\subsection{Transport on a spherical grid}
\arj{Most 3D Monte Carlo RT codes, at least for supernova simulations, deploy a Cartesian grid \citep[e.g.,][]{Lucy2005,Kasen2006, Kromer2009}. However, the hydrodynamic simulations used for our input models deploy a Yin-Yan grid \citep{Kageyama2004, Wongwat2010}, and the ejecta are stored on a spherical coordinate system. As such, it is natural and desirable to carry out the radiative transfer in the same spherical coordinate system. This has several advantages. It avoids a remapping of the composition grid to a Cartesian grid, in which cell compositions would experience numeric smearing. Further, it allows for larger cell sizes at larger radii, which allows high-resolution small-cell transfer in the core where much fine-structure exists, but more efficient lower-resolution transfer at large radii where less emission or gas interaction occurs. The downside is that it is more expensive, per cell, to calculate the various cell wall distances needed in the transport. However, we have devised a scheme which is only a factor few more expensive per cell, and the more efficient gridding means that fewer cells are needed in total and overall similar or even better run times compared to the Cartesian case can be obtained. We describe the details and test the algorithm in Appendix \ref{sec:transportcodedetails}.} 

\subsection{Gamma-ray interactions}
\arj{For the gamma-ray transport we implement standard Compton scattering.} 
\arj{The two main \co~gamma decay channels at 846.77 keV and 1238.28 keV have Klein-Nishina cross section $\sigma_{\rm KN}$ of  0.34 and 0.29 times the Thomson cross section $\sigma_{\rm T}=6.65\e{-25}$ cm$^2$ , respectively. 
For the  \ti~X-ray lines the corresponding cross section is $0.8 \sigma_{\rm T}$, with just a few percent difference between the 68 and 78 keV lines, and $0.3\sigma_{\rm T}$ for the 1157 keV line.} 

\subsection{Flight-time effects}
\arj{Flight-time effects can have a small to moderate impact on the line profiles.  An unscattered photon from the very far side reaches the observer with a delay $t_{\rm delay}=\left(2v_{\rm max}/c\right)t$ relative to a photon emitted from the closest approaching point, where $v_{\rm max}$ is the maximum velocity of emitting material. Here we have $v_{\rm max}/c \sim 0.01$ for the H-rich models and $v_{\rm max}/c \sim 0.02$ for the stripped-envelope model, so $t_{\rm delay} \lesssim \left(0.02-0.04\right) t$. The relevant time-scale to compare this to is the radioactive decay time-scale, which is 111d for \co~and 85y for \ti}. 

\arj{For \co~lines in H-rich models, flight-time effects remain small ($<10$\% difference between extreme ends of line profile) for $t < 111\mbox{d} \times 0.10 / 0.02 \sim 600$d. At later times, ignoring transport-time effects can lead to non-neglegible distortions if the Compton optical depth is low. In practice, the Compton optical depth at 600d is still significant for all H-rich models (optical depths in range $0.3-2.4$), which will limit such errors also beyond 600d.
For a stripped-envelope SN, $v_{\rm max}$ is about a factor 2 higher and the 10\% accuracy threshold for \co~is reached around $\sim$300d. Modelling the \co~decay lines for times $\gtrsim$300d, correction for transport-time becomes increasingly important in this case}.

\arj{For \ti~lines, the effect stays below 10\% up to 440y post-explosion for the H-rich case, and 220y for the stripped-envelope case. For Cas A, with an age of 340y, the diametric flight time is about 14y, involving a 15\% different \ti~mass at emission time between the opposite sides. Here the time-dependence should be included}.

\arj{For the case of decay radiation, flight-time effects can be straightforwardly accommodated by a correction factor in non-timedependent runs as those done here. For optically thin or ray-tracing modes (see below), a correction to the emissivity}
\begin{equation}
\color{black} f(v_{\rm p}^{\rm emiss},t) = \exp{\left(\frac{v_{\rm p}^{\rm emiss}t /c}{\tau_{\rm decay}}\right)}, 
\end{equation}
\arj{can be used, where $v_{\rm p}$ is the projected (line of sight) velocity of the emission point (positive away from observer). For Compton scattering simulations, a correction factor is applied instead at escape time, as}
\begin{equation}
\color{black} g(v_{\rm p}^{\rm esc},t) = \exp{\left(\frac{t_{\rm travel} - |v_{\rm p}|^{\rm esc}t/c}{\tau_{\rm decay}}\right),} 
\end {equation}
\arj{where $t_{\rm travel}$ is the total travel time between emission and escape (including scatterings). When calculating depositions, this formula can also be used with $|v_{\rm p}|^{\rm esc}=0$.}

With this treatment we ignore the effects of ``live expansion'' of the ejecta during the photon transport (whereas \citet{Alp2019} includes this), but effects due to this should be truly small as $\Delta R/R = v_{\rm max}/c$.

\subsection{Optically thin mode}
\arj{For the case $\tau \rightarrow 0$, the line profile can be obtained by direct integration of the emissivities (also considering flight-time effects if desired as described above). The contribution from each cell $i$ to the observed luminosity $L_{\rm i,j,k}$ in energy bin $j$, in observer direction $k$, is}
\begin{equation}
\color{black} \Delta L_{\rm i,j,k}^{\rm thin} = j_i\left(\textbf{x}\right) \Delta \Omega_k f(v_{\rm p,i}^{\rm emiss},t)  \left(1+\frac{v_{\rm p,i}^{\rm emiss}}{c}\right)^{-1},
\end{equation}
\arj{where $j_{\rm i}$ is the emissivity (erg s$^{-1}$ cm$^{-3}$ ster$^{-1}$), $\Delta \Omega_{\rm k}= \sin{\theta_{\rm k}} \Delta \theta_{\rm k} \Delta \phi_{\rm k}$ is the solid angle, and bin $j$ is found from $E_{\rm j} = E_0 \left(1+\frac{v_{\rm p,i}^{\rm emiss}}{c}\right)^{-1}$}, where $E_0$ is the decay energy. \arj{As the spatial grids used here are quite refined, one may associate each cell with a single energy bin, for coarser grids one may need to subgrid the cells before this mapping}.

\arj{In post-processing, one in principle divides by the solid angle of the viewing direction $\Delta \Omega_{\rm k}$ to get the specific luminosity. In practice, one can omit $\Delta \Omega_{\rm k}$ at both emission and postprocessing stages since they cancel. Each emission then involves a packet of energy $j_{\rm i}=1/4\pi L_{\rm i}/N_{\rm viewers}$. For $L_{\rm i}$ we use the fraction of the total radioisotope mass in that cell. In the post-processing stage physical units are obtained by multiplying by the total radioisotope mass at the calculated time, and the decay power per unit mass. Multiplication of $\sum_i L_{\rm i,j,k}$ by the solid angle of a unit area at the distance of the observer, $1/d^2$, gives the observed flux}.

\subsection{Ray-tracing mode}
\arj{We now consider treatments when opacity is not neglegible. For decay lines, one can to good approximation deploy a pure attenuation formalism in the transport as the vast majority of Compton scatterings will lead to significant energy loss and removal of the photon from the line (for any ejecta with $v/c \ll 1$). Then,}
\begin{equation}
\color{black} \Delta L_{\rm i,j,k}^{\rm ray-trace} = \Delta L_{\rm i,j,k}^{\rm thin} \times \exp{\left(-\tau(\textbf{x}_{\rm i},\Omega_{\rm k})\right).}
\end{equation}
\arj{As the Compton optical depths decline as $t^{-2}$, one could compute the $\tau_{\rm i,k}$ matrix once and for all. However, a $10^7$ cell ejecta grid and of order $10^3$ viewing angles would require storage of a $10^{10}$ element matrix, and this would also be expensive to do I/O on. In addition, the matrix has to be recomputed every time the ejecta resolution or viewing angle grid is changed. In practice, we found that an implementation that instead computes the optical depths cell-by-cell on-the-fly was fast enough and the improved flexibility made us choose this implementation}.

\arj{In ray-tracing mode, the energy packets are propagated on straight paths, experiencing attenuation with a factor $\exp{\left(-\Delta \tau\right)}$ for each cell traversed, where $\Delta \tau$ is the optical depth of that cell between entry and exit point. At high optical depth, this may save computing time compared to calculating the full path optical depth, if one allows a termination of the trajectory above a certain attenuation.}

\subsection{Compton scattering mode}
\arj{To obtain the internal gamma-ray field, line profiles to the highest accuracy \citep[see e.g.][for 1d cases]{Wilk2019}, and emergent broad-band SEDs and light curves \citep[see][]{Alp2019}, the full Compton scattering process needs to be followed. For this standard Monte Carlo formalism is used, with comoving frame energy transformations \citep[see e.g.,][]{Lucy2005, Jerkstrand2011, Jerkstrand2014} cell-by-cell until a scattering occurs. We do not treat photoelectric absorption as its effects have been shown to be minor for the energies considered here ($>68~\mbox{keV}$) \citep{Alp2018} and does not affect the observables we are focusing on}. 

\arj{The packeting details are now somewhat different. The energy of each packet is $L_{\rm i}/N_{\rm i}$, and random draws determine the direction of send-off (which is isotropic).  At postprocessing, division by the solid angle of the escape direction bin, $\Omega_{\rm k}$ now gives the specific intensity}. 

\arj{The differential Compton scattering cross section is \citep{Rybicki1979}} 
\begin{equation}
\color{black} \frac{d\sigma(E_{\rm in},\theta)}{d\Omega} = \frac{r_0^2}{2}\frac{E_{\rm out}^2}{E_{\rm in}^2}\left(\frac{E_{\rm in}}{E_{\rm out}} + \frac{E_{\rm out}}{E_{\rm in}} - \mbox{sin}^2 \theta\right), 
\label{eq:angular}
\end{equation}
where 
\begin{equation}
\color{black} \frac{E_{\rm out}}{E_{\rm in}} = \frac{1}{1 + \frac{E_{\rm in}}{m_ec^2}\left(1-\cos{\theta}\right)}. 
\label{eq:eloss}
\end{equation}
\arj{The scattering angle is drawn by solving $A(\theta) = z$, where $A$ is the cumulative probability distribution, $A = \int_0^\theta p(\theta') d\theta'$. This in general depends on $E_{\rm in}$. For $z \lesssim 0.3$, $A$ is well approximated by $c_1 \theta^{c_2}$, where $c_1=0.41 + 0.15R$ and $c_2 = 1.89-0.068R$, where $R=E_{\rm in}/m_ec^2$. Thus we solve $\theta = (z/c_1)^{(1/c_2)}$. For $z>0.3$ we instead solve the second order polynomial $\theta = -b/(2c) - \sqrt{(b/2c)^2 - (a-z)/c}$, where $a=-0.18+0.077R$, $b=0.63+0.0097R$, $c=-0.078-0.012R$. These fits are accurate to within a few percent. Combined with a randomly drawn azimuthal scattering angle, the new direction vector in the stellar rest frame can be found by standard transformations.} 

\arj{Consider the first scattering of a $E_{\rm in}=847$ keV photon. From Eq. \ref{eq:eloss}, if the angle is less than 0.16 rad (9.2$^\circ$), the energy loss is less than 2\% and the photon may stay within the line (whose width is $2v_{\rm ej}/c \sim 0.02$). From Eq. \ref{eq:angular}, one can calculate that about 1\% of the photons experience such small-angle scatterings. While this is small, photons emitted from the receding side of the ejecta have high optical depth early on, and less than 1\% of them emerge if $\tau \gtrsim 4$. In this situation, the small fraction of small-angle scattered photons emitted on the approaching side may influence the red side of the line profile}.  

\begin{figure}
\includegraphics[width=1\linewidth]{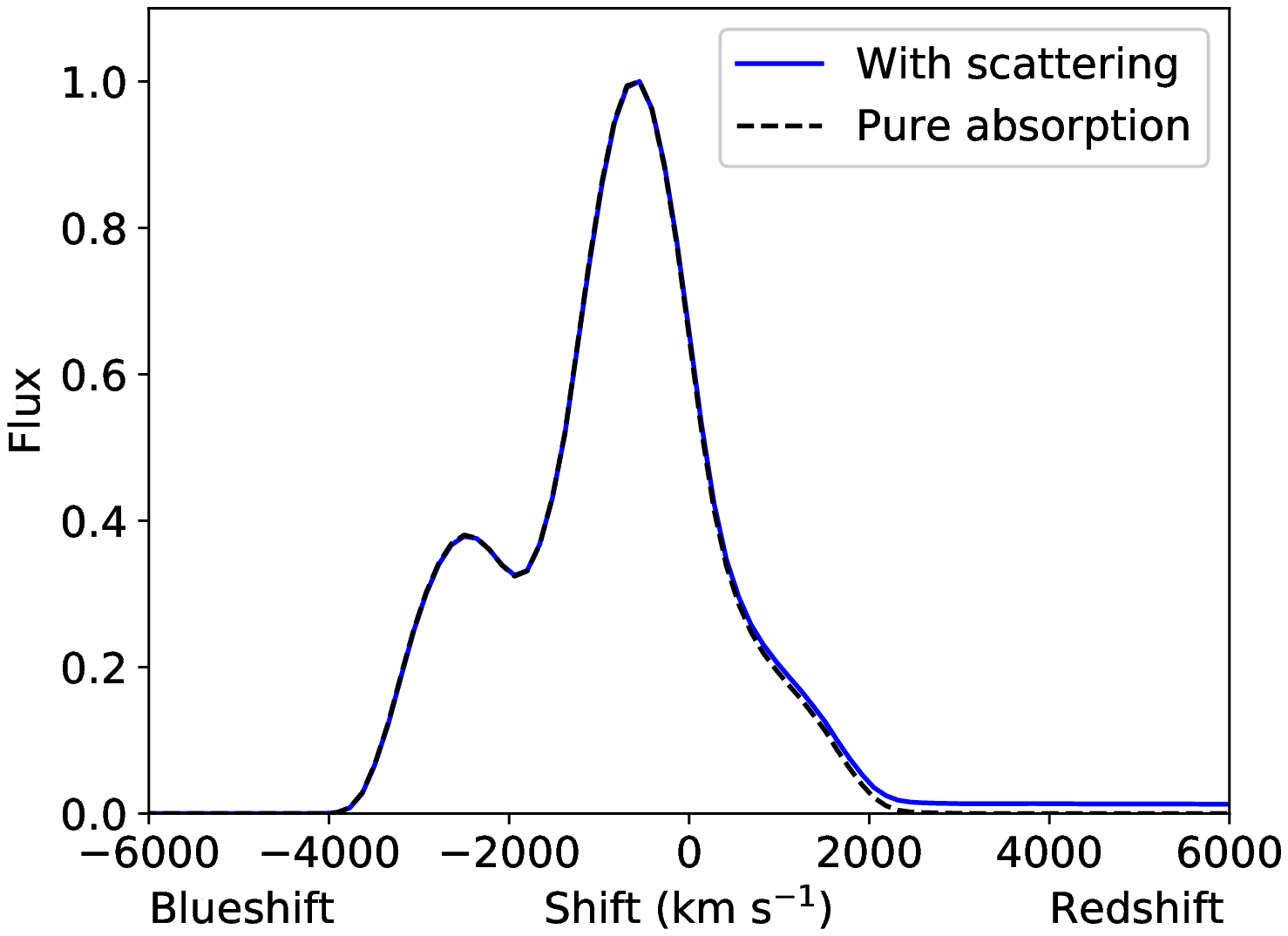}
\includegraphics[width=1\linewidth]{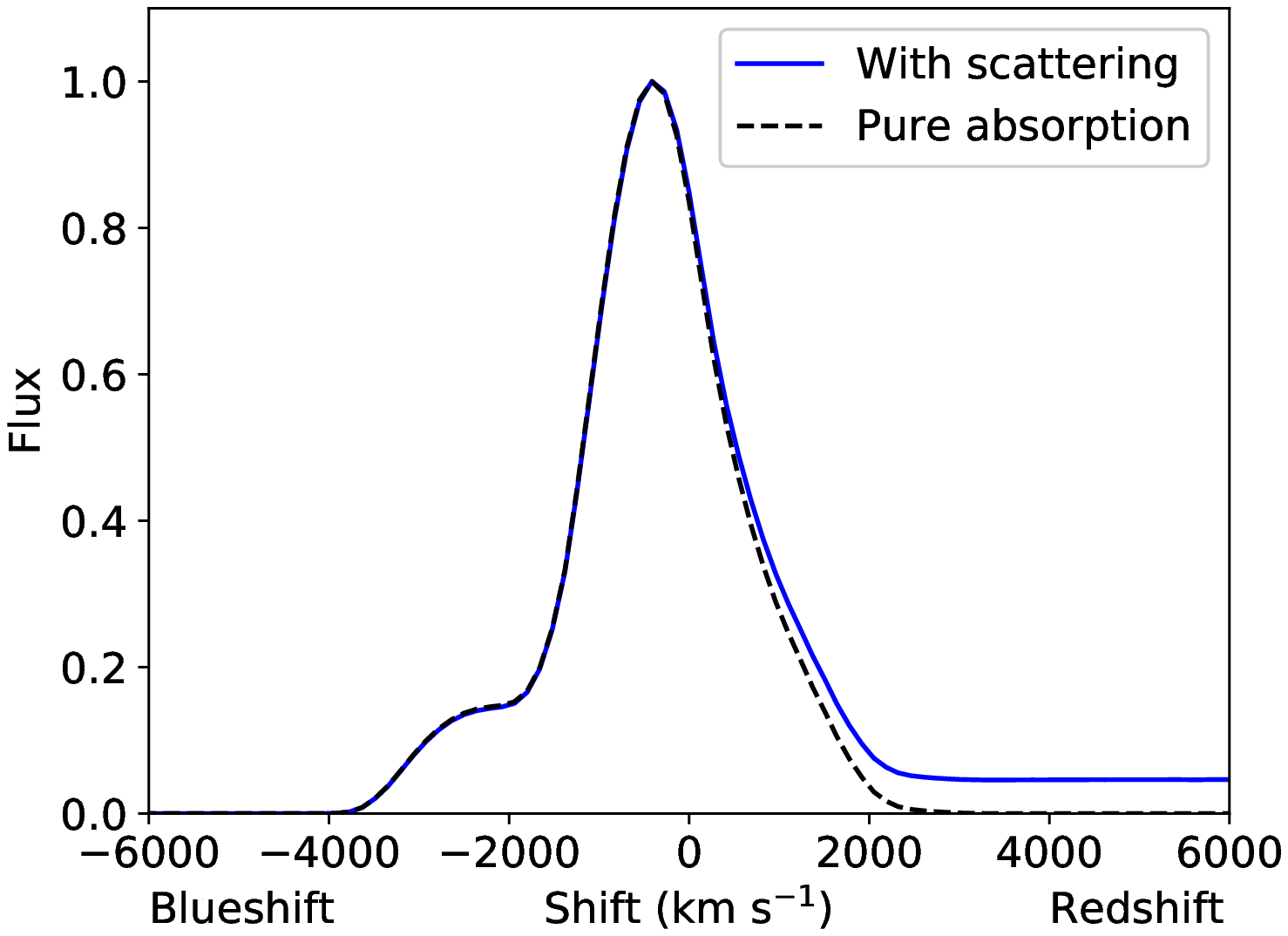}
\caption{\arj{\textit{Top}: Comparison of \co~line profiles at 847 keV using pure absorption treatment (black, dashed), or Compton scattering (blue, solid), where small-angle scatterings lead to a contribution to the red wing. Model B15-1L at 400d ($\tau=5.5$).  \textit{Bottom}: \ti~decay line at 3y post-explosion ($\tau=2.1$), also B15-1L.}}
\label{fig:tails}
\end{figure}

\arj{Testing whether these small-angle scatterings have any influence is straightforward, involving just comparing runs with or without them. We carried out a few such comparisons. Fig. \ref{fig:tails} shows two examples. The effect is mostly minor, adding a red excess with flux levels of a few percent of the peak flux. As the flux is slowly varying over the velocity scale of the line, for practical purposes it will probably be taken as part of the background continuum in data reduction (and in fact it is the continuum of scattered photons).
The effect on line centroid shift will depend on practical considerations how the line profile is observed. Formally, including the red tail (up to a few thousand \kms) can produce an additional redshift of several hundred \kms. But fitting a Gaussian to the line profile typically gives a much smaller effect, of order 20 \kms. Given the characteristics of gamma-ray line observations and data analysis including background and line fitting, this dim red tail would probably not influence any measurements. For purposes of calculating line centroids, it is even a possibility to use simulations without this tail (ray-tracing) as there is no fully natural way to decide a velocity cut-off if it is included.
When we do include it, we prescribe this velocity interval to $\pm$6000 \kms.}

\arj{For the \ti~lines at 68/78 keV, their lower energies mean that larger scattering angles are allowed to keep the scattered photon within the line (the energy range of the scattering emissivity is compressed to $\{0.79-1\}E_0$ compared to $\{0.24-1\}E_0$ for the \co ~lines. At 68/78 keV, scattering within 0.55 radians (32$^\circ$) keeps the photon within the line. Of order 10\% of scatterings will occur within such an angle and the effect can be larger than for \co, adding a red tail at the 10\% level}. In this paper we extract \ti~line properties only in the optically thin limit, so are therefore not affected by this problem, but if one needs to model \ti~lines with non-neglegible opacity care has to be taken for this effect. 

\subsection{Emergent spectrum} \arj{From the specific luminosity in each bin, we can construct any kind of flux quantity. By dividing by the bin width in wavelength units we get $F_\lambda$ (erg s$^{-1}$ \AA$^{-1}$). By defining $\Delta v/c = \Delta \lambda/\lambda$, we can also plot velocity shift on the x-axis instead of wavelength, and $F_v$ has the same shape as $F_\lambda$.} 

\subsection{Angular sampling}
\arj{We use throughout a uniform grid of viewing angles with 20 angles in the $\theta$ direction and 20 in the $\phi$ direction, for a total of 400 viewing angles. In this simple setup (constant $\theta$ and $\phi$ steps), the angle between viewing directions is not constant but smaller toward the poles. These polar viewing angles will therefore have more Monte Carlo noise in the Compton scattering simulations}.

\section{Analysis} 
\label{sec:analysis}

\subsection{Metrics}
\arj{From the emergent model line profiles, we calculate the following quantities. Below $C_{\rm E}$ denotes the spectral count rate (counts s$^{-1}$ cm$^{-2}$ keV$^{-1}$) for observed energy $E$, and the velocity shift is $V(E) = c \times (E_0-E)/E_0$, where $E_0$ is the rest energy of the line}.
\begin{enumerate}

\item \arj{\textit{Line shift-direct, $V_{\rm shift}^{\rm direct}$}. We measure this as}
\begin{equation}
\arj{V_{\rm shift}^{\rm direct} = \frac{\int C_{\rm E} V(E) dE}{\int C_{\rm E} dE}}.
\end{equation}
\arj{This quantity is invariant under convolutions of $C_{\rm E}$ with symmetric functions (e.g. Gaussian-like telescope PSFs), because $V(E)$ is a linear function.}

\item \arj{\textit{Line shift-fit, $V_{\rm shift}^{\rm fit}$}, the line center obtained from Gaussian fitting of the line profile within $\pm$6000 \kms.}

\item  \arj{\textit{Line width-direct, $\Delta V^{\rm direct}$}. We measure this as}
\begin{equation}
\arj{\Delta V^{\rm direct} = 2.35 \left(\frac{\int (V(E)-V_{\rm shift}^{\rm direct})^2 C_{\rm E} dE}{\int C_{\rm E} dE}\right)^{1/2}}
\end{equation}
\arj{The factor 2.35 ensures that $\Delta V^{\rm direct}$ corresponds to FWHM for a Gaussian profile.}

\item  \arj{\textit{Line width-fit, $\Delta V^{\rm fit}$}, the FWHM obtained from the same Gaussian fitting as in point (ii).}

\end{enumerate}
\arj{For our analysis we will generally use $V_{\rm shift}^{\rm direct}$ and $\Delta V^{\rm direct}$, unless otherwise stated. The line profiles are often quite far from Gaussian and fitting Gaussians has therefore no intrinsic meaning per se. It is also likely that the differences in line shifts we obtain by using Gaussian fits instead of direct integrations has no or weak correlation with the corresponding shift on observed lines, given large statistical uncertainties and low resolution typical for gamma-ray line observations. We will sometimes show both to give an idea of what kind of differences the distinction involves.}

\subsection{General properties of the decay lines}
\arj{We use the L15-1L model to study how line profiles can appear at different viewing angles. We limit ourselves initially to the \co~lines, and describe later in the section how the \co, \ti, and X distributions relate to each other}.

\begin{figure*}
\includegraphics[width=0.33\linewidth]{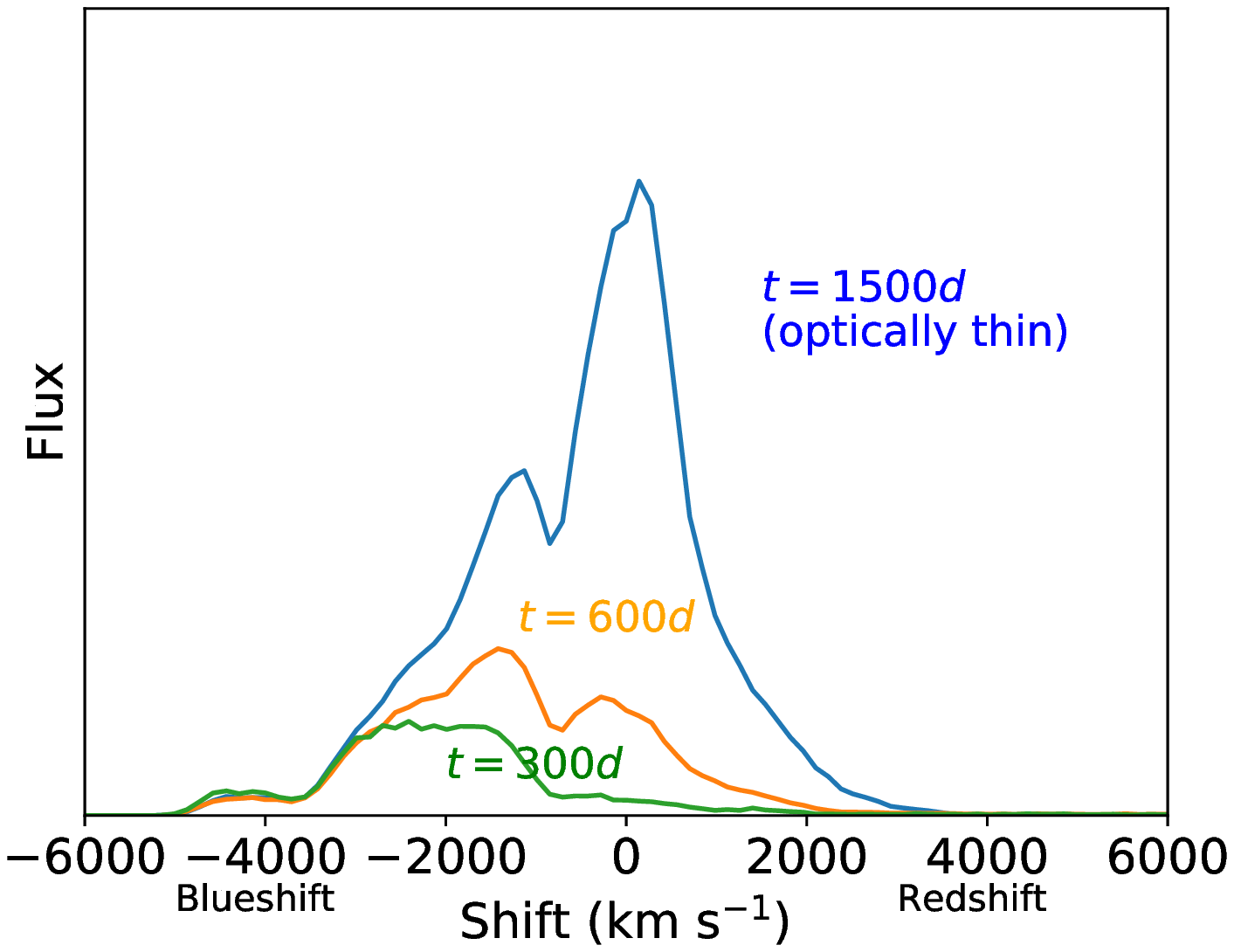}
\includegraphics[width=0.33\linewidth]{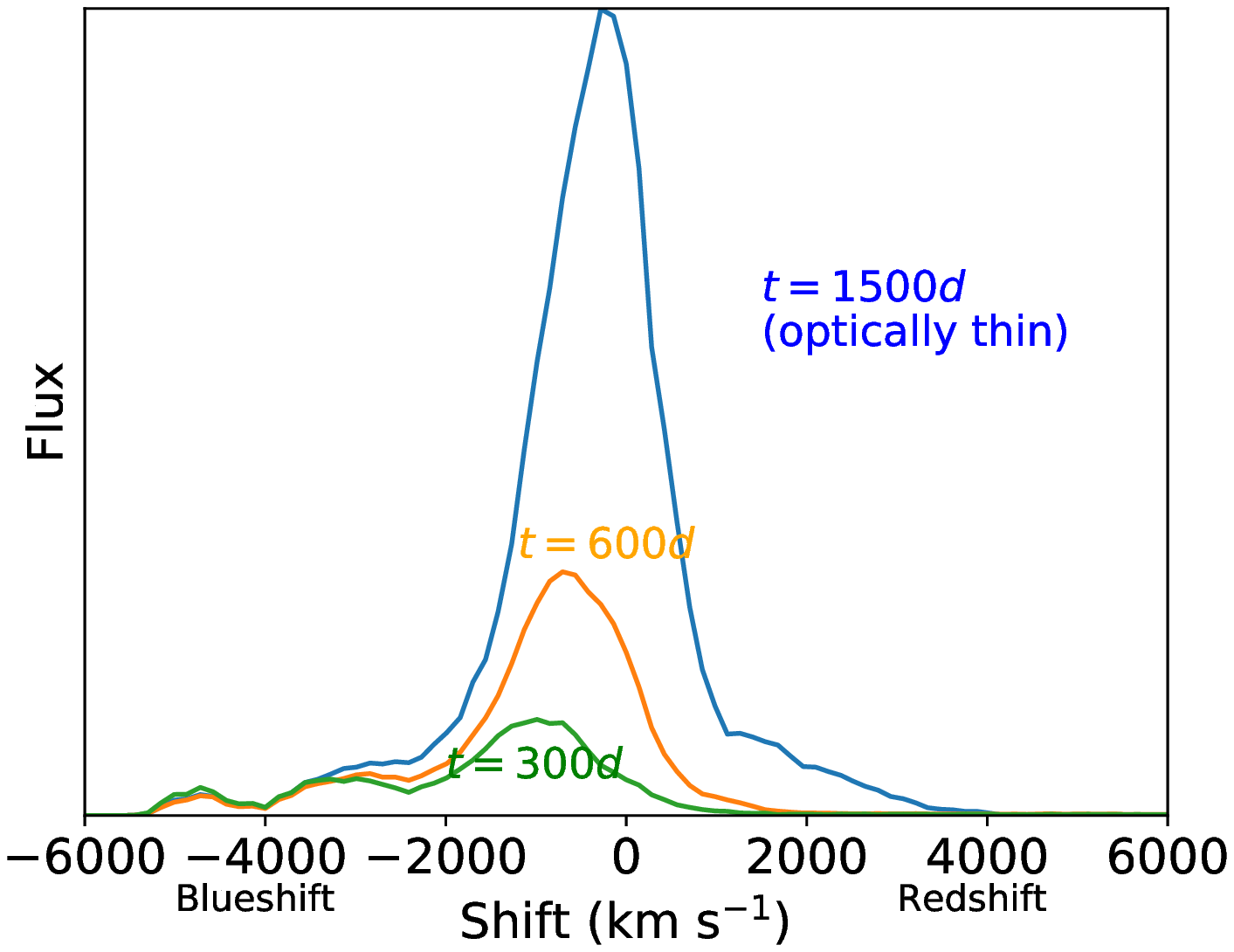}
\includegraphics[width=0.33\linewidth]{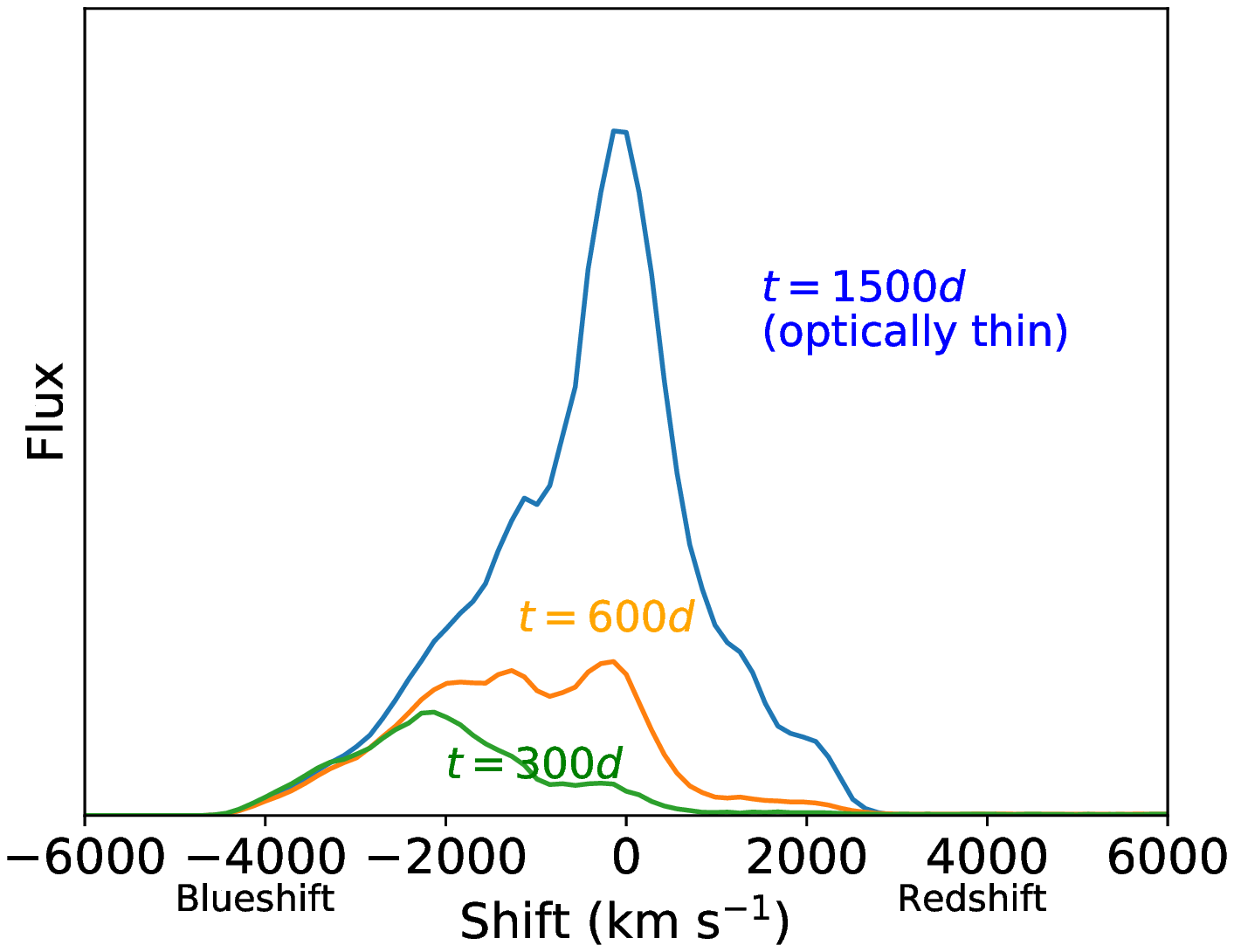}
\caption{Examples of \co~decay line profiles, using model L15-1L.  The three panels are for three different viewing angles, and in each panel the line profile is plotted for three different epochs. The lines are normalized to the decay luminosity, so the time evolution is due to increasing degree of escape only.}
\label{fig:examplesL151}
\end{figure*}

\arj{Figure \ref{fig:examplesL151} shows emergent \co~line profiles for model L15-1L at three different epochs for three different viewing angles. The line profiles can show complex and varying morphologies depending on the viewing angle. The line shapes range from roughly symmetric smooth or Gaussian-like types to more asymmetric profiles sometimes with double or even triple peaks. The figure illustrates how, as one goes to earlier times, the Compton scattering losses cause the line profile to be eaten away preferentially on the redshifted side. The profile can change significantly with this process, and viewing angle variations tend to be larger for earlier times}. 

\begin{figure*}
\includegraphics[width=0.265\linewidth]{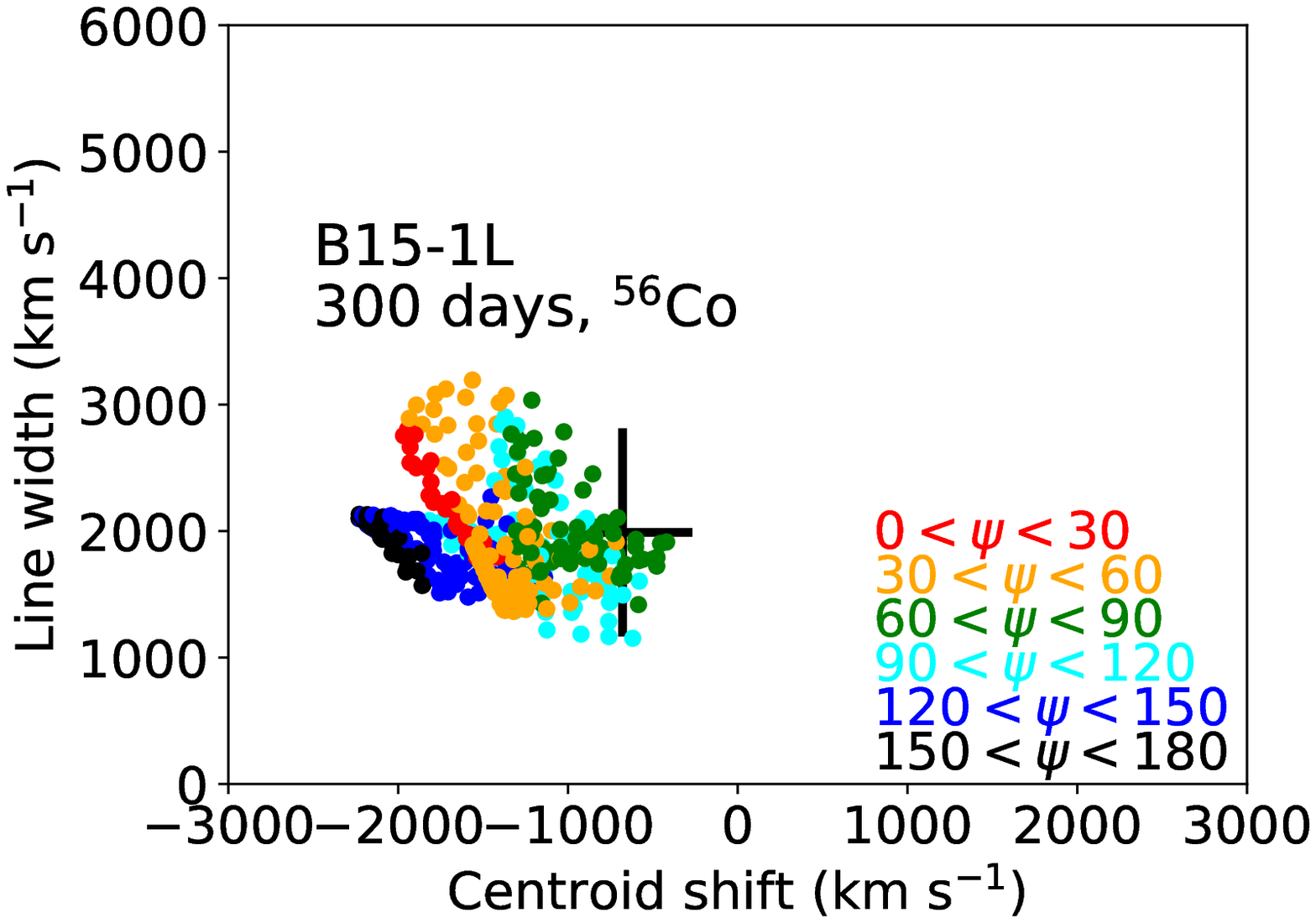}
\includegraphics[width=0.24\linewidth]{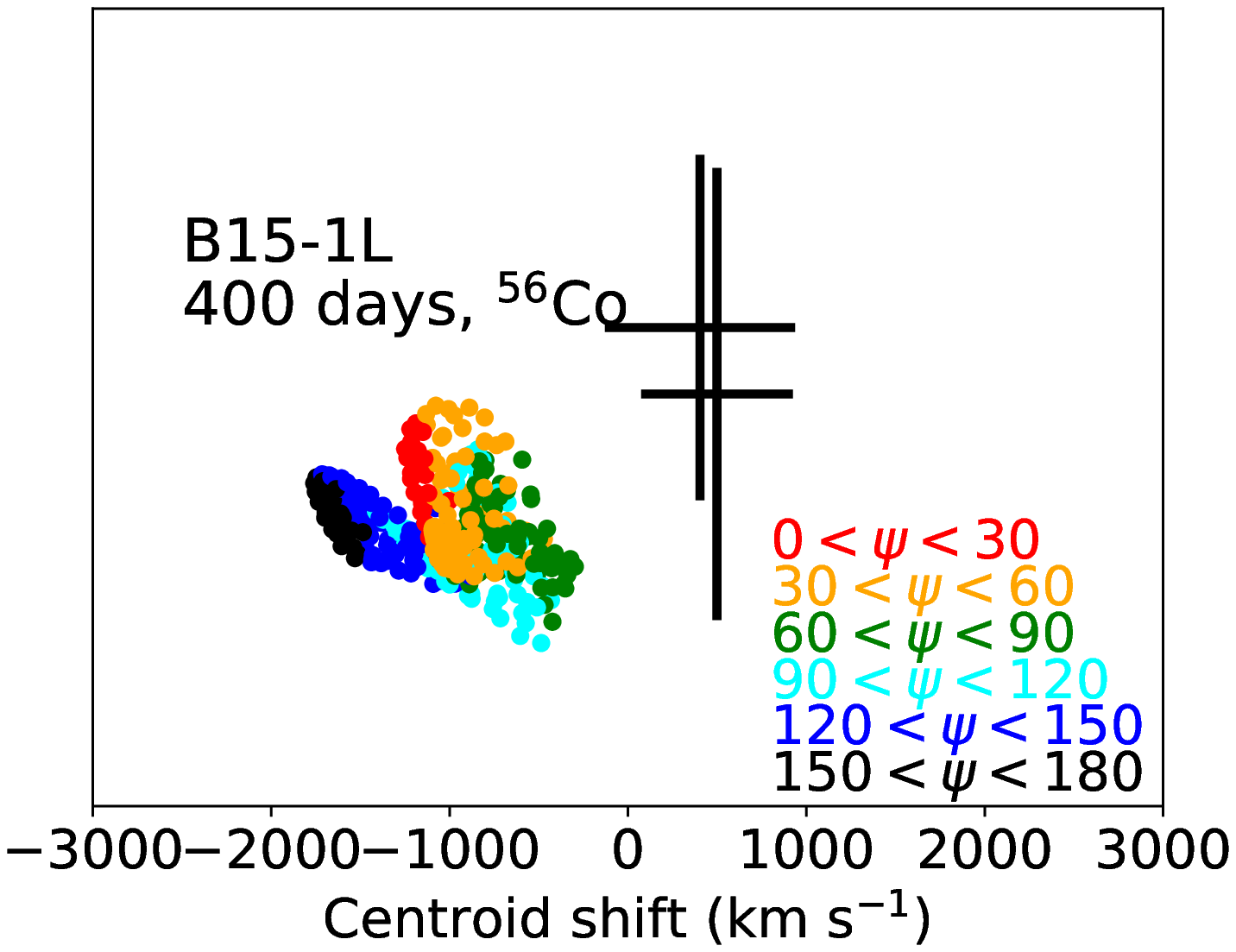}
\includegraphics[width=0.24\linewidth]{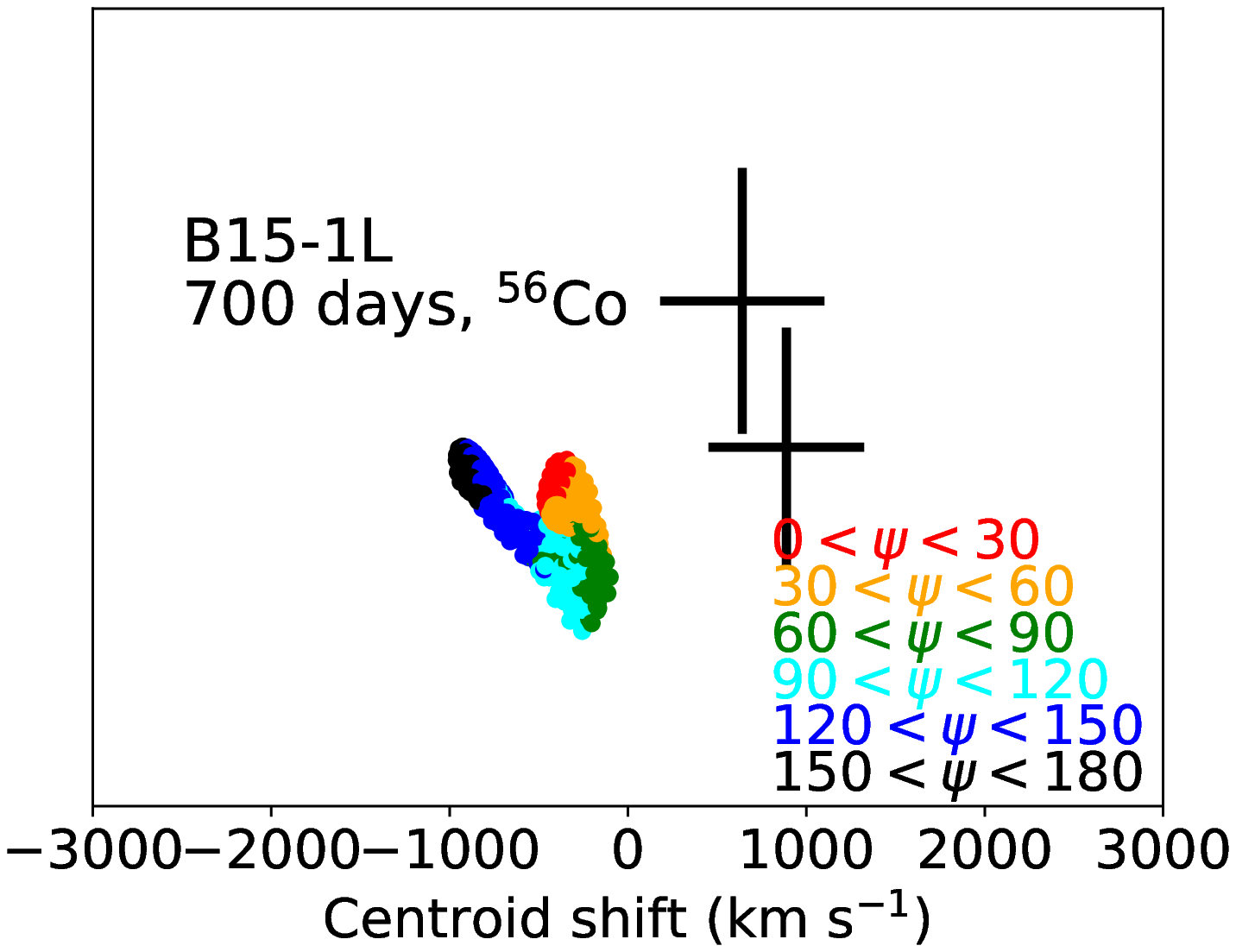}
\includegraphics[width=0.24\linewidth]{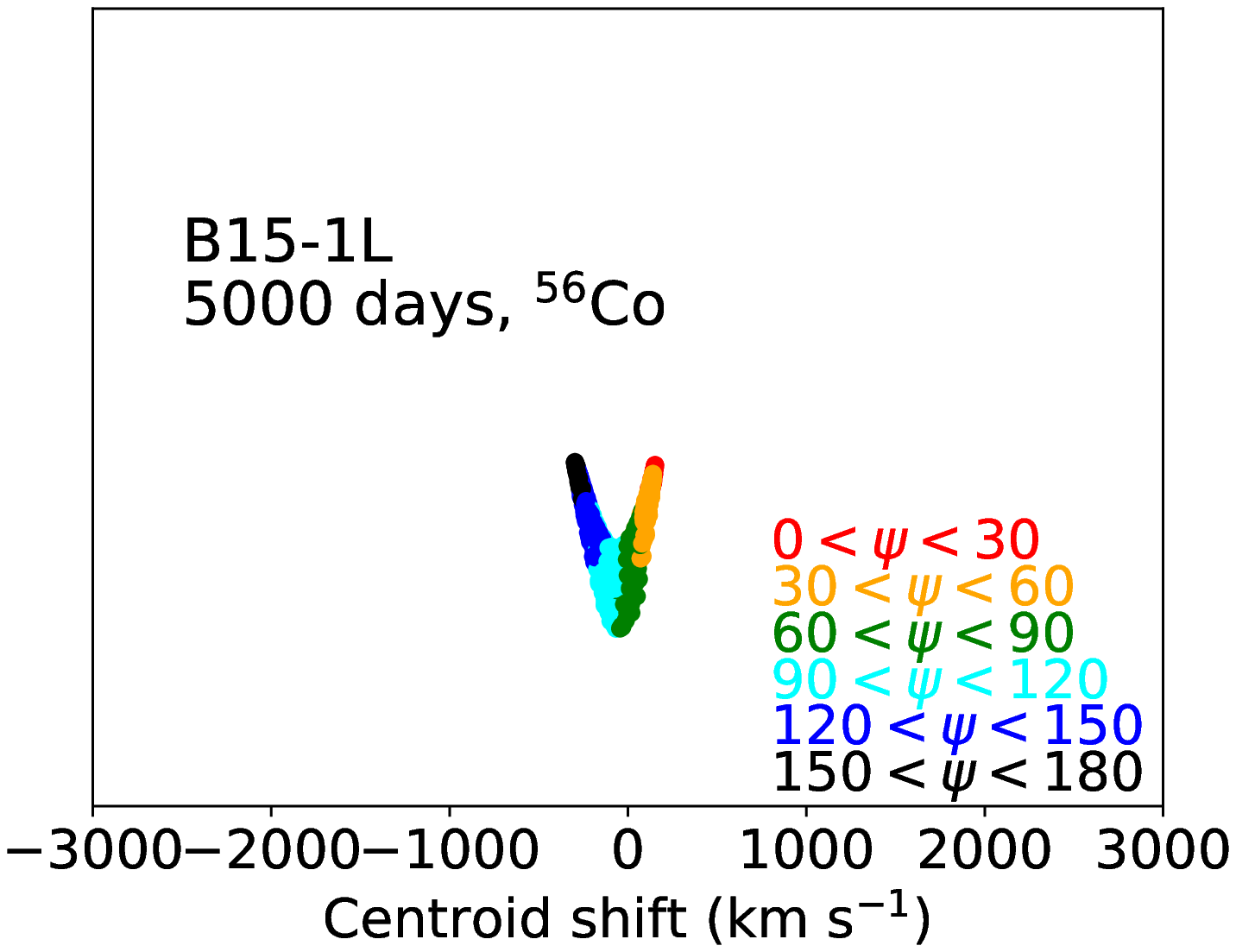}\\
\includegraphics[width=0.265\linewidth]{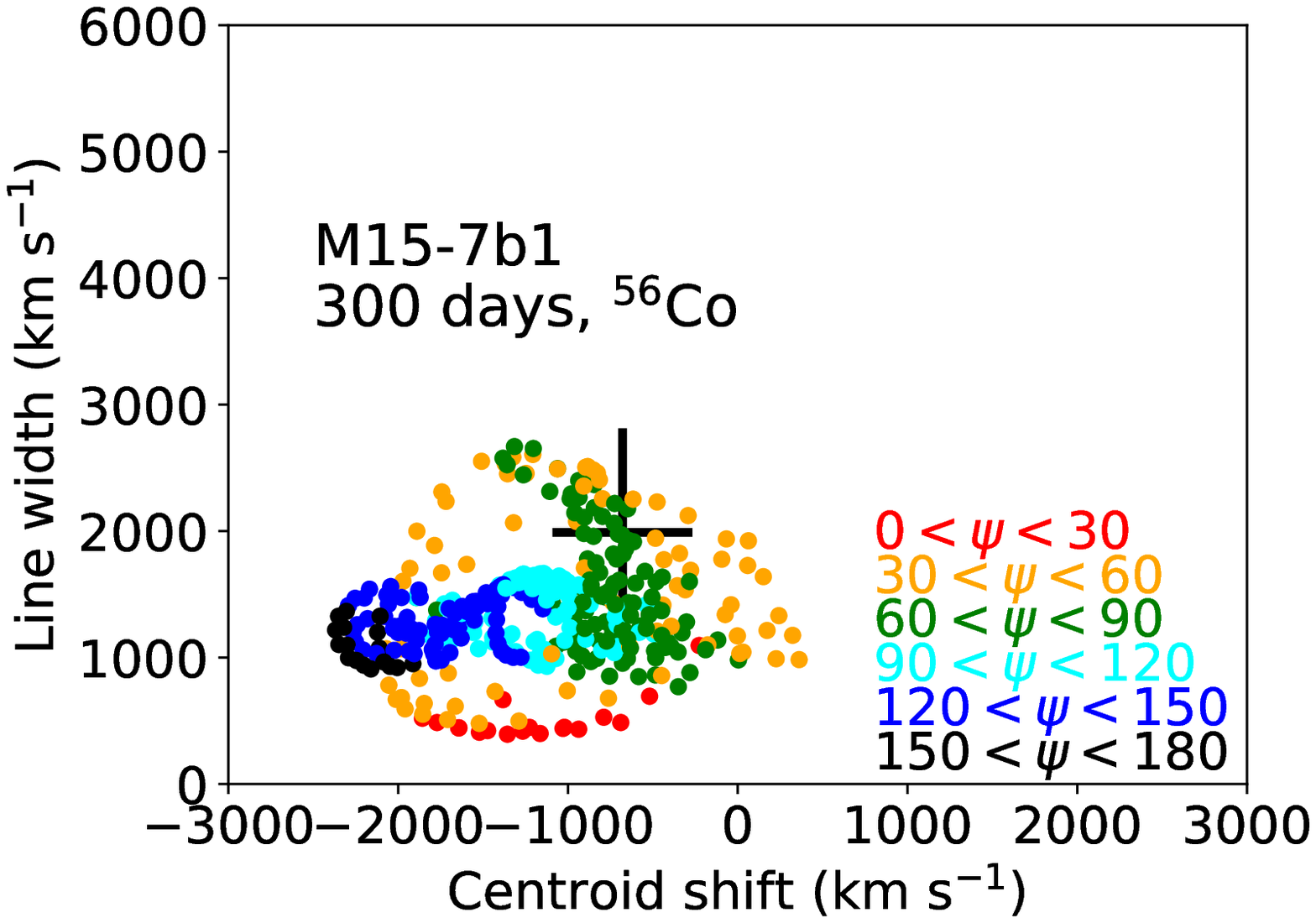}
\includegraphics[width=0.24\linewidth]{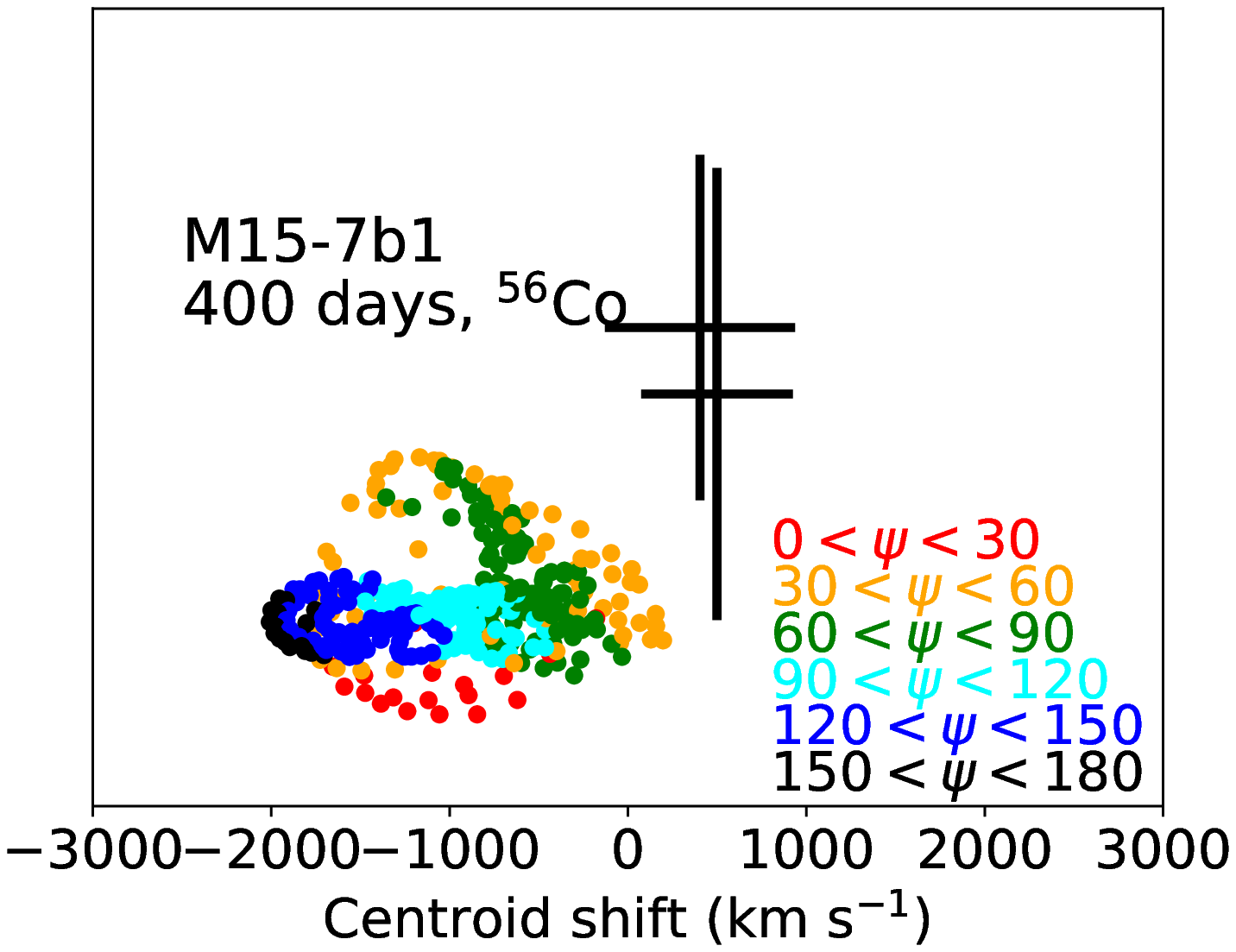}
\includegraphics[width=0.24\linewidth]{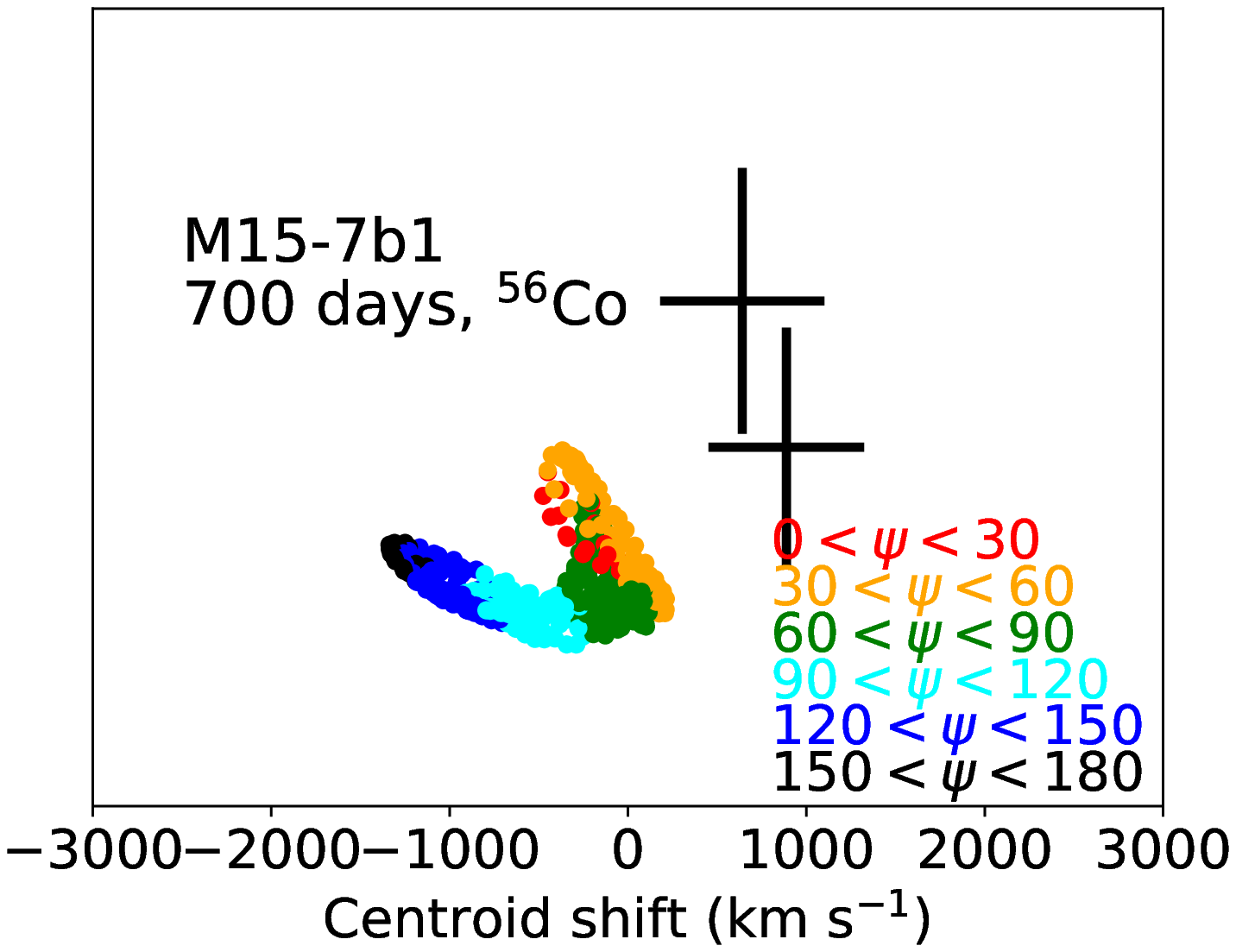}
\includegraphics[width=0.24\linewidth]{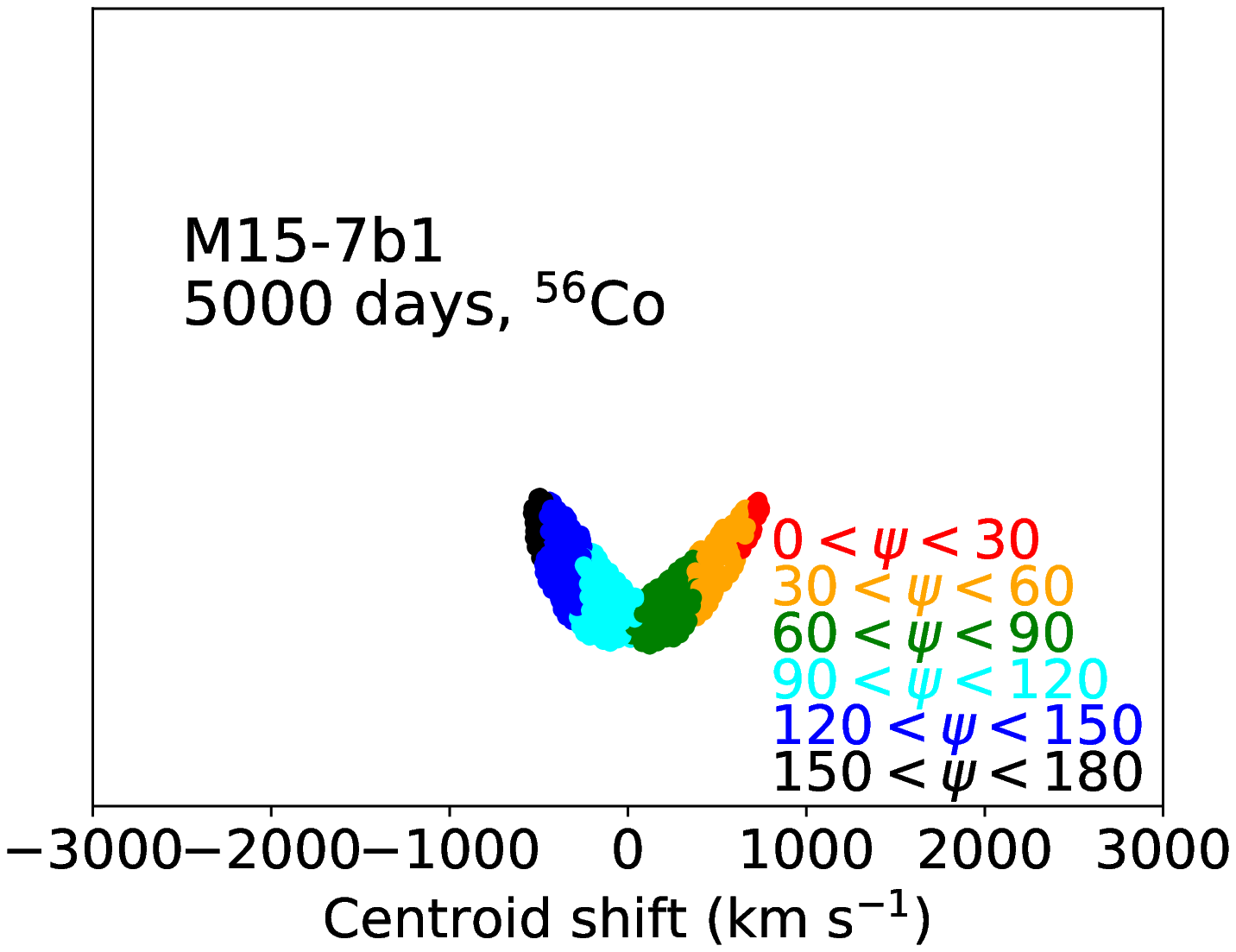}\\
\includegraphics[width=0.265\linewidth]{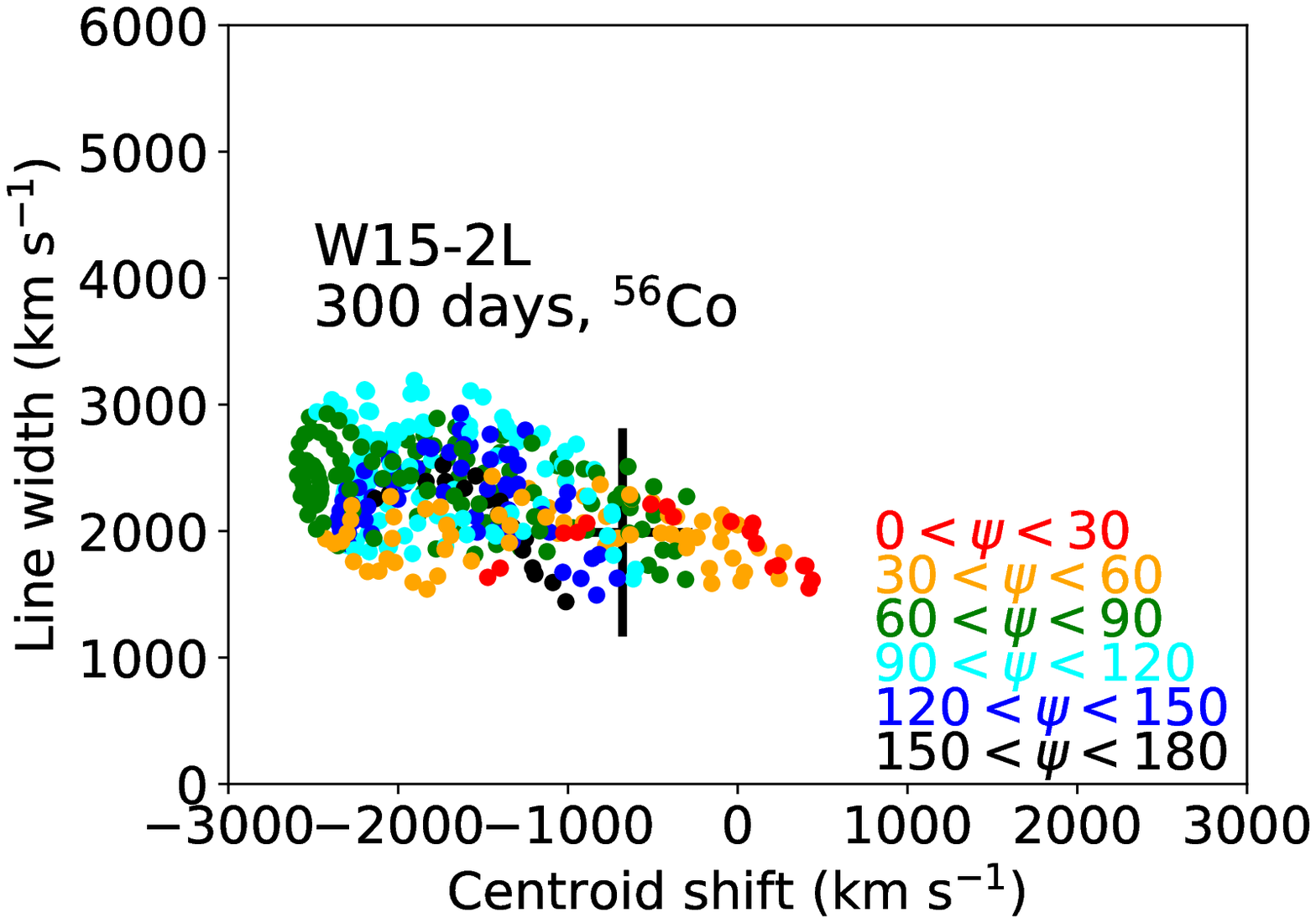}
\includegraphics[width=0.24\linewidth]{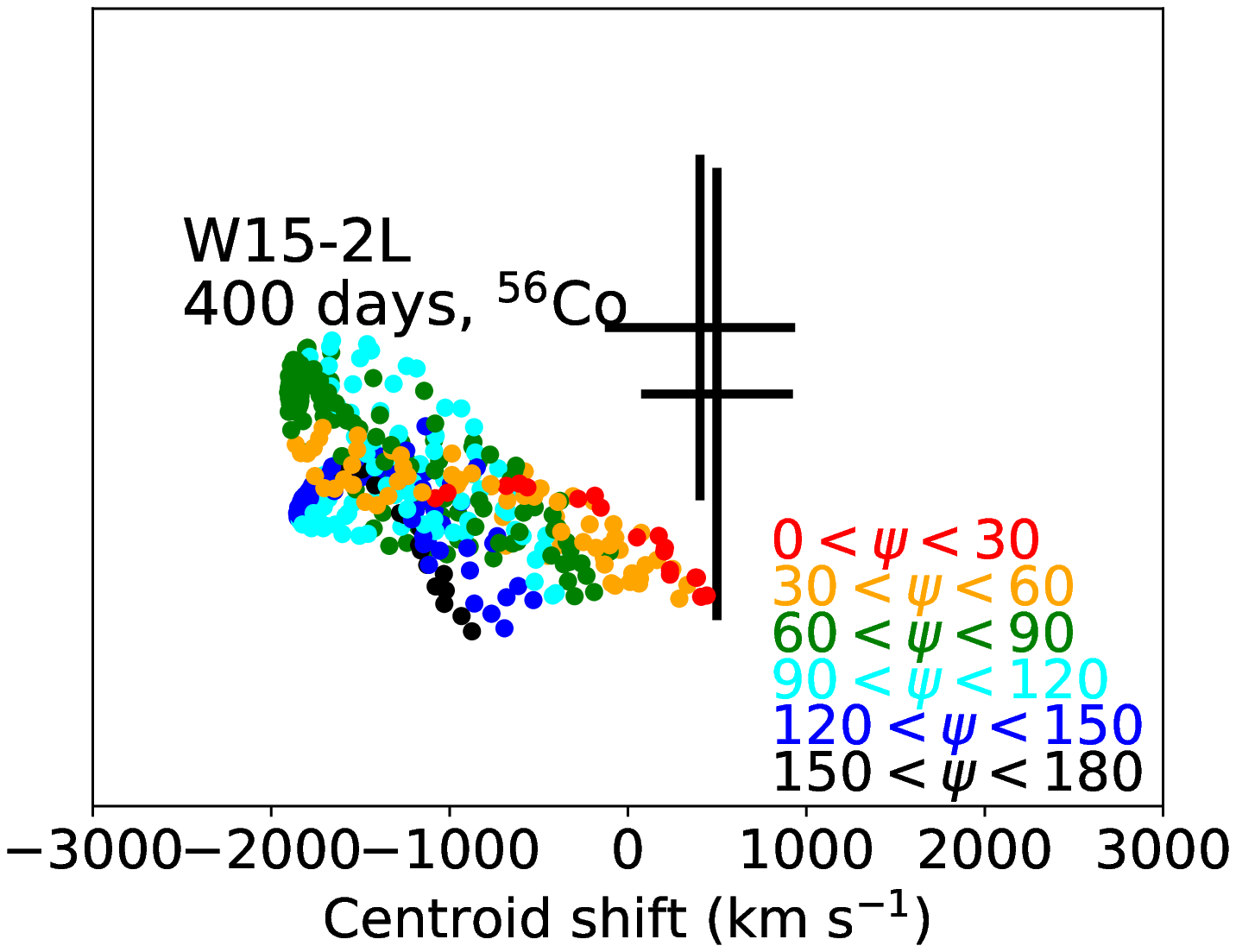}
\includegraphics[width=0.24\linewidth]{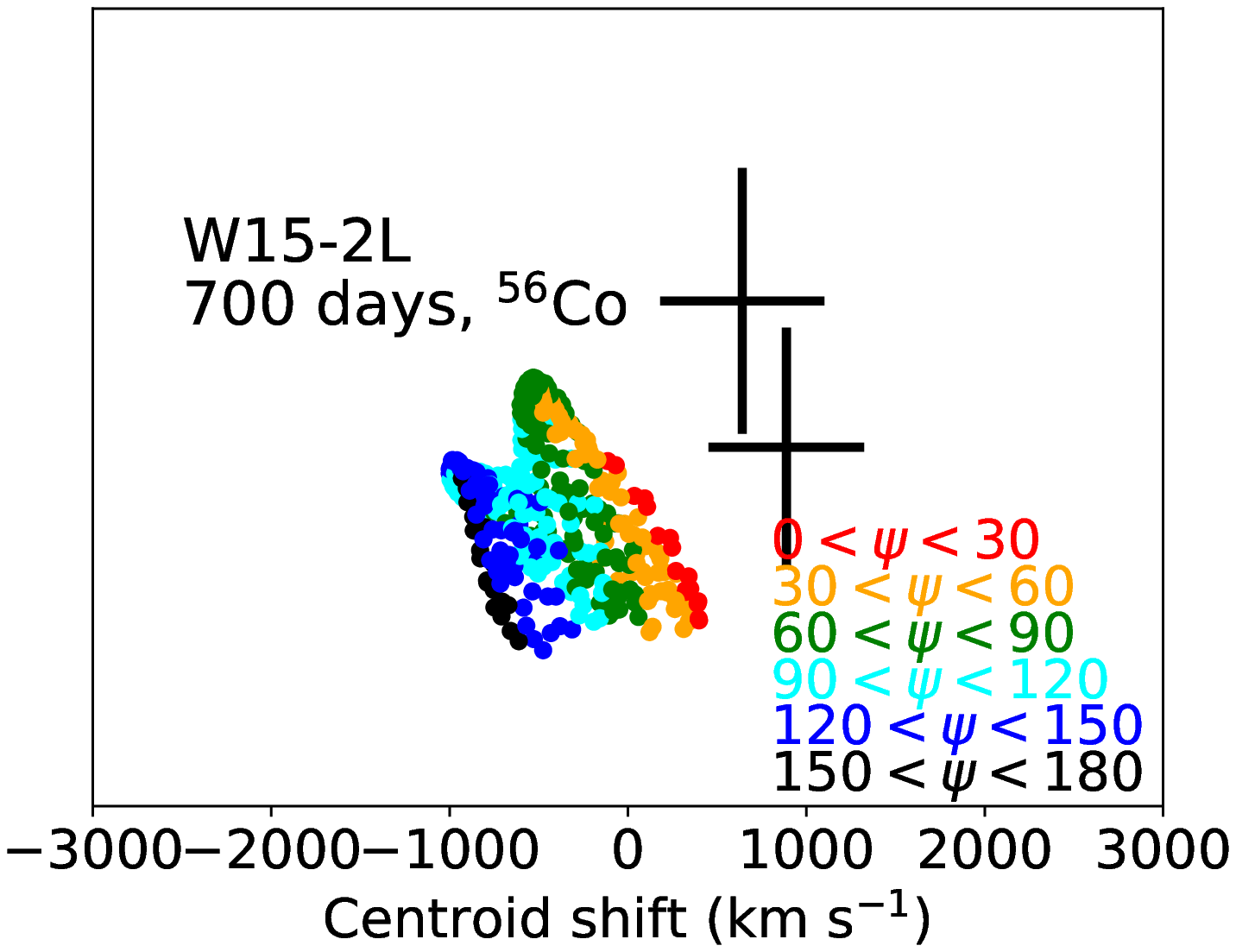}
\includegraphics[width=0.24\linewidth]{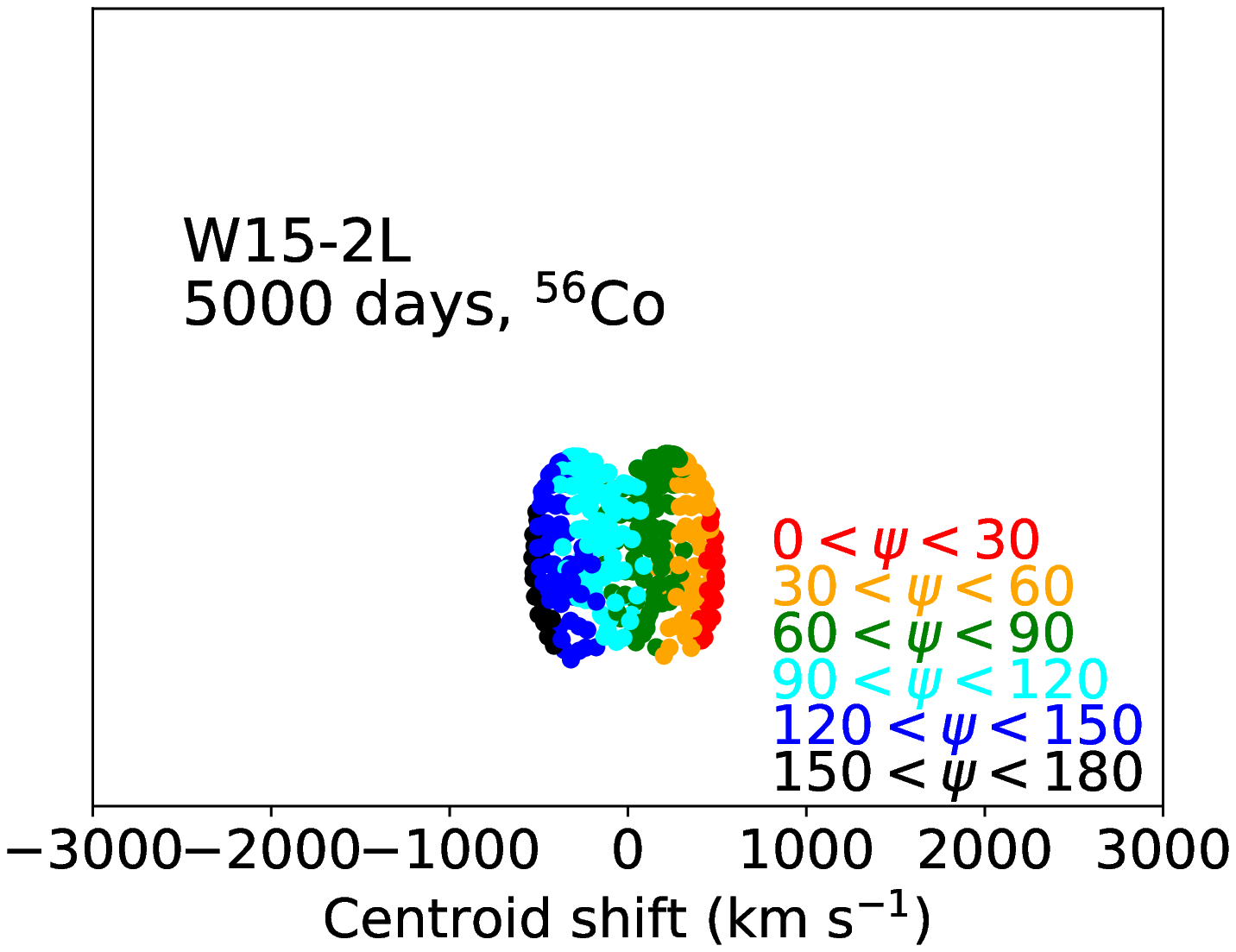}\\
\includegraphics[width=0.265\linewidth]{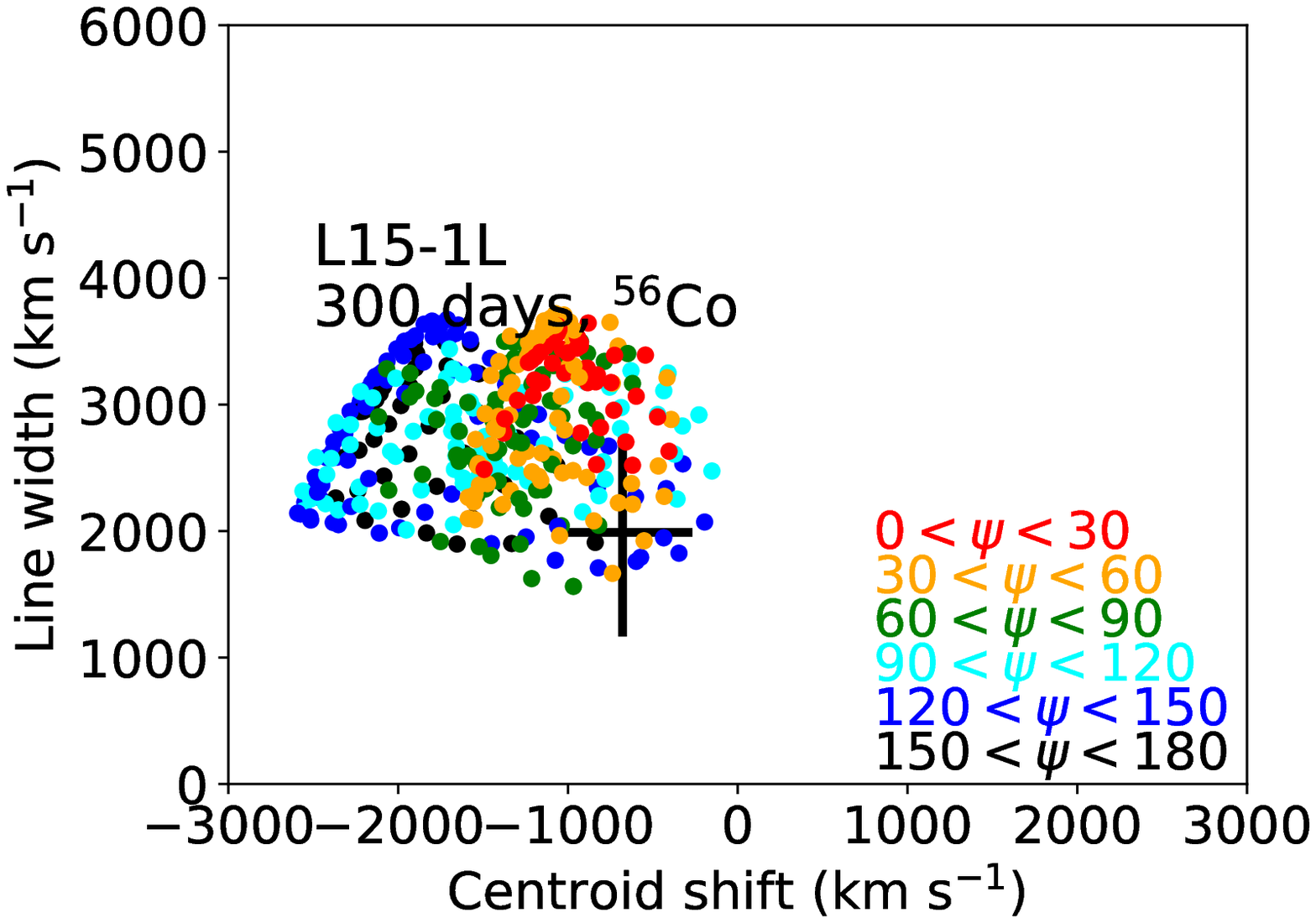}
\includegraphics[width=0.24\linewidth]{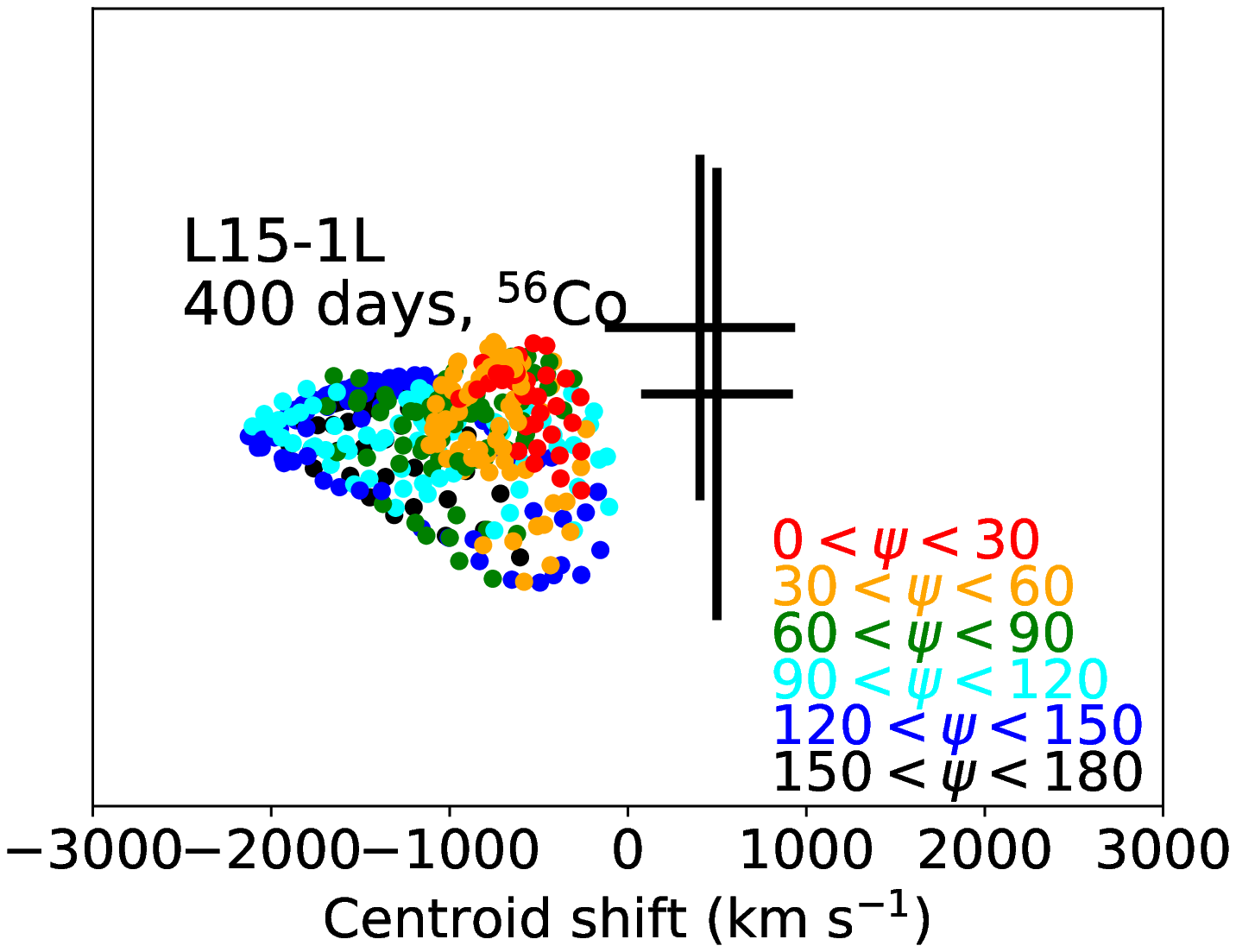}
\includegraphics[width=0.24\linewidth]{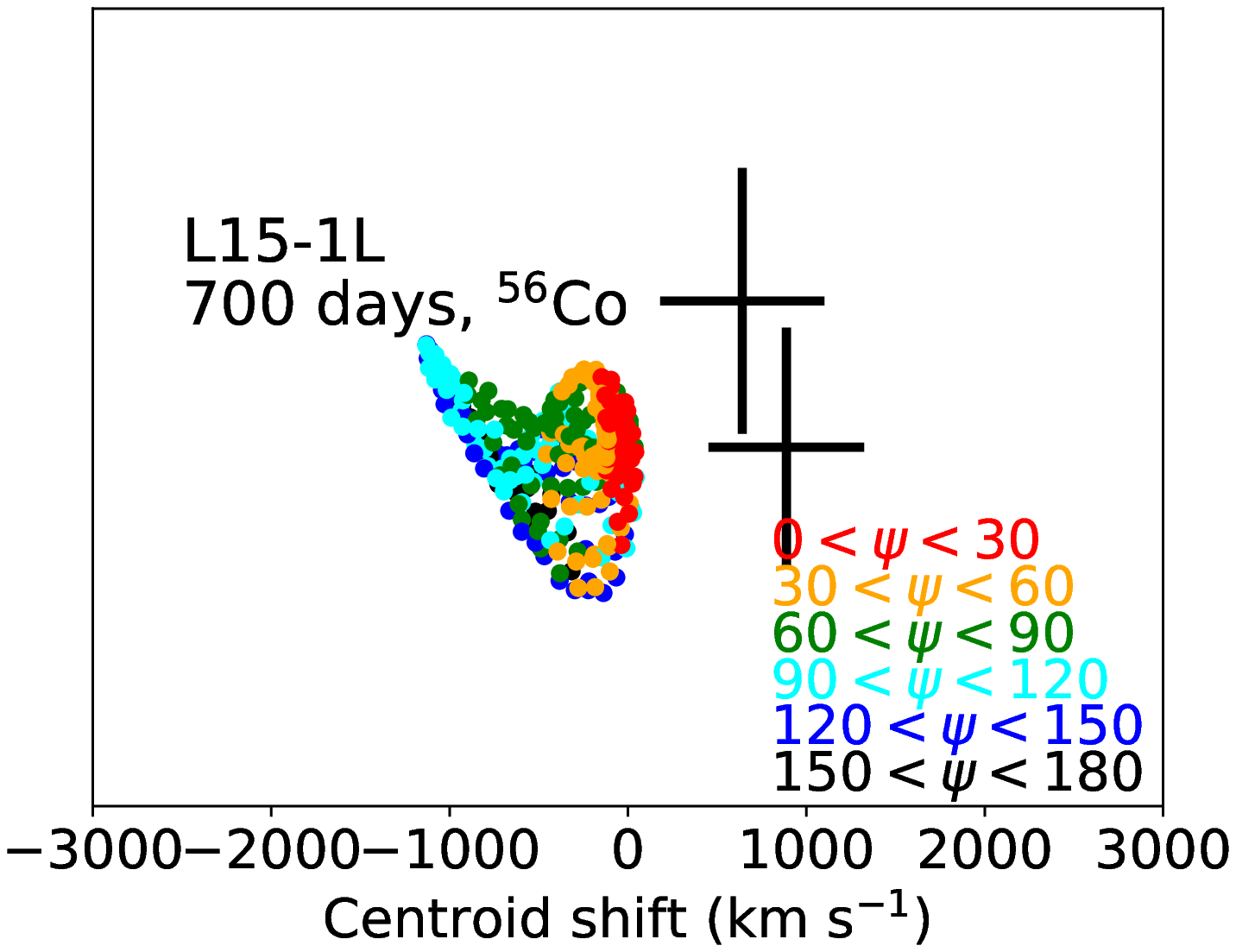}
\includegraphics[width=0.24\linewidth]{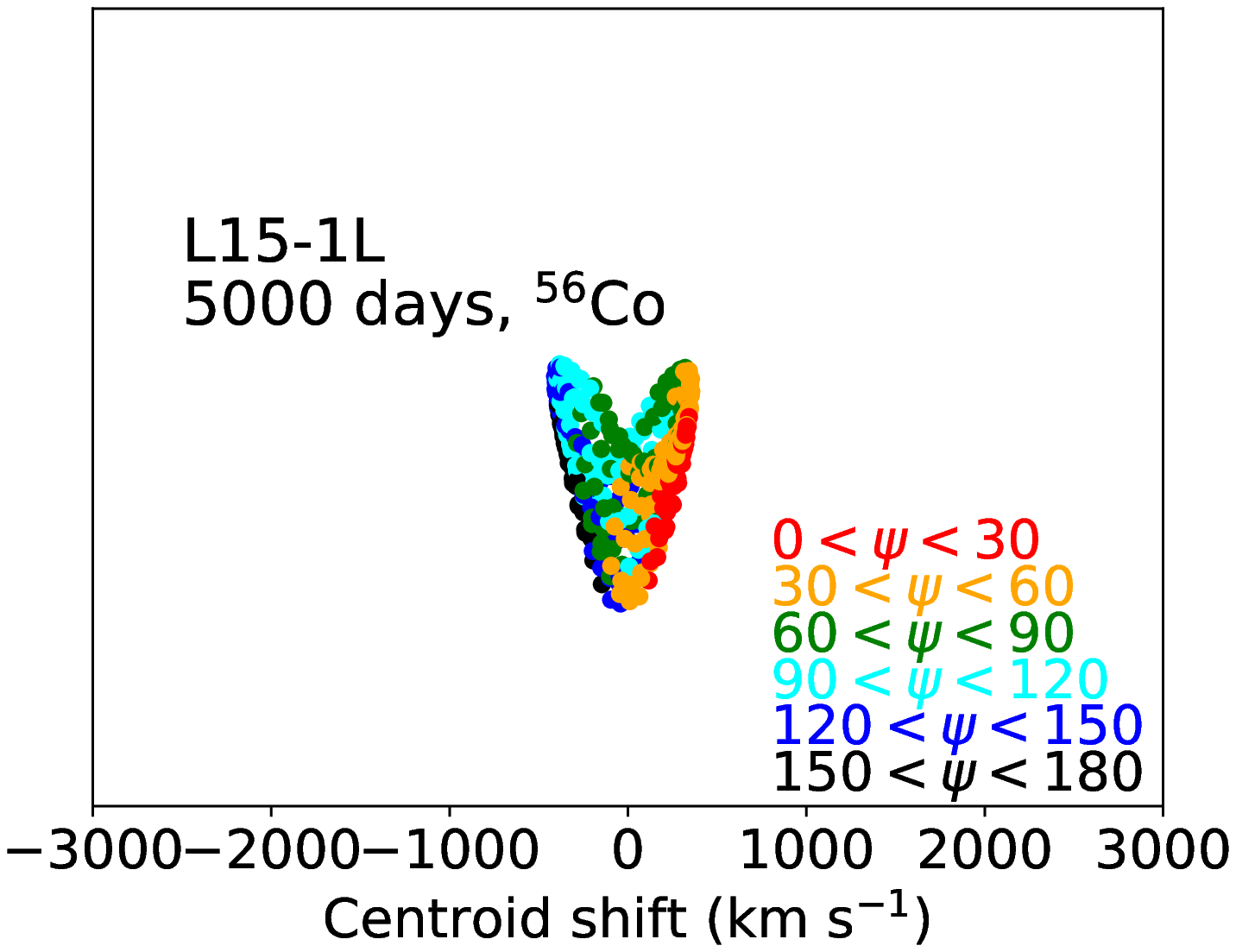}\\
\caption{\arj{Line shifts for \co~decay lines ($V_{\rm shift}^{\rm direct}$, x-axis, \textbf{redshifts are positive}) and widths ($\Delta V^{\rm direct}$, y-axis) for each viewing angle in models B15-1L, M15-7b1, W15-2L, and L15-1L at 300d, 400d, 700d and 5000d. Each viewing direction is colored according to the angle $\psi$ between the NS kick vector and the direction to the observer}. The evolution with time is due to increasing transparency. Also plotted are the observed points in SN 1987A corrected for LMC redshift \citep{Mahoney1988,Tueller1990}. Note that the models are here not convolved by any telescope response. However, as the observations are in the limit that the observed line width is much larger than the instrument broadening, the instrumental broadening has little effect, $\lesssim 100$ \kms.}
\label{fig:LP_all}
\end{figure*}

\arj{Figure \ref{fig:LP_all} shows line shifts ($V_{\rm shift}^{\rm direct}$) versus line widths ($\Delta V^{\rm direct}$) for each viewing angle, at four epochs for each model. The viewing directions with respect to the NS kick are color coded. Also plotted here are observations of SN 1987A, although those will be discussed only in Sec. \ref{sec:87Acomp}. From this figure we make the following observations.}
\begin{enumerate}
\item \arj{The typical spread in line width with viewing angle is around 1500 \kms. Thus comparing observations against 1D models to an accuracy more refined than this has limited meaning}.
\item \arj{The spread in line shift with viewing angle is of order 3000 \kms~at early times, dropping to around 1000 \kms~at late times.}
\item At early times the vast majority of models and viewing angles give blueshifted line centroids due to Compton scattering preferentially blocking the far side of the ejecta. However, some models, at some viewing angles, allow redshifted lines already at 300d.
\item \arj{With decreasing degree of Compton scattering (going to later times), line centroids move towards lower energies, but the line widths do not change much. The maximum widths can decrease somewhat ($\lesssim 20\%$), whereas the minimum values stay almost constant. This means that the line width is quite independent of uncertainties in the degree of Compton scattering, e.g. the mass and morphology of the overlying ejecta}. 
\item \arj{There exists no meaningful correlation between line shift and line width. There is a slight preference in some models that viewing angles giving larger shifts also give larger widths (especially in the optically thin limit), but the scatter is large and in other models no such association is seen}.
\end{enumerate}

\paragraph*{Differences between \co, \ti, and X line profiles.}
\label{sec:all3}
\arj{We explored the differences between \co, \ti, and X line profiles. A summary view is that there are typically no large differences between these, but they are still different enough that one should treat them separately. This also implies that our results are not strongly sensitive to the uncertainty in what part of the X distribution to use (see discussion in Sec. \ref{sec:models})}.

\section{Comparison with SN 1987A} 
\label{sec:87Acomp}
In this section we show comparisons between the model calculations and observations of SN 1987A. Note that ideally for direct comparisons it is desirable to run model outputs through telescope response functions. For the data for SN 1987A taken over 30 years ago these are, however, not readily obtainable. In addition, errors introduced by approximating response functions as Gaussians are not likely dominant sources of uncertainty in the analysis. The same comments hold for the comparison with \ti~decay line observations; here the NuSTAR response function is known, but we do not draw any conclusions based on such fine level detail that such detailed processing is warranted.

Whereas SN 1987A is known to have had a BSG progenitor \citep{Walborn1987}, we include here both BSG and RSG models in the comparisons. The principal motivation for this is that we simply have too few BSG models to make an exploration of the variations in fit quality when fundamental parameters such as mass and degree of explosion asymmetry (arising from core structure variations) vary. Including the RSG models allows us to make broader comments on what kind of ejecta morphologies are required for SN 1987A. How to get to such morphologies, enforcing a stricter requirement of a BSG envelope, is discussed in section \ref{sec:discussion}.

\subsection{\co~lines}
\arj{The \co~decay lines at 847 and 1238 keV were observed by several satellite and balloon instruments \citep[see][for an overview]{Teegarden1991}. Unresolved observations, i.e. where instrumental resolution is modest so no broadening of the line in excess of the instrumental spectral width could be measured, were presented by \citet{Matz1988, Cook1988} and \citet{Leising1990}. In principle such observations can still be used to estimate line shifts. Resolved line profiles were presented by \citet{Mahoney1988, Sandie1988, Rester1989} and \citet{Tueller1990}. Of particularly good quality are the +613d observations of the 847 keV line by GRIS \citep{Tueller1990}. GRIS had an energy resolution corresponding to $\sim$700 \kms, and the line profiles are resolved over a few energy bins, with flux error bars of order $\Delta F/F = 0.5$}. 

\begin{figure*}
\includegraphics[width=0.43\linewidth]{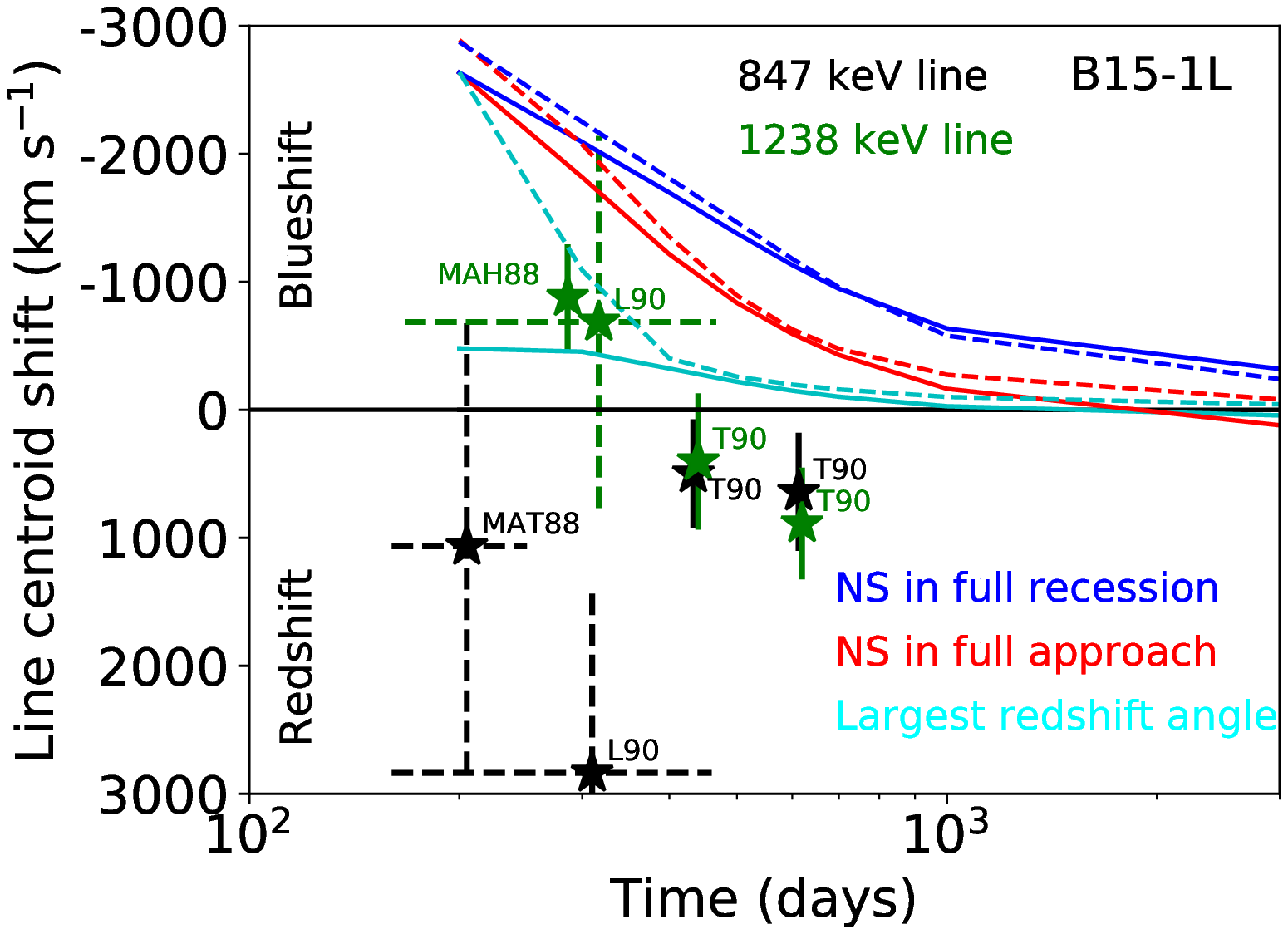}
\includegraphics[width=0.42\linewidth]{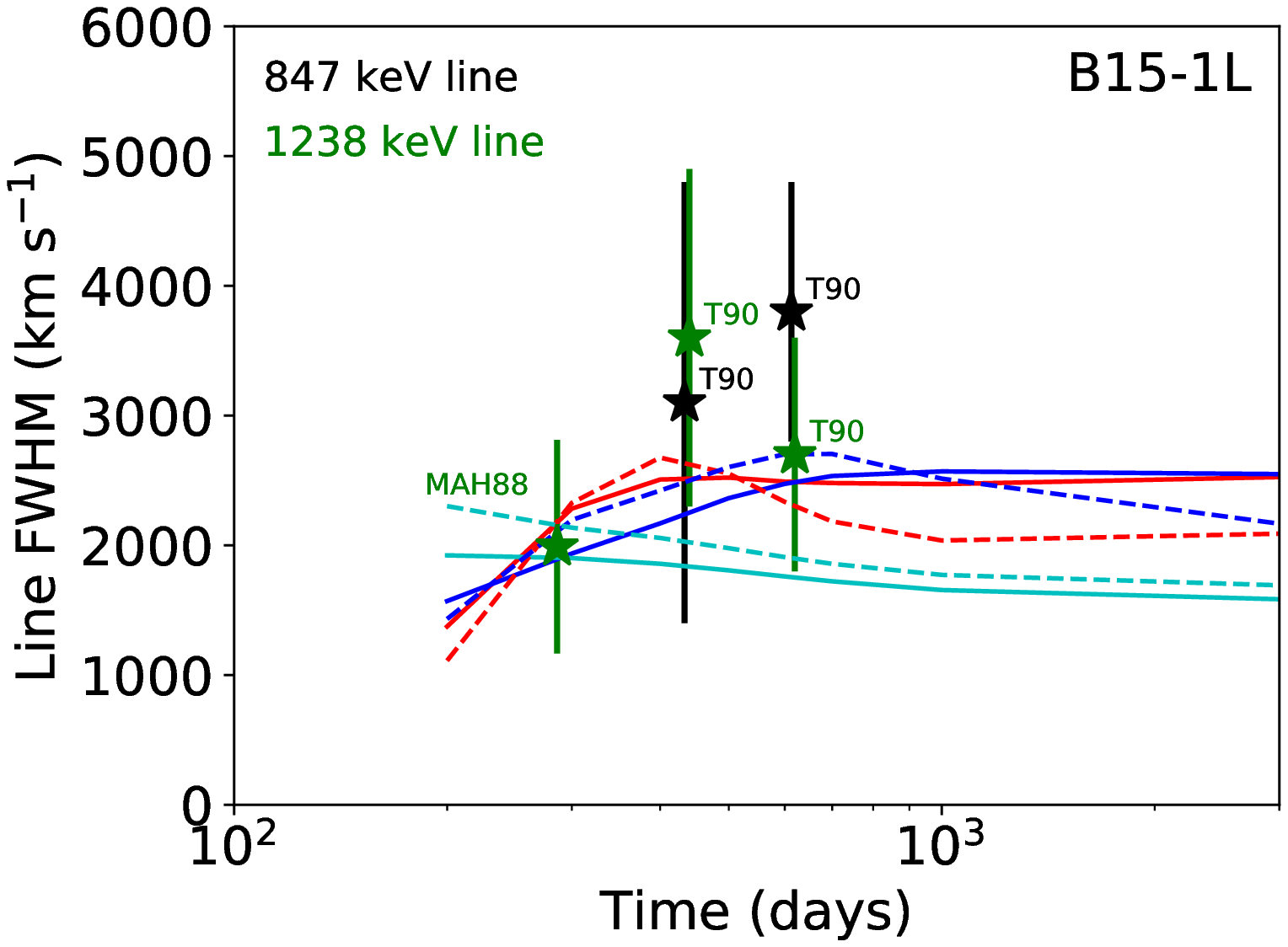}\\
\vspace{-0.1cm}
\includegraphics[width=0.43\linewidth]{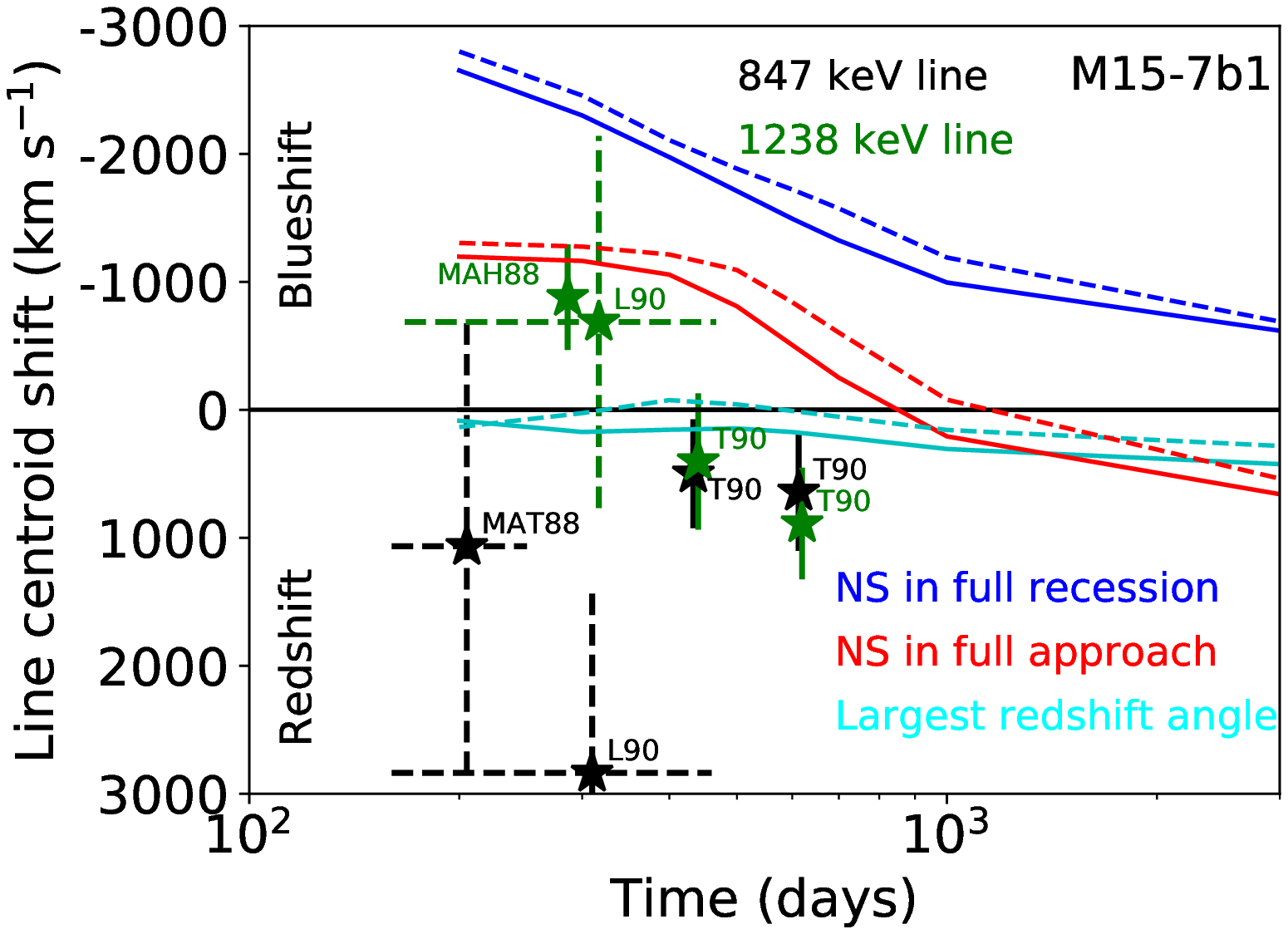}
\includegraphics[width=0.42\linewidth]{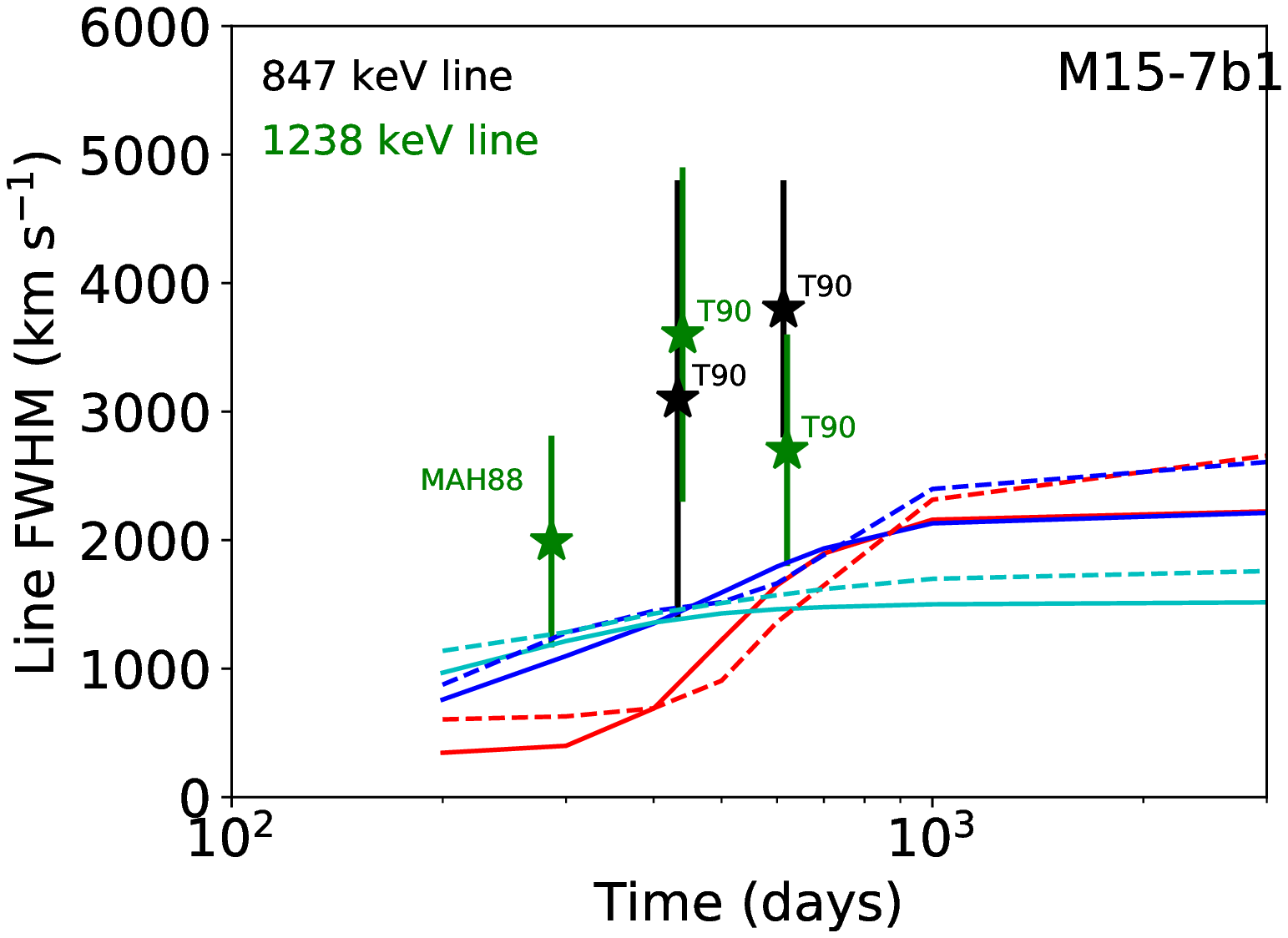}
\vspace{-0.1cm}
\includegraphics[width=0.43\linewidth]{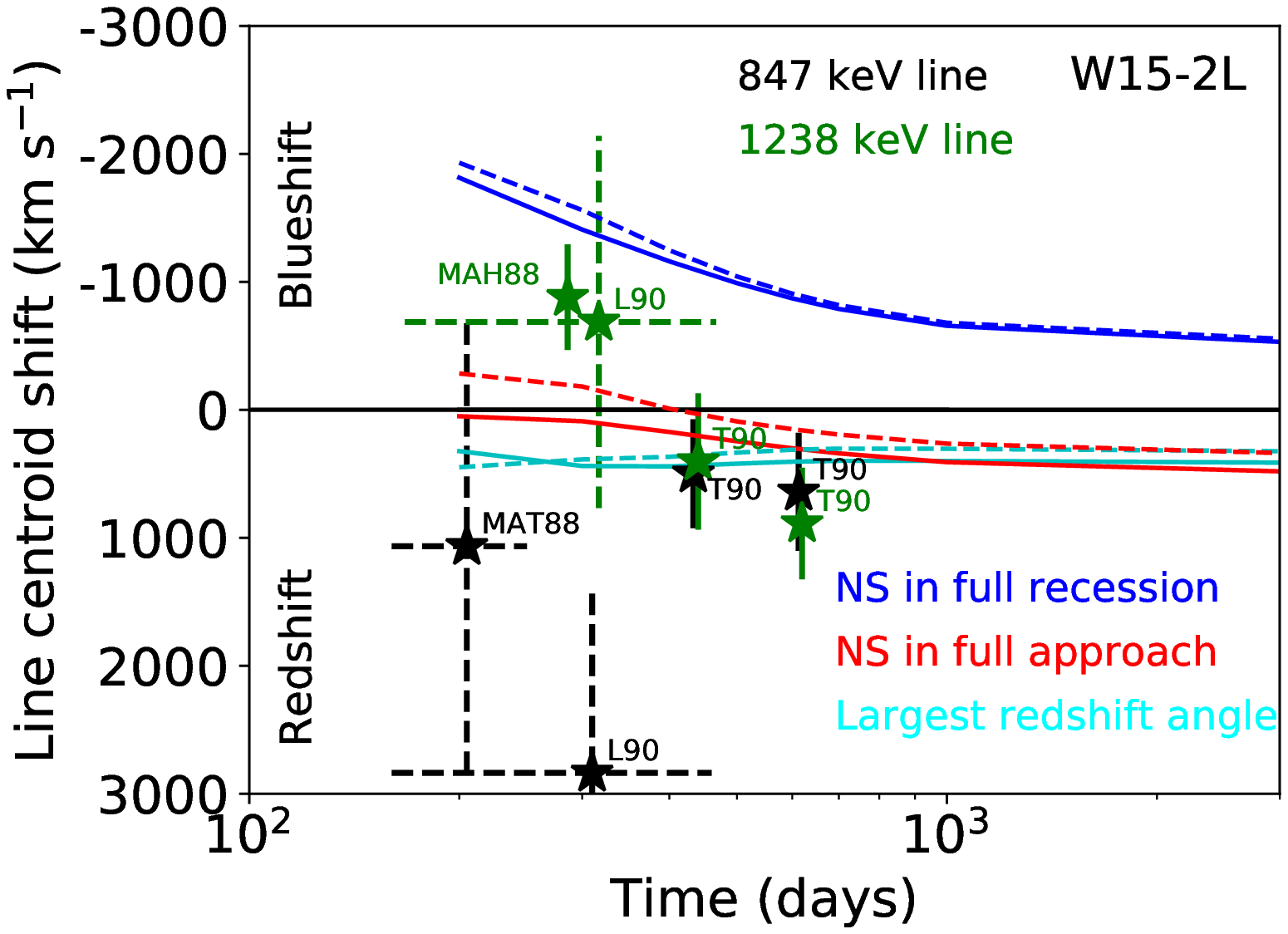}
\includegraphics[width=0.42\linewidth]{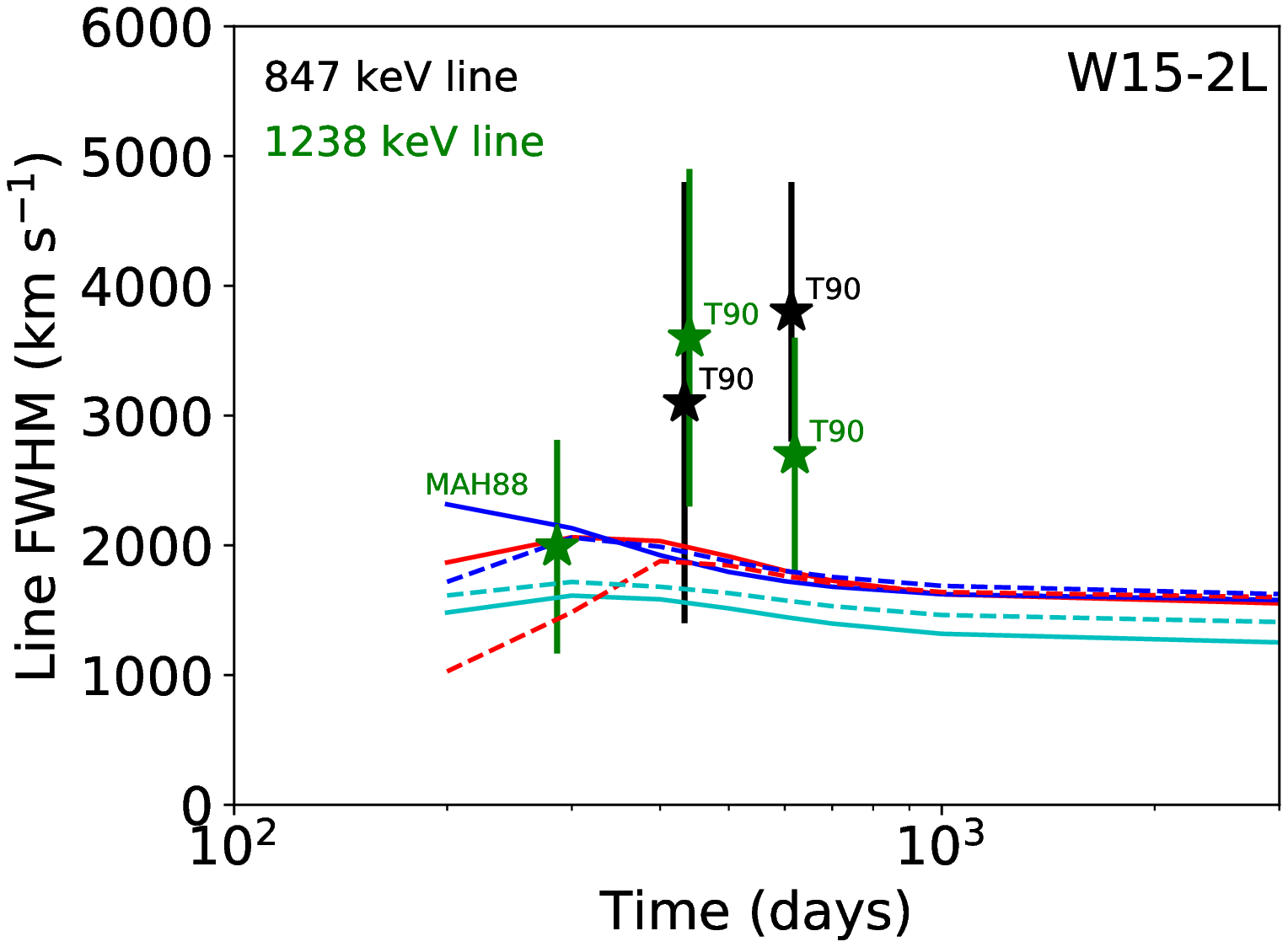}\\
\vspace{-0.1cm}
\includegraphics[width=0.43\linewidth]{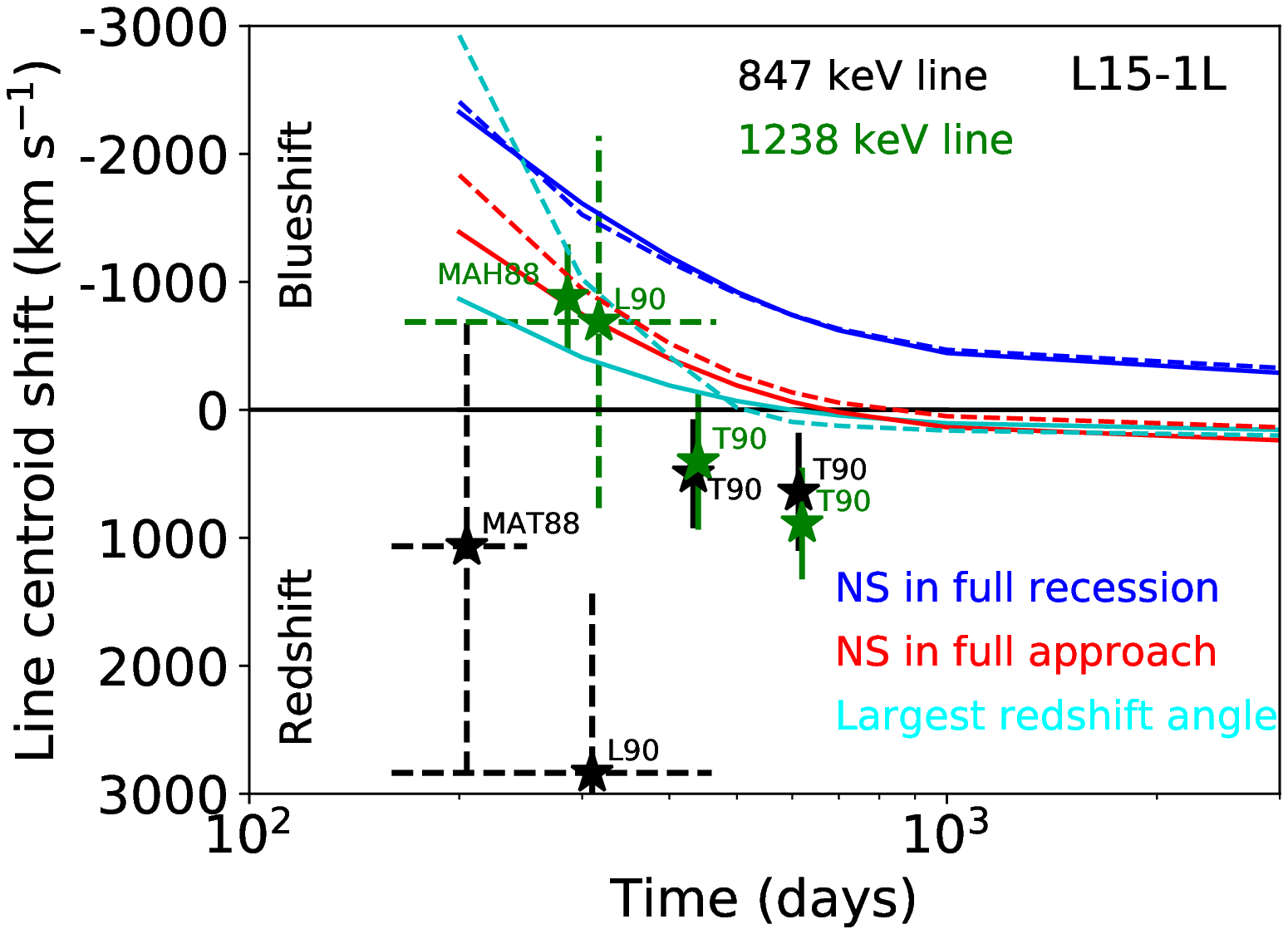}
\includegraphics[width=0.42\linewidth]{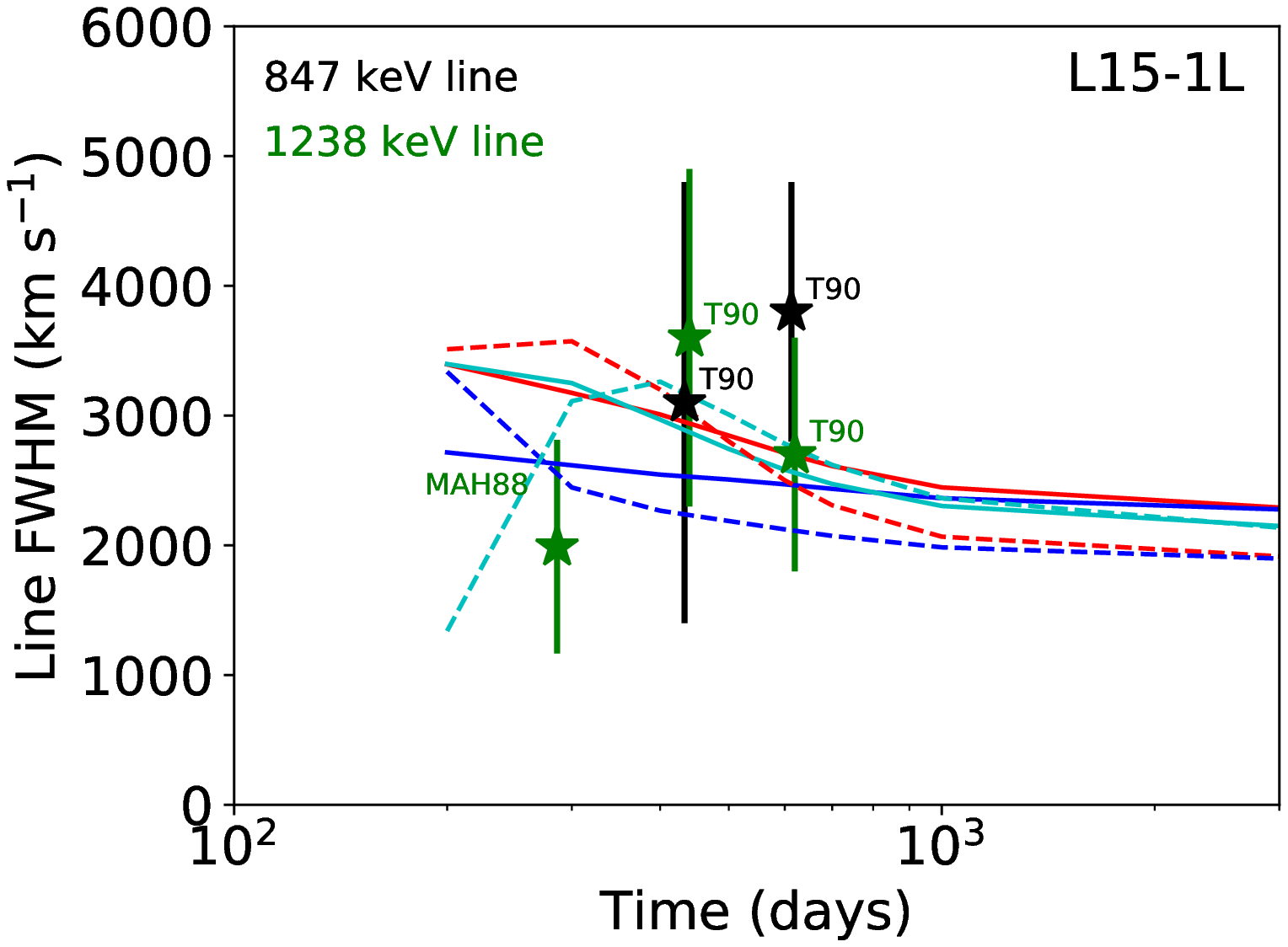}\\
\caption{\arj{Evolution of \co~line shifts (left column) and widths (right column) in SN 1987A compared to the models. Measurements for the 847 keV line are marked in black, and for the 1238 keV line in green, with resolved observations having solid error bars and unresolved having dashed. The model tracks are for 847 keV, but the 1238 keV ones are almost identical. Solid lines are for direct values ($V_{\rm shift}^{\rm direct}$ and $\Delta V^{\rm direct}$), dashed are for Gaussian fits ($V_{\rm shift}^{\rm fit}$ and $\Delta V^{\rm fit}$). The models have been convolved with a Gaussian of 700 \kms. The red model lines are for the viewing angle along the NS kick (NS approaches us), the blue lines are for the anti-parallel viewing angle (NS recedes from us), and the cyan lines are for the viewing angle that gives the largest redshift (or smallest blueshift) at 600d. The acronyms are MAT88 for \citet{Matz1988}, MAH88 for \citet{Mahoney1988}, L90 for \citet{Leising1990} and T90 for \citet{Tueller1990}.}}
\label{fig:shifts}
\end{figure*}

\begin{figure*}
\includegraphics[width=0.44\linewidth]{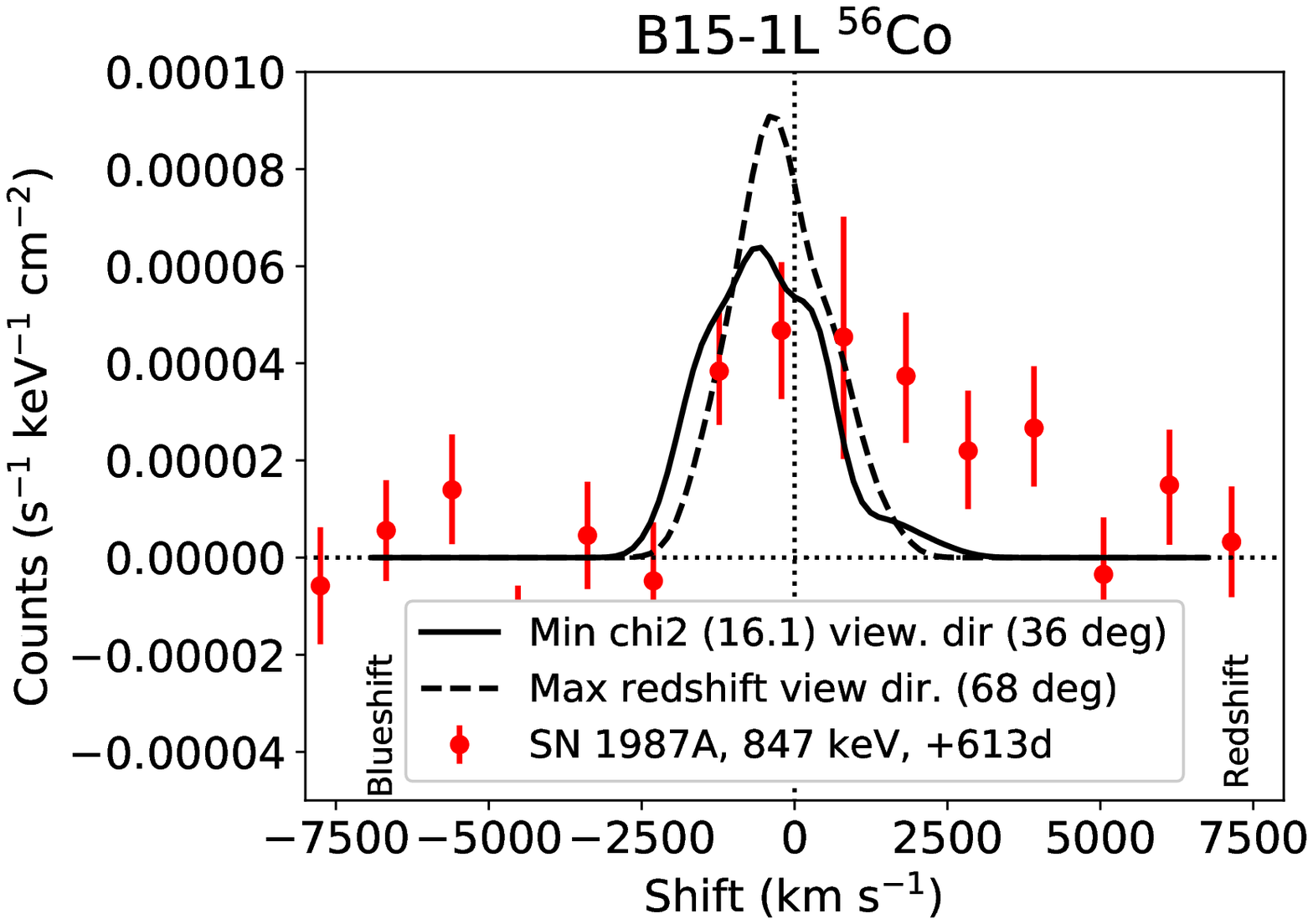}
\includegraphics[width=0.44\linewidth]{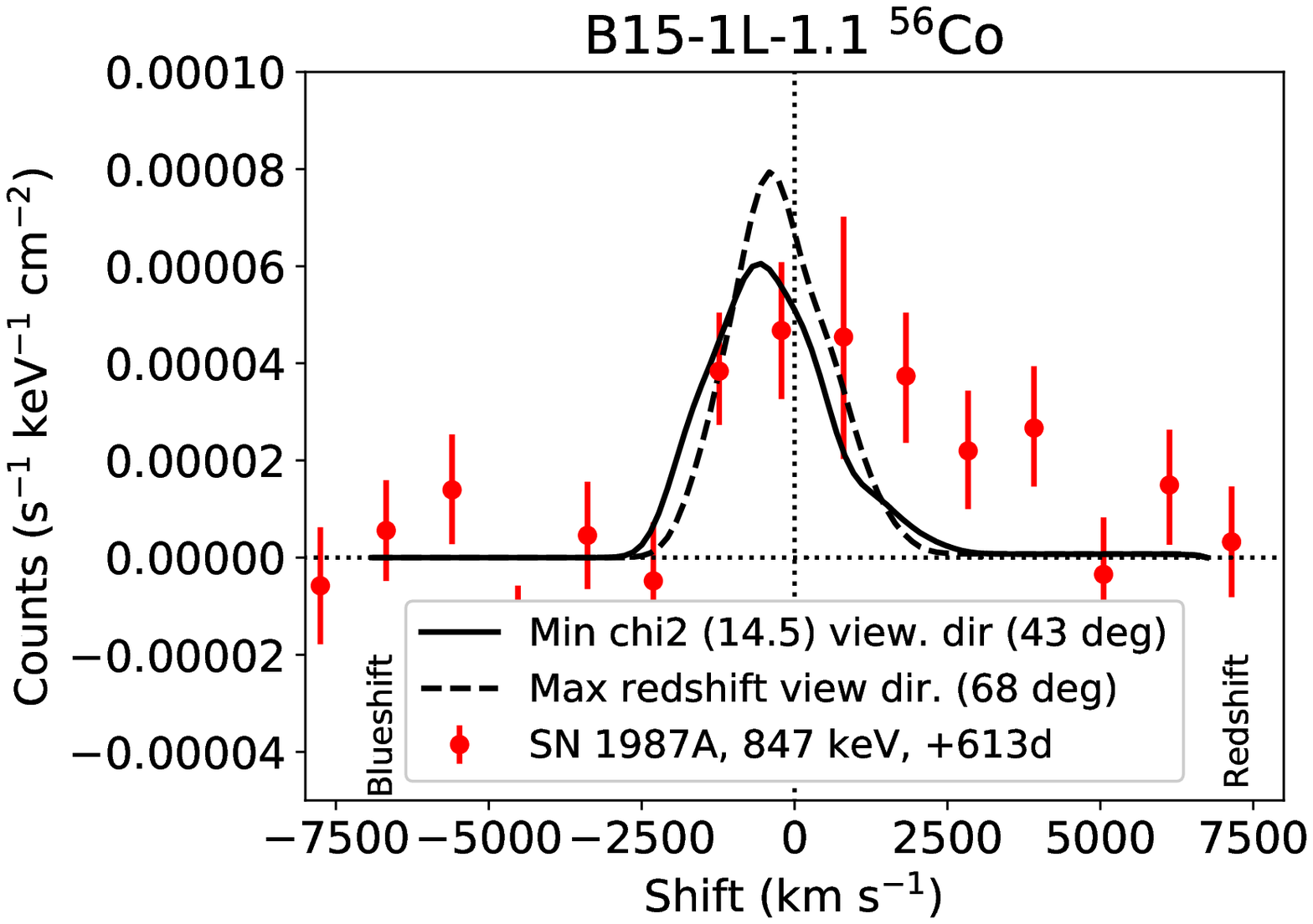}\\ 
\includegraphics[width=0.44\linewidth]{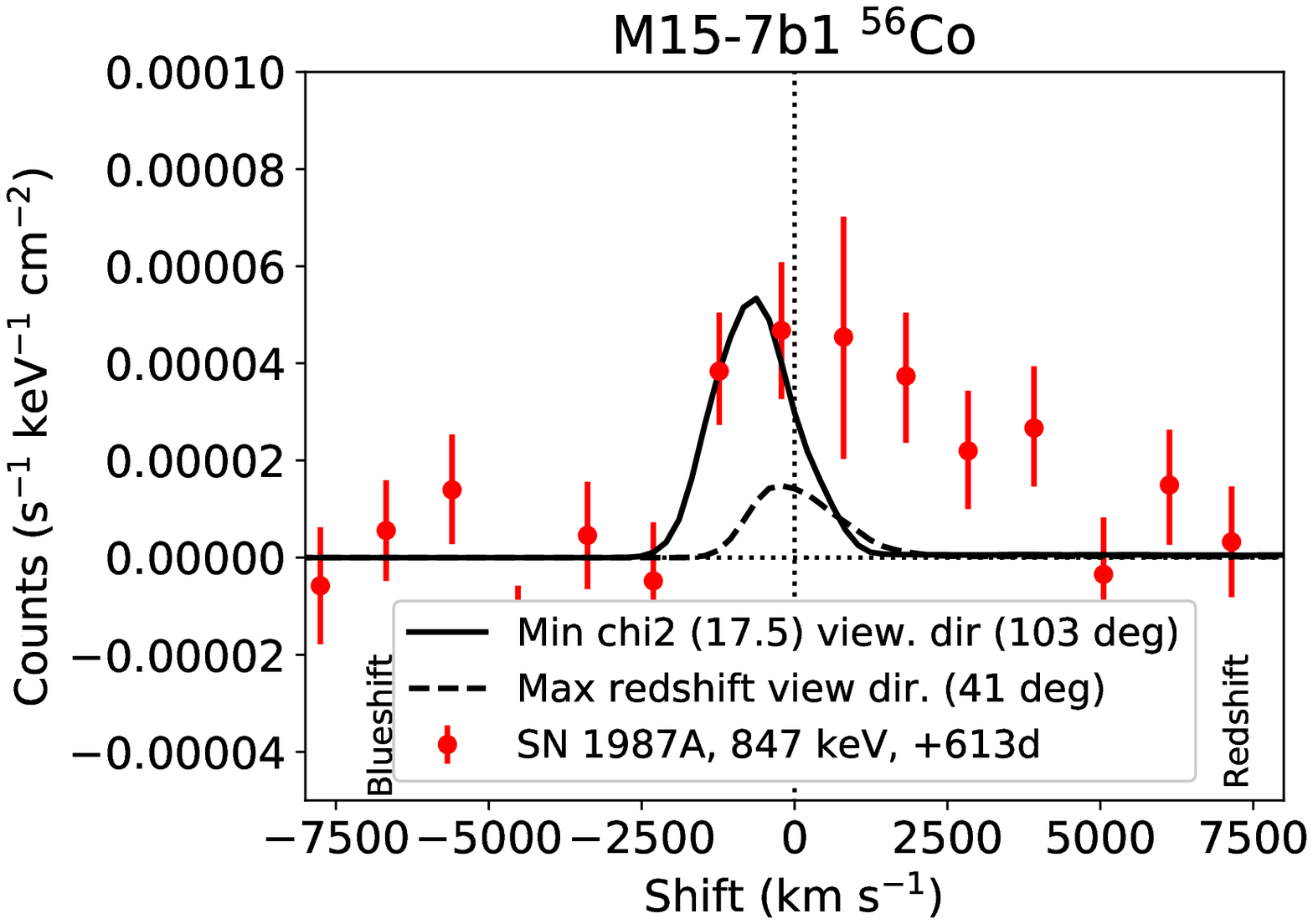}
\includegraphics[width=0.44\linewidth]{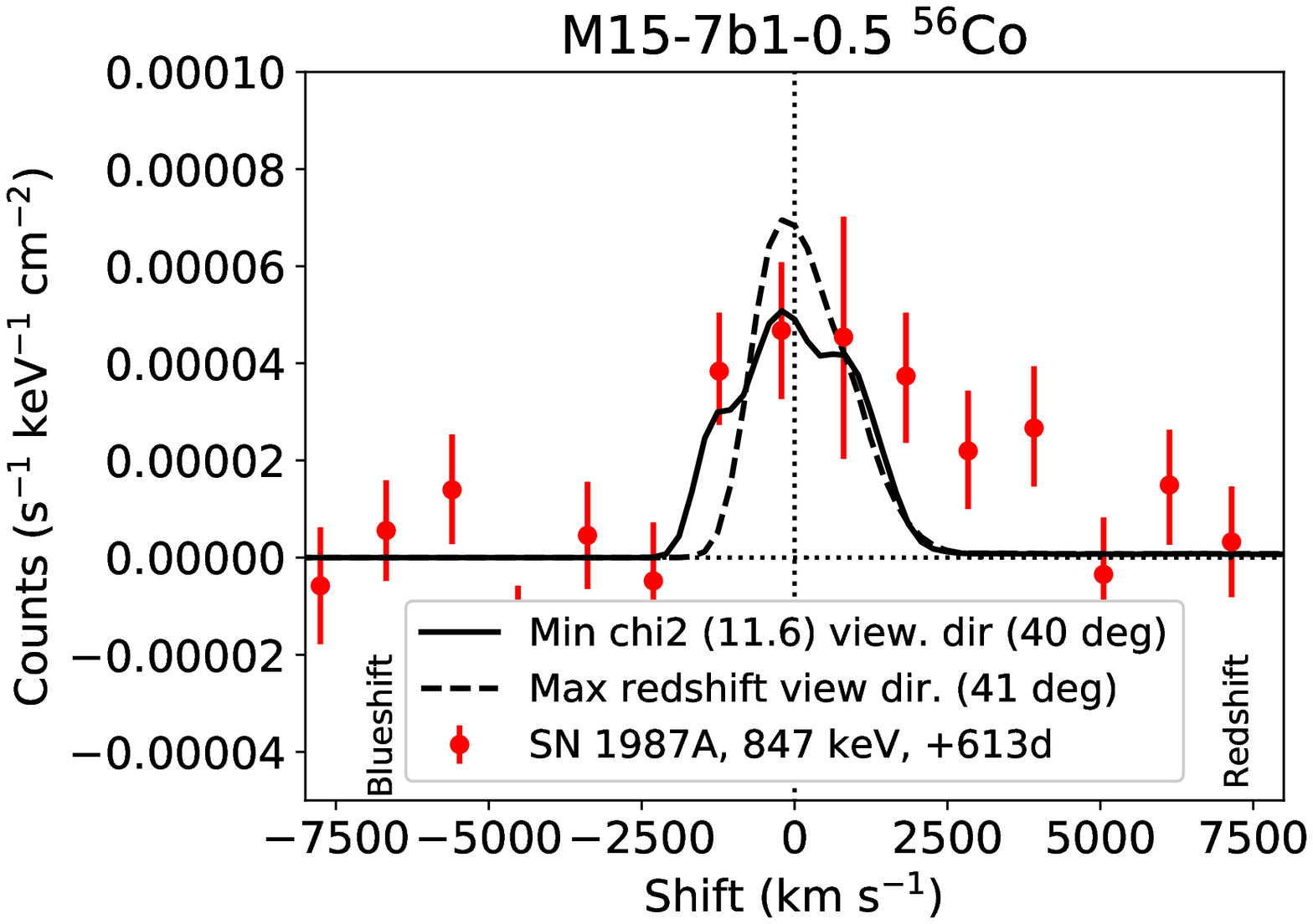}\\ 
\includegraphics[width=0.44\linewidth]{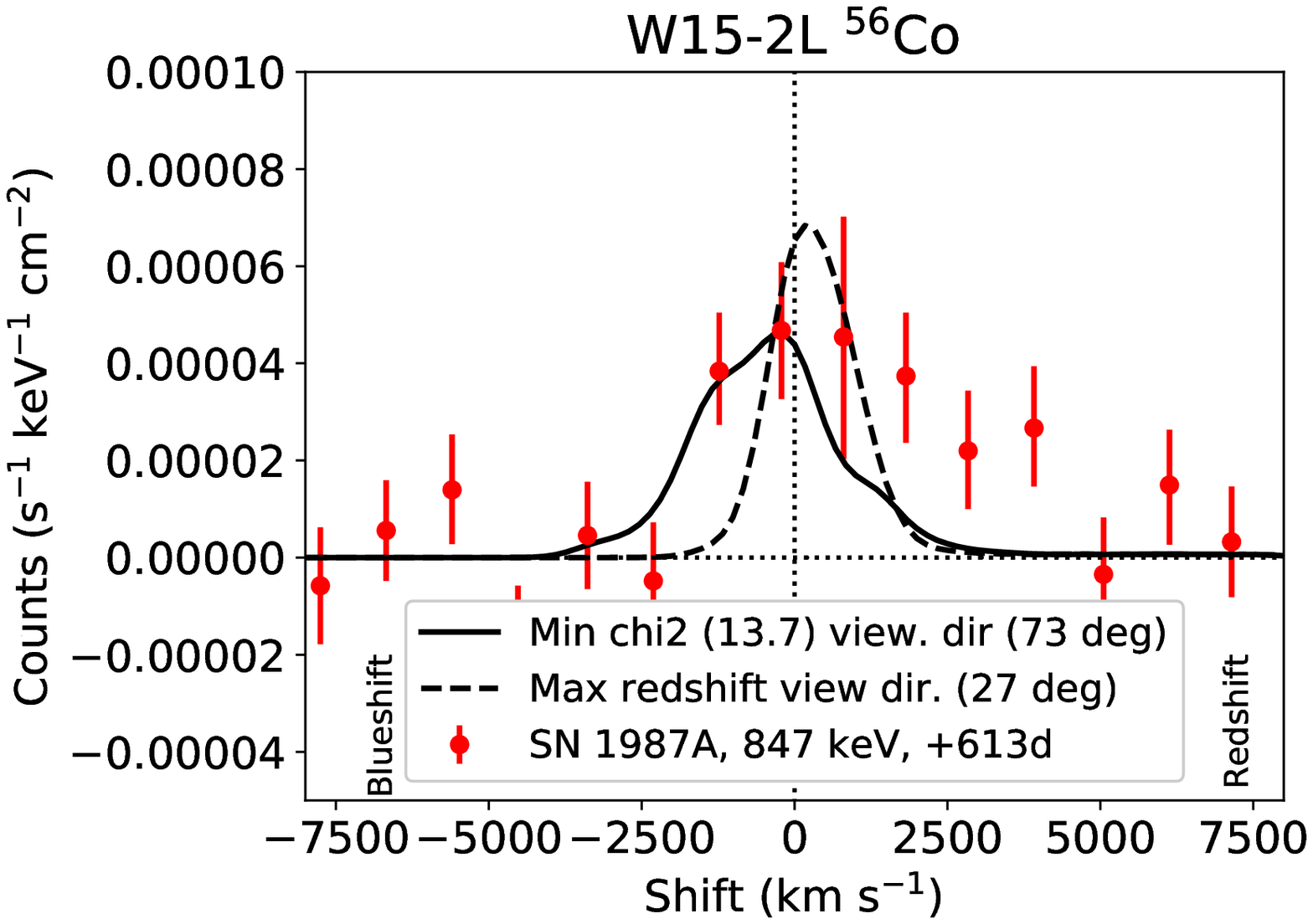}
\includegraphics[width=0.44\linewidth]{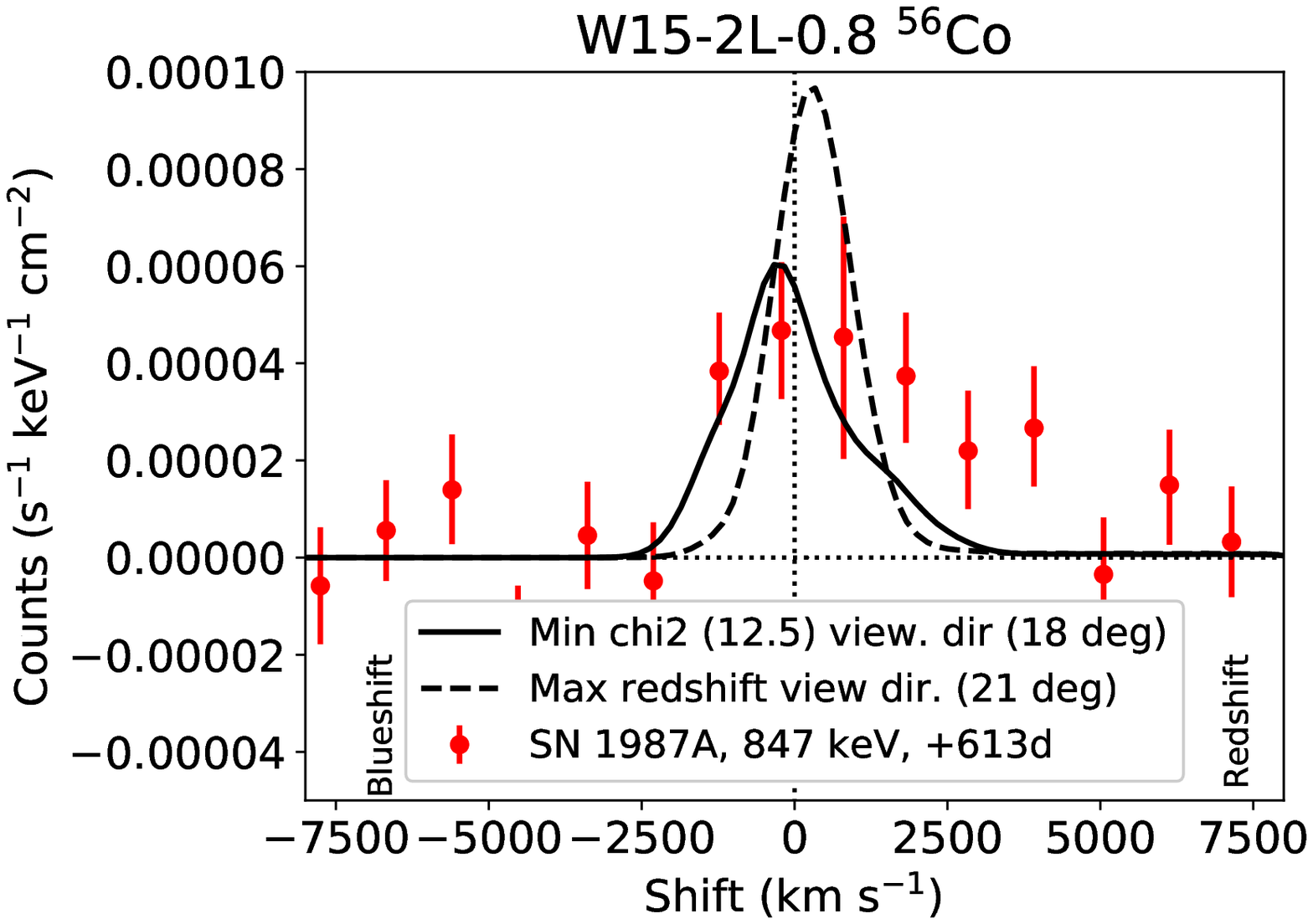} 
\includegraphics[width=0.44\linewidth]{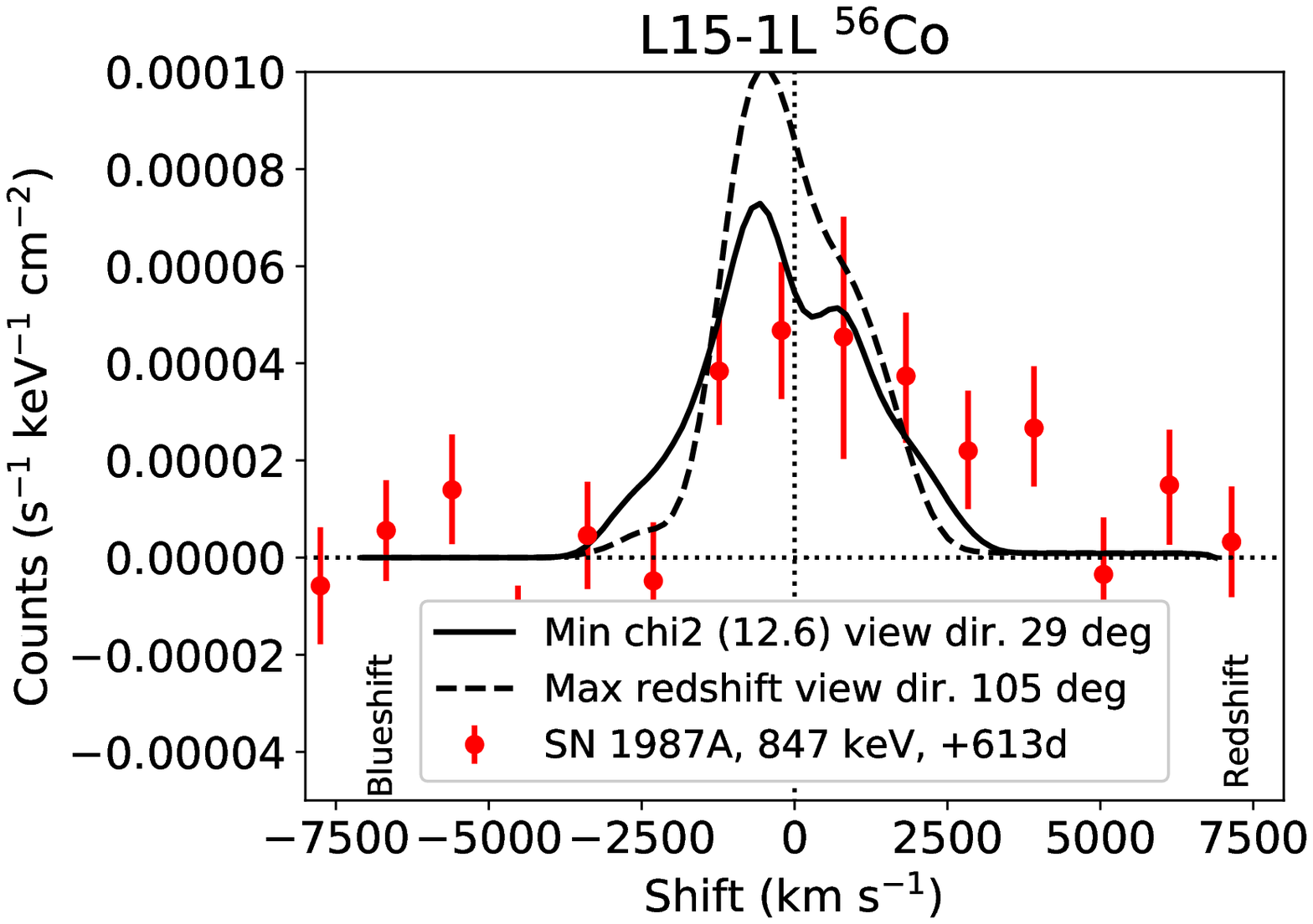}
\includegraphics[width=0.44\linewidth]{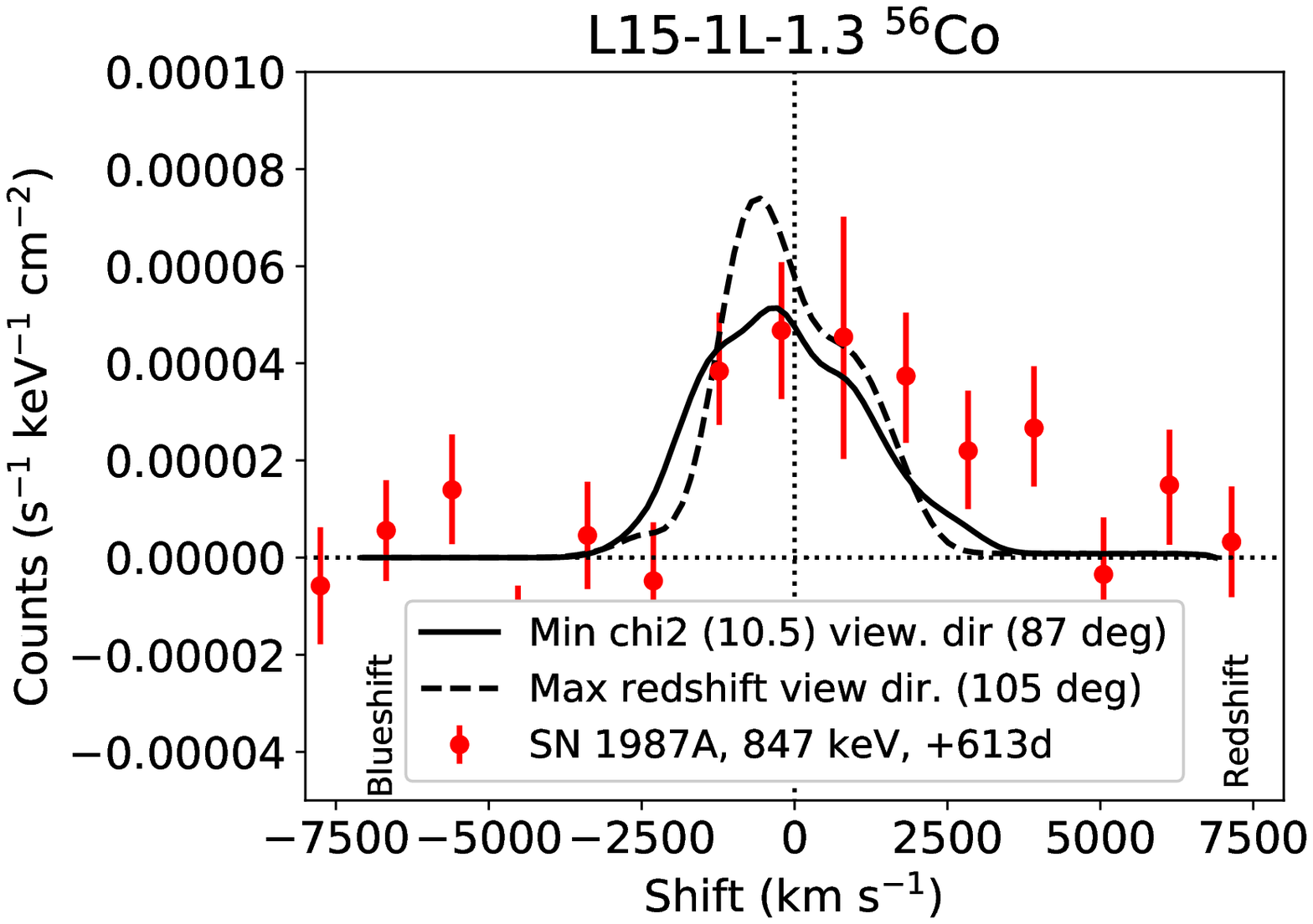}\\ 
\caption{\arj{Line profile fits to the 847 keV line in SN 1987A at +613d. The left column is for models without any modifications, whereas the right column is for the case where all optical depths are rescaled by a factor $s$ that gives the lowest $\chi^2$ (for any viewing angle). The listed viewing directions are relative to the NS motion (0 degrees for the case of the NS moving towards us along the LOS). The models have been convolved with a Gaussian of FWHM 700 \kms.}}
\label{fig:bestfits}
\end{figure*}

\arj{The observational papers present Gaussian fits to the $C_{\rm E}$ quantity (counts s$^{-1}$ cm$^{-2}$ keV$^{-1}$). Operations on lines (centroid calculation, etc) using other quantities such as flux per unit energy ($F_{\rm E}$) or wavelength ( $F_\lambda$) may in general give slightly different results, so it is important to have consistency in what quantity is used. Fig. \ref{fig:shifts} compares observed line shifts and widths to the model values, using $C_{\rm E}$ for both, and Fig. \ref{fig:bestfits} shows the line profiles for the best-fitting viewing angle for each model compared to the observed line at +613d. We have taken account of the telescope broadening effects by convolving the model lines with a Gaussian of FWHM 700 \kms. This convolution in general has quite a small effect as $\Delta V_{\rm obs} = \sqrt{\Delta V_{\rm source}^2 + \Delta V_{\rm res}^2}$, and $\Delta V_{\rm res} \ll \Delta V_{\rm source}$.}

\arj{The error bars on the observational results are in general quite large due to large instrumental background levels, and the compilation of measurements needs some commenting.  The Compton opacity at 847 keV is slightly higher (15\%) than at 1238 keV. This higher attenuation should give a slightly larger blueshift for the 847 keV line compared to the 1238 keV line. The (unresolved) \citet{Leising1990} measurements  indicate the opposite, with the 847 keV line being more redshifted, although the measurements could be marginally compatibile given the stated uncertainties. Also the unresolved \citet{Matz1988} measurement gives a surprising redshift already at 200d}.

\arj{We have marked the unresolved observations with dashed error bars, to highlight that they may involve larger systematic uncertainties than the resolved ones. We focus our model comparisons to the resolved line observations of \citet{Mahoney1988} and \citet{Tueller1990}. These show a more expected behaviour, with blueshifts at early times that then abate as optical depths decrease with time. The 847 and 1238 keV lines are at roughly the same shift, as expected (the expected difference is smaller than the error bars)}.

\arj{There are three main properties of the (resolved) data that a satisfactory model should fulfil: \textit{(i) A significant redshift of the line ($\sim 700$ \kms) at late times, $t\gtrsim$1y. (ii) A FWHM width of the line of around 3000 \kms~throughout the evolution. (iii) A relatively rapid change in the line center between 300-600d, of order 2000 \kms~towards the red}. One may add a possible fourth requirement; that $^{56}$Ni with receding line of sight velocity of $3000-4000$ \kms~should contribute emission at 613d, although this would rely on 1-2 specific data points, each with significant uncertainties}.

\arj{Starting with the centroid shifts, any 1D model would give a curve that begins with a blueshift, as emission from the far side is blocked to a larger degree compared to emission from the approaching side, and then relax towards zero as the optical depth reduces with time. At no point can a redshift be achieved (ignoring the small time-dependent effects which would slightly boost redshifted emission). The starting point to obtain a redshift is therefore to have a 3D distribution where the \ni~bulk momentum is oriented towards the receding hemisphere. A good model then also needs to have the right distribution of mass to produce the observed gamma ray escape as function of time}. 

\arj{Figure \ref{fig:shifts} shows that for most of the 3D models and viewing angles, the time evolution behaviour is similar to the 1D case; material on the far side sees higher column densities and gains more by the reduced optical depths coming about as time goes by, reducing the blueshifting of the line. However, morphology and viewing angle combinations exist where this time evolution is weak, flat, or even reversed (the cyan lines give examples). To understand this behaviour, consider a case where a localized blob has a relatively clear line of sight (a ``tunnel''), and comes to dominate the emission. As in this case there is no differential effect from different regions evolving differently in time, there is no time evolution of the line shift. In the limiting case of all \co~emission coming from a single such clump, the track is flat. By adding a second clump on the approaching side, the line can even blueshift with time as this second clump becomes increasingly visible.}

\arj{If data property (iii) is believed (and granted this cannot be fully established due to uncertainties in the observations), it therefore  puts an important constraint on the ejecta morphology; that emission must come from an extended region of \co~rather than from one or a few localized regions. Another way to try to constrain this is to consider the FWHM; models where a localized region dominates will necessarily give quite a narrow line profile.}

\arj{We now discuss the models one by one, from Figs. \ref{fig:shifts} and \ref{fig:bestfits}. For each model we discuss the fits it produces without any modifications, and also whether allowing for a rescaling of all optical depths by a factor $s$ can improve the fits, and if so for which $s$}.

\textit{\underline{B15-1L.}} \arj{As can be seen from Fig. \ref{fig:shifts} by the asymptotic line shifts as $t\rightarrow \infty$, B15-1L has a too small \ni~asymmetry to give a sufficient redshift even without Compton scattering effects (which blueshift). The \ni~shift velocity is only of order 150 \kms, compared to an observed 700 \kms~already at 600d when there is in addition blueshifting still affecting the line. At 600d all viewing angles give a blueshift of at least several 100 \kms. Fig. \ref{fig:bestfits} shows that viewing angles exist giving reasonable fits to the central and blueshifted parts of the line profile, however there is lack of model emission from the redshifted side. This is also reflected in somewhat too low line widths. Notable is that the optimal opacity scaling is $s=1.0-1.1$, so the line profile cannot be improved by free scalings of the optical depths.}

\textit{\underline{M15-7b1.}} \arj{M15-7b1 has more shifted line asymmetries as $t\rightarrow \infty$, and thus fullfils a first requirement for a model match, that the \ni~is asymmetric enough. Most viewing angles, however, still give too much Compton blueshifting at 600d. Fig. \ref{fig:bestfits} shows that the unscaled model fits are worse than for B15-1L, as for M15-7b1 even more flux on the red side and even in the central line region is blocked away. This happens because the model has an ejecta mass a factor 1.5 higher than B15 (and the other models), and the Compton optical depth scales with $M^2$ (in a 1D/angle-average sense)}

\arj{Fig. \ref{fig:shifts} shows that there are viewing angles that yield redshifted line centroids at 600d and even earlier. The plot of the corresponding line profile in Fig. \ref{fig:bestfits} shows, however, that there is too low flux at these directions. Furthermore, the emission comes from a too restricted radial velocity range, resulting in too low line widths, as discussed above. These are viewing angles where ``tunneling'' of quite localized \ni~clumps takes over, but they fail to reproduce the correct line strength, width, and time evolution.}

\arj{The M15-7b1 model thus appears to have too much mass. The optimal rescaling here is $s=0.5$, which would correspond to a factor 1.4 lower ejecta mass. However, even after rescaling the optical depths this models produces insufficient flux on the red side of the line.}

\textit{\underline{W15-2L.}} \arj{This model has a similarly large \ni~asymmetry as M15-7b1. The lower ejecta mass now brings the tracks of the line centroid shifts down towards the observed values for viewing angles where the \ni~momentum vector is oriented away from us}.

\arj{Inspection of the line profiles show a bit more flux is now emerging on the red side, especially when we apply a small rescaling factor of $s=0.8$. The best $\chi^2$ line profile starts to approach the observed one, with red-side line fluxes up towards 3500 \kms~for the first time. The shifts are still marginally on the low side, and the line widths appear to be somewhat too low. Nevertheless, the combination of a significant \ni~asymmetry and a smaller ejecta mass compared to M15-7b1 clearly works towards an improved agreement with the observations.}

\textit{\underline{L15-1L.}} \arj{This model has somewhat less \ni~asymmetry than M15-7b1 or W15-2L. It has overall faster \ni, however, which leads to general improvements in the line profile fits and FWHM compared to SN 1987A. The optimal rescaling factor is $s=1.3$, suggesting that the model has somewhat too low trapping of the gamma rays. This model has formally the lowest $\chi^2$ in the set, both with and without $s$ rescalings.}

\subsubsection{Discussion} 
\label{sec:optimal}
None of the four models achieve a fully satisfactory fit to SN 1987A. However, from the combined comparisons, one can sketch out an idea of what would be needed to achieve an improved fit. These ingredients, to first order, involve i) a sufficient bulk velocity of \ni, something similar to model L15-1L (wheras B15, W15, and M15 are on the low side), ii) a sufficient degree of asymmetry of the \ni, something similar to models M15-7b1 or W15-2L (whereas B15 and L15 are on the low side), and iii) a suitable ejecta mass/energy combination (which sets the degree of Compton scattering), something similar to models B15 and W15, whereas M15-7b1 is on the high side and L15 on the low side.

\arj{To expand on the last point, if one estimates an optimal ejecta mass from each fit by the optimal rescaling parameter $s$ as $M_{\rm optimal} = M \sqrt{s\times \left(1.4~\mbox{B}/E\right)}$, one obtains a quite clustered set of results at $M_{\rm optimal}=14.1, 11.0, 12.0$ and 14.1 \msun~for B15-1, M15-7b1, W15-2L and L15-1L, respectively. This expression comes from $\tau \propto s\kappa M^2/E$, and thus changing $s/E$ from the default ratio has the same effect on the optical depth as changing $M$ by a factor $\sqrt{s/E}$. This analysis would therefore suggest the ejecta mass is in the ballpark of $13 \sqrt{E/1.4~\mbox{B}}$ \msun~(taking the average of the $M_{\rm optimal}$ values listed above), with good consistency. 
An independent viewpoint on this comes from the UVOIR bolometric light curve of SN 1987A, which we will look into in section \ref{sec:gammafield}.}

\subsection{\ti~lines}

\arj{\citet{Boggs2015} presented NuSTAR observations of the \ti~decay lines (68 and 78 keV) in SN 1987A at a remnant age of 27y. An instrumental cut-off prohibited a full capture of the 78 keV line, but the 68 keV line was observed in full. The lines were unresolved, with an upper limit to the intrinsic line width $\mbox{FWHM}<4100$ \kms~derived. The 68 keV line has a measured centroid redshift of $730\pm 400$ \kms~(90\% errors) after correction for the LMC movement \footnote{The relevant quantity is the measured redshift of the 67.87 keV line which is $0.23\pm 0.09$ keV.}. This is similar to the late-time measurements of the \co~lines \citep{Tueller1990} and strengthens the picture that the explosive burning ashes in SN 1987A have an overall net momentum away from us. Note that INTEGRAL observations \citep{Grebenev2012} are of lower resolution and are not used here.} 

\begin{figure*}
\includegraphics[width=0.45\linewidth]{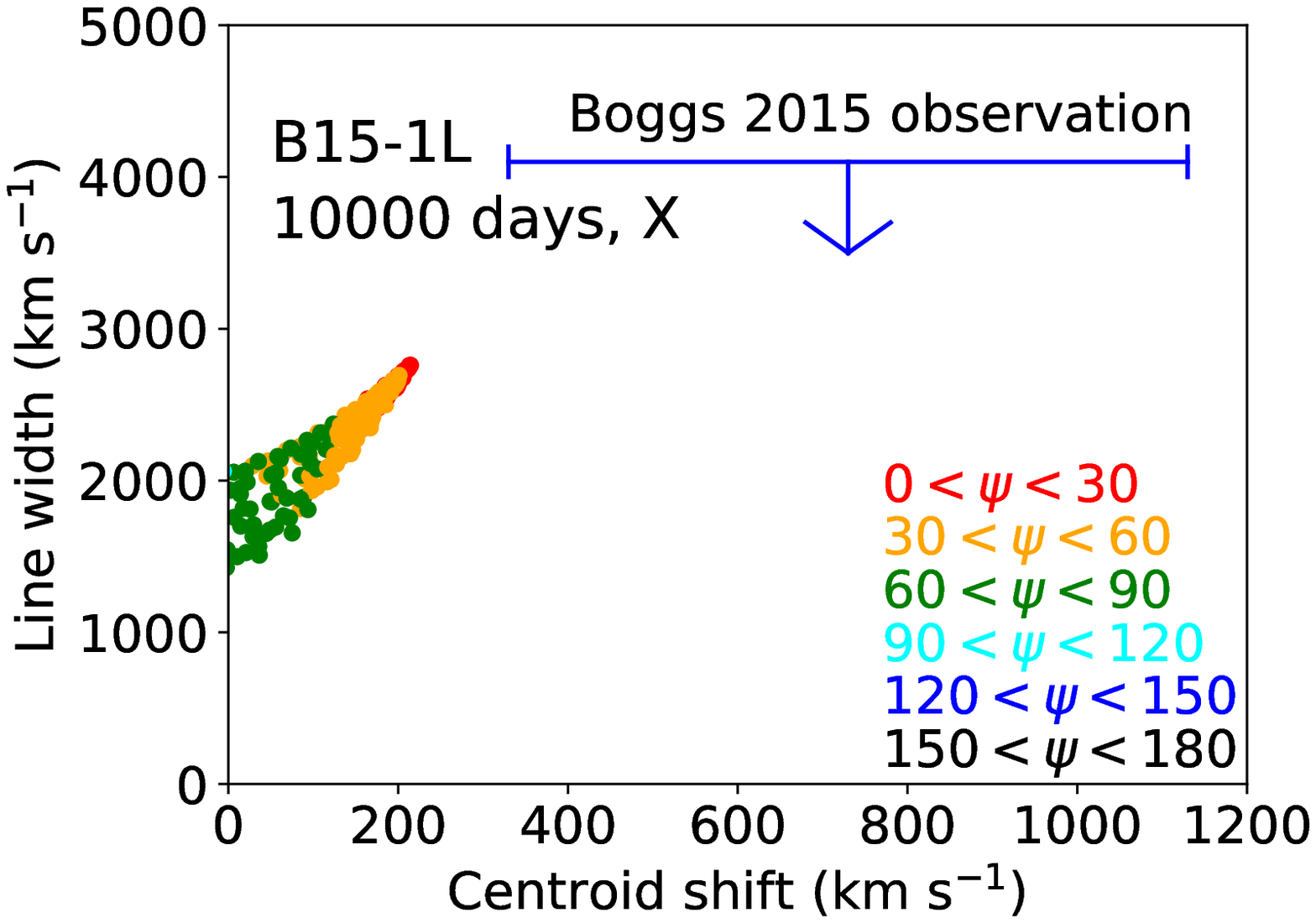}
\includegraphics[width=0.45\linewidth]{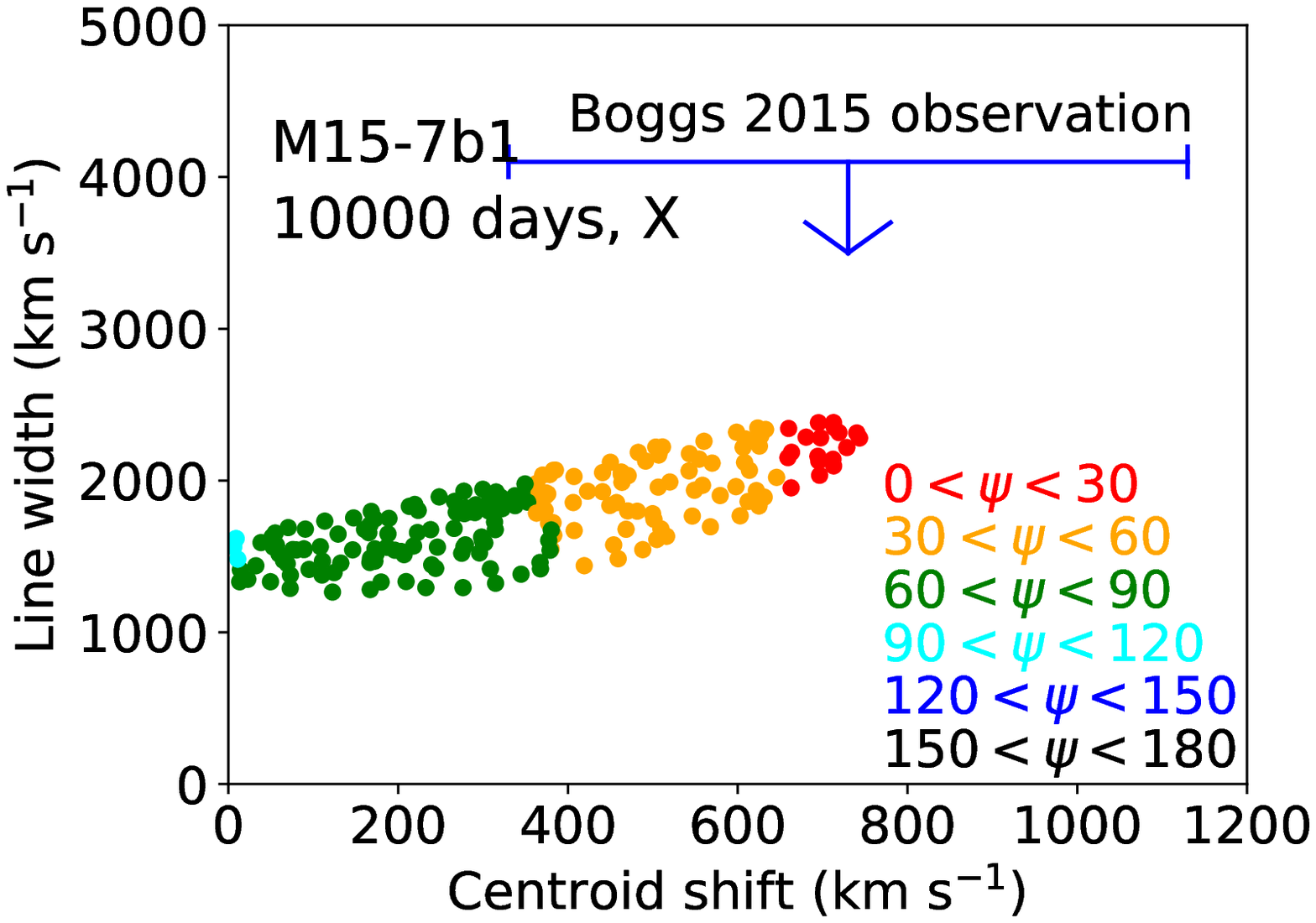}\\
\includegraphics[width=0.45\linewidth]{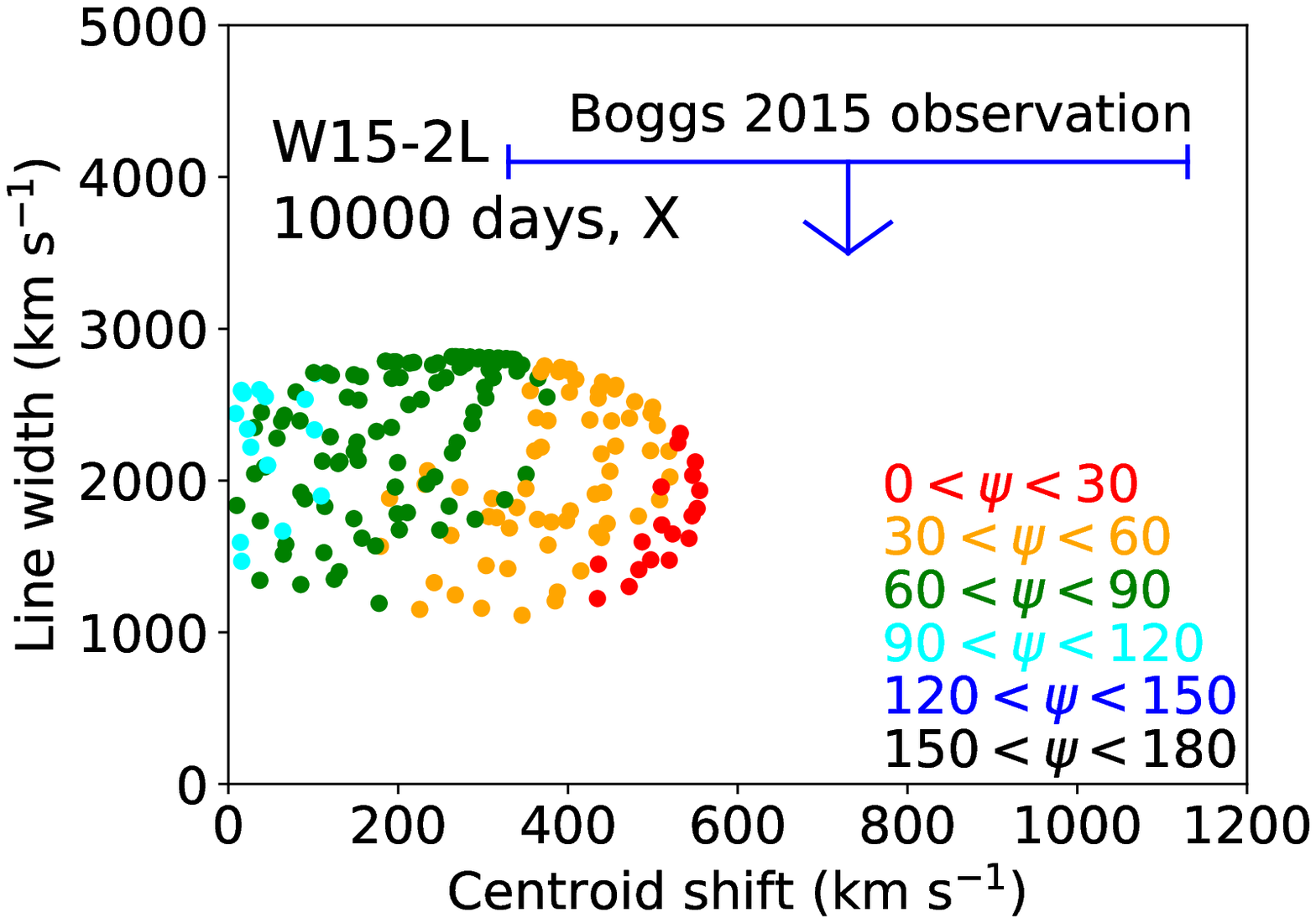}
\includegraphics[width=0.45\linewidth]{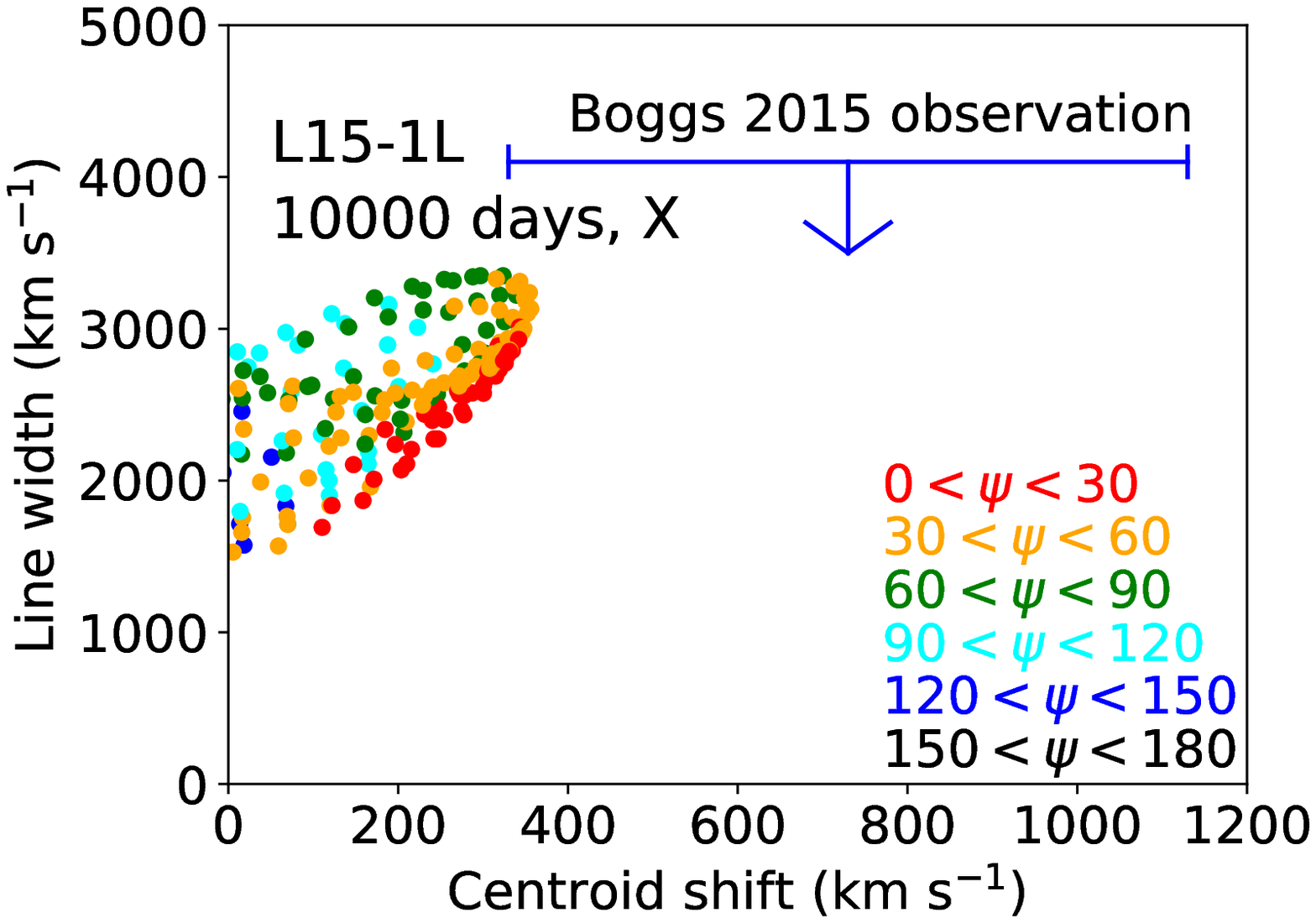}
\caption{\arj{Line shifts and widths of the \ti~lines (using the X distribution) in the models, in the optically thin limit, compared to SN 1987A observations. The viewing directions are color coded according to the angle $\psi$ between the NS kick vector and the observer direction. Models M15-7b1 and W15-2L give agreement with the observed line shifts for $\psi \lesssim 60^\circ$. Model B15-1L has insufficient line asymmetry for all viewing angles, whereas model L15-1L reaches marginal agreement for a small fraction of angles.} }
\label{fig:ti_Hmodels}
\end{figure*}

\arj{Figure \ref{fig:ti_Hmodels} compares the four models to these observations. All models comply with the upper limit to the line width. B15-1L has too little asymmetry in the \ti~(taken as X)~distribution (290 \kms, Table \ref{table:ejecta}), giving a similar line shift (Fig. \ref{fig:ti_Hmodels}), similar to its insufficient \ni~asymmetry. Models M15-7b1 and W15-2L reach into the observed shift domain for a significant fraction of viewing angles (angle $\psi$ between NS kick vector and viewing direction $\lesssim$60$^\circ$), which means the \ti~is sufficiently asymmetric in these models (790 and 580 \kms, respectively, using the X distribution in Table \ref{table:ejecta}). Finally, L15-1L has a marginally sufficient asymmetry (390 \kms, X distribution). Note that L15-1L has a more complex behavior between line shift and $\psi$ than the other models. This stems from a stronger deviation from exact anti-alignment between NS kick and element momentum vectors in this model (Table \ref{table:dircos}, see also Fig. 15 in \citet{Wongwat2013}). Models with somewhat more asymmetry than this are preferred, especially when combined with the \co~results. In general the \co~and \ti~results paint a consistent picture that an asymmetry of the explosive ashes of order 500-1000 \kms~is needed. Less than this would result in dissonance with respect to both \co~and \ti~ observations, and this lower limit should therefore be robust. More asymmetry than this leads to dissonance with the \ti~data, whereas the \co~data does not give further constraints as different degrees of Compton blueshifting of the \co~lines can be called upon at that epoch}.

\section{Comparison with Cas A} 
\label{sec:CasAcomp}

\arj{The \ti~decay lines from Cas A were observed with NuSTAR and reported in \citet[][G14]{Grefenstette2014} (1.2 Ms data) and \citet[][G17]{Grefenstette2017} (2.4 Ms data). In contrast to SN 1987A, the \ti~lines are for this source resolved due to the higher expansion velocities. Observations have also been carried out by COMPTEL, Beppo-Sax, and INTEGRAL-SPI. Of these only the SPI measurements are also resolved, and give similar results as from NuSTAR \citep{Siegert2015}.} 

\arj{G14 measure a bulk redshift of $2050\pm 950$ \kms~(90\% error bars), and a line width $\mbox{FWHM}=7600$ \kms \citep[well above the detector energy resolution of 4000 \kms,][]{Harrison2013}. The systematic calibration uncertainty is of order 200 \kms (G17), which can be ignored for our purposes. G17 report similar values ($1950\pm 440$ \kms~and $7060\pm 1560$ \kms, respectively) using a factor two more data, but do not plot the integrated line profile.
Note that these widths are including the telescope broadening, removing these gives a value $\sim$5800 \kms. INTEGRAL/SPI finds a bulk redshift of 1800 km s$^{-1}$ $\pm$ 800 km s$^{-1}$, and a line width of 6400 $\pm$1900 km s${^-1}$, when results from the 78 and 1157 keV lines are combined (Weinberger et al. 2020, submitted). This broadly supports the constraints from NuSTAR, with consistent results from three different lines and observed with two different instruments.}

A difficulty with Cas A is that some of the \ti~may already have been affected by the reverse shock (G17). According to G17, the reverse shock is currently at a position corresponding to 4500 \kms~free expansion. The shock will decelerate the ejecta and make the line profile more narrow compared to free-expansion models. There is, however, no obvious way of treating this effect. Note that the shocking has been assessed at least not to ionize the \ti~so high that its decay properties are affected. See however \citet{Mochizuki1999}~for a in-depth discussion.

\begin{figure*}
\includegraphics[width=0.48\linewidth]{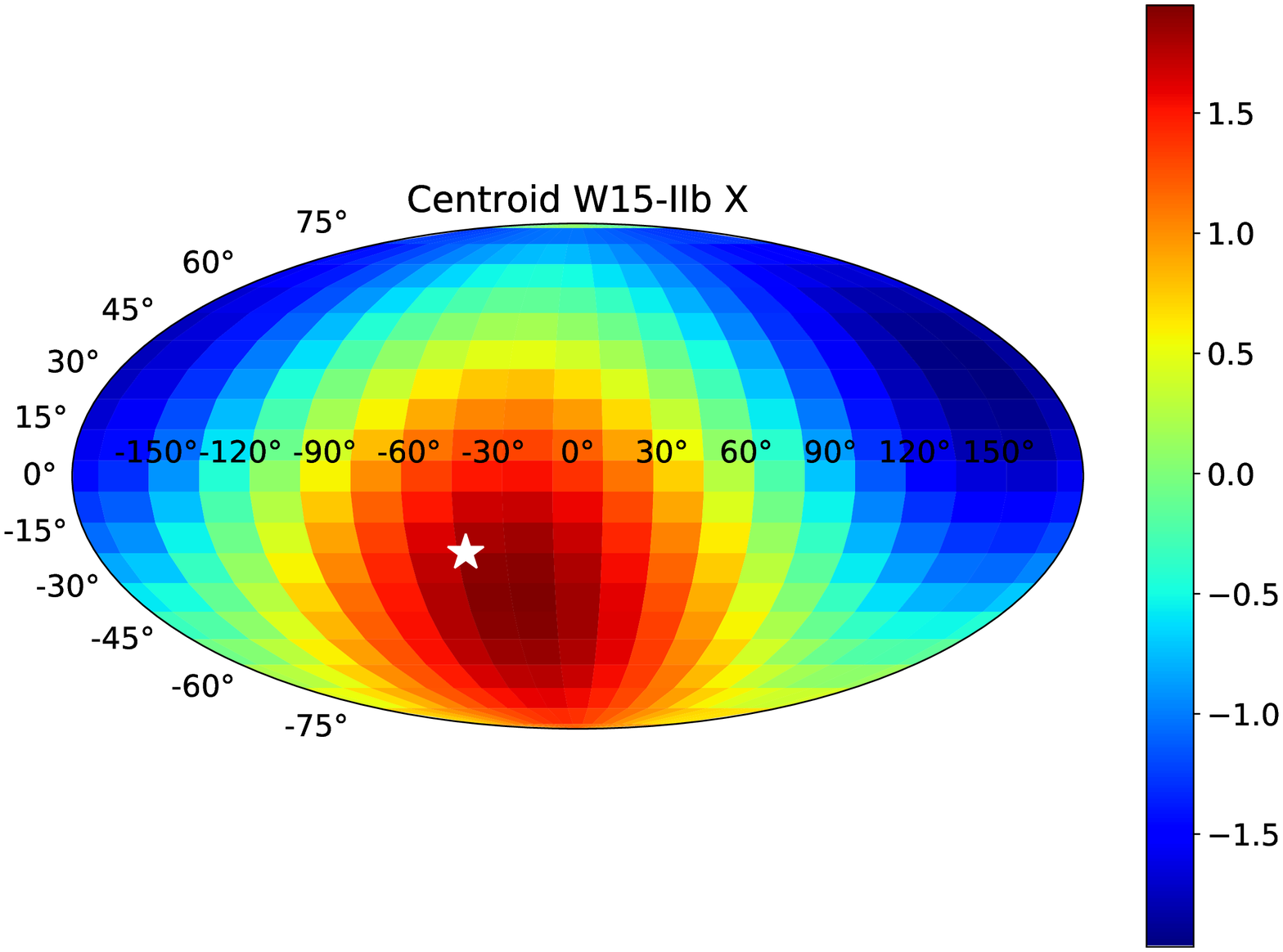}
\includegraphics[width=0.48\linewidth]{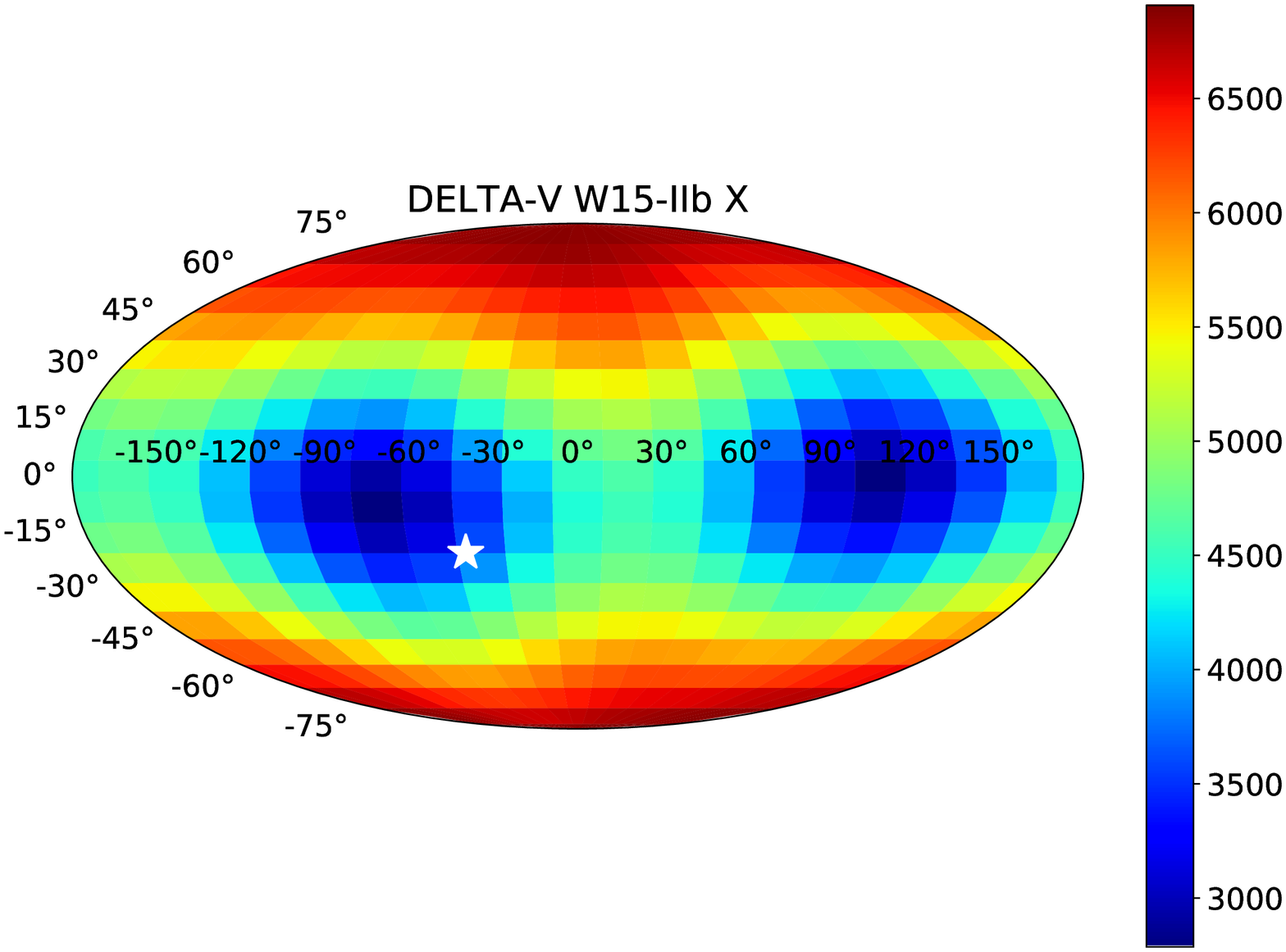}
\caption{\textit{Left}: \arj{Centroid shift as function of viewing angle (units of NS kick = 719 \kms), for \ti~in model W15-IIb in the optically thin limit. \textit{Right}: Line width (in \kms, unconvolved) as function of viewing angle for the same model. The direction of the NS (where it is moving straight towards us) is marked with a white star.}}
\label{fig:eggplots_W15-3}
\end{figure*}

\arj{For our comparison we used the X distribution in the single stripped-envelope model W15-IIb (see discussion in Sec. \ref{sec:models}). 
This model gives a maximum centroid shift of $\sim$1300 \kms, which is marginally too low compared to the observed value  ($1950\pm 440$ \kms). The (unconvolved) FWHM in the model is about  4000 \kms~when the \ti~momentum vector is aligned along the line-of-sight, and Fig.  \ref{fig:eggplots_W15-3} (right) shows that this is among the lowest values in the range 3000-6500 \kms~spanned by different viewing directions. The observed value of 5800 \kms~is covered by this range}.

\begin{figure*}
\includegraphics[width=0.48\linewidth]{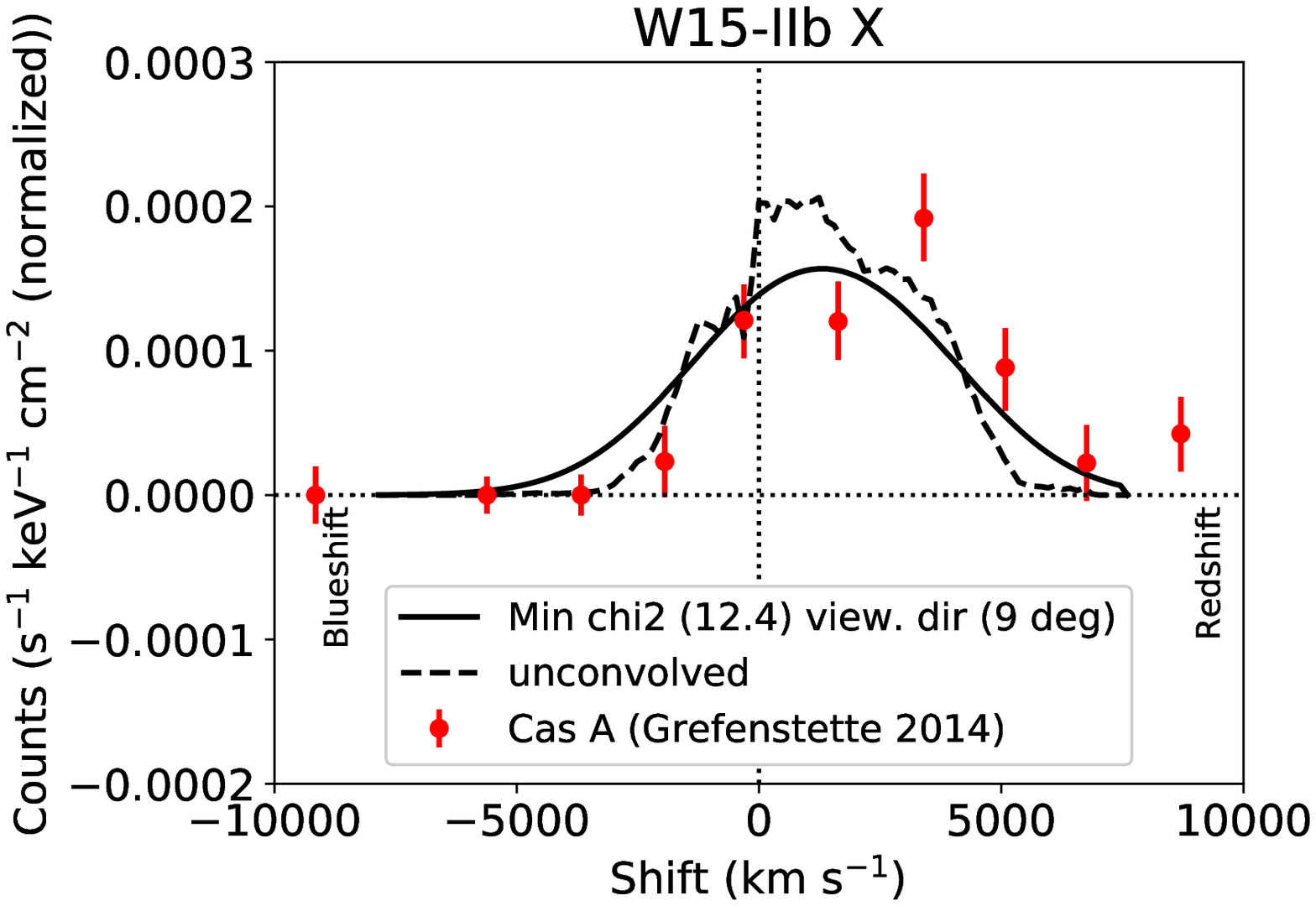}
\includegraphics[width=0.48\linewidth]{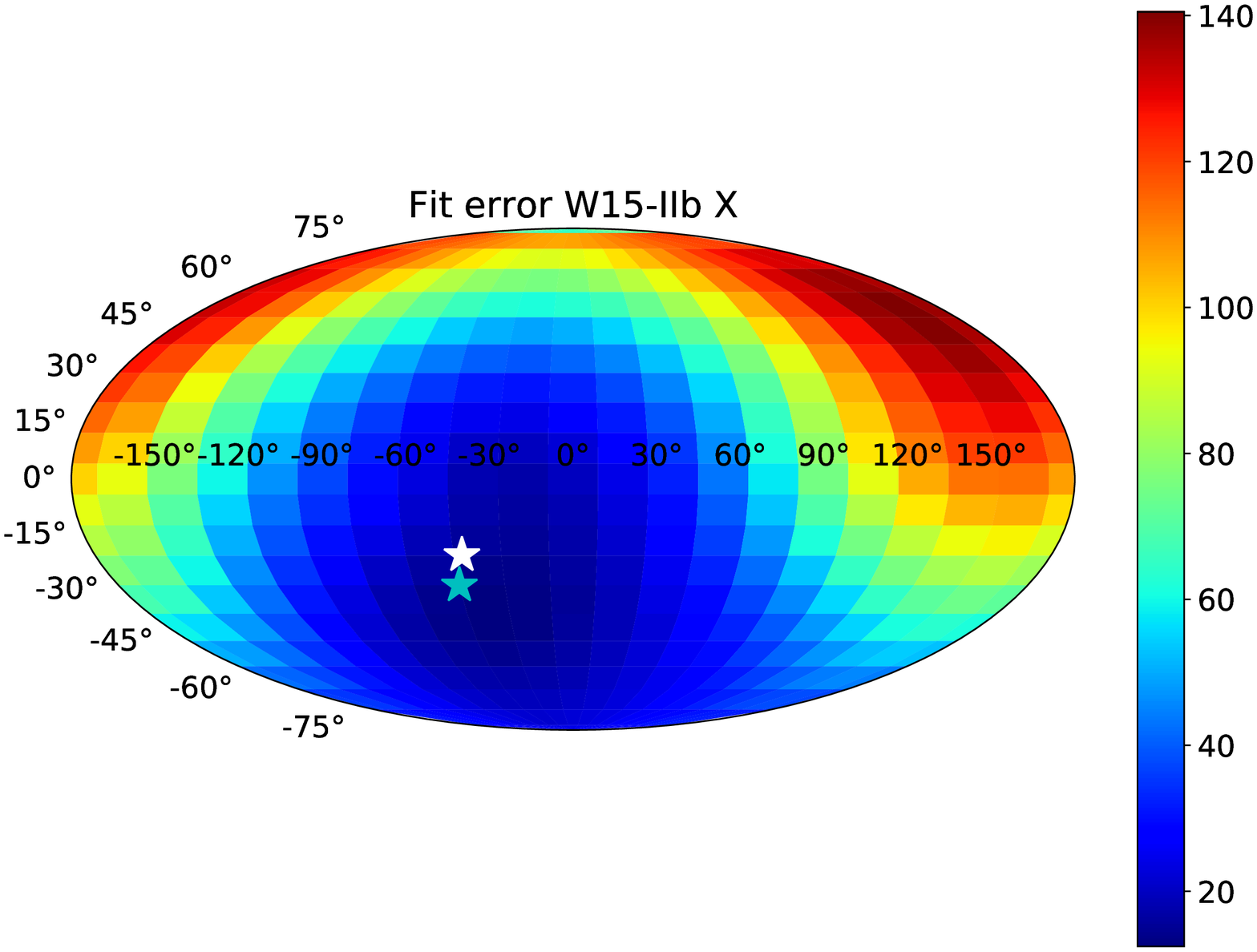}
\caption{\textit{Left}: \arj{Comparison between the observed \ti~line profile of Cas A (G14) and the best fitting line profile from model W15-IIb (using the X distribution), both normalized. We use convolved models (4000 
\kms~Gaussian) for the fitting, but the unconvolved line is also shown (dashed). \textit{Right}: $\chi^2$ map. The $\chi^2$ minimum is marked with a cyan star and the NS motion by a white star.}}
\label{fig:CompCasA_LP}
\end{figure*}

\arj{Fig. \ref{fig:CompCasA_LP} shows the observed line profile from G14 compared to the best fitting model line profile, which occurs for a viewing angle 
close to antiparallel with the NS kick direction (NS moves towards us).
As the models provide no reliable \ti~masses (more detailed post-processing estimates are available in \citet{Wongwat2017} but still depend on choices of $Y_e$), we renormalized all line profiles when fitting to the data, seeking only to fit the shape of the line. Basically all viewing angles within $\lesssim$60$^\circ$ of the NS direction give a $\chi^2$ within a factor 2 of the minimum value}.

\subsection{Discussion} \arj{The observed NS plane-of-sky motion is 330 \kms~(G17). G17 estimate a viewing angle of $58\pm20$$^\circ$  between the observer viewing angle and the \ti~momentum vector, and the angle to the NS kick would be the same (but approaching). This gives a line-of-sight NS motion of 205 \kms~and a total speed of 390 \kms. This is only about half the value in the W15-IIb model (719 \kms). For this model to give a plane-of-sky proper motion of 330 \kms, the viewing angle has to be about 30 degrees. Such an angle gives a $\chi^2$ well within factor 2 of the $\chi^2$ minimum (Fig. \ref{fig:CompCasA_LP}, right) and shows that a satisfactory solution for both line profiles and the NS plane-of-sky motion can be achieved by the model}.  

\arj{One should keep in mind that there is no reason why this particular model, the He core of a 15 \msun~star (plus a small H envelope) exploded with 1.5 B, should be the perfect match for Cas A. Improved fits would require yet some more asymmetry in the \ti~distribution; even less material on the approaching hemisphere and more on the receding one. There is currently no known correlation between the degree of asymmetry and properties such as He core mass or explosion energy, so we cannot speculate which combination of these could give better agreement. Rather, the main conclusion we can draw at this point is that the neutrino-driven explosion paradigm seems to pass another observational constraint; to (roughly) reproduce the \ti~decay line properties (and NS motion) in Cas A. This builds and expands upon the results in \citet{Wongwat2017} who found agreement with the image of the \ti~distribution.}

\arj{Finally, we mention that we compared line profile simulations with and without time-dependence, and found that this had only a minor impact, changing the line profile by order 5\%, which did not significantly affect the fits or results}.

\section{Correlation between line centroid shifts and neutron star kick}
\label{sec:thecorrelation}
\arj{An interesting question is whether the NS kick can be constrained by observed shifts of the decay lines. The 3D explosion simulations have shown that the \ni~and \ti~bulk motions tend to be in opposite direction to the NS kick motion \citep{Wongwat2013}. We further quantify that here by calculating the direction cosines for the NS kick and the \ni~and \ti~momentum vectors (Table \ref{table:dircos}). These are typically close to -1 meaning almost complete anti-alignment. Observations of SNRs give evidence for such anti-correlation between intermediate mass elements and NS kicks \citep{Katsuda2018}.}

\begin{table}
\centering
\begin{tabular}{|c|c|c|c|c|}
\hline
Element\textbackslash Model & B15-1L & M15-7b1 & W15-2L & L15-1L \\
\hline
$^{56}$Ni & -0.9920 & -0.9996 & -0.9775 & -0.7420 \\
X       & -0.9808 & -0.9998 & -0.9750 & -0.7301\\
$^{44}$Ti   & -0.9790 & -0.9983 & -0.9815 & -0.7950\\
O       & -0.9899 & -0.9988 & -0.9984 & -0.9533\\
\hline
\end{tabular}
\caption{Direction cosines of NS kick and element momentum vectors.}
\label{table:dircos}
\end{table}

\arj{The possibility of inferring the NS kick depends on the quantitative link between the velocity of the NS and the radioactive element bulk shift. To establish the full framework, the line profiles need to be simulated to also include the Compton scattering effects and understand the full relation between NS velocities and line centroid shifts. We explore this by studying also the W15-1 and L15-2 RSG models evolved to 1 day, to make the sample larger. We tested that remapped 1-day RSG models (Sect. \ref{sec:homology}) did not significantly differ from the long-term simulations in cases where both are available, for the purposes here.} 

\begin{figure*}
\includegraphics[width=0.6\linewidth]{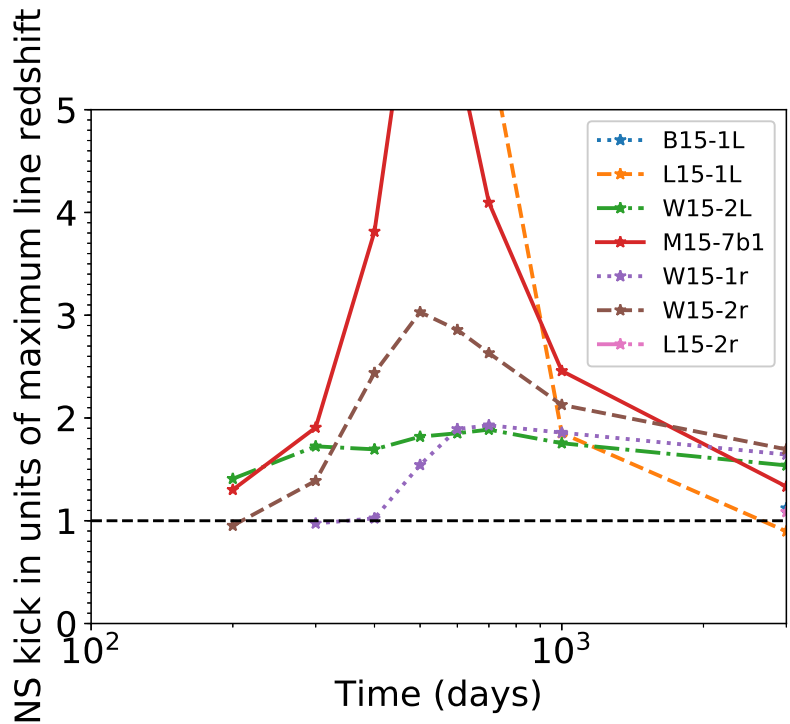}
\caption{\arj{Relations derived from the model grid of what minimum NS kick can be inferred from a measured \co~line centroid shift, in the case that the line is redshifted. The relations indicate that for observations at $t \lesssim$ 2000d, one can infer $V_{\rm NS} > V_{\rm redshift}$, and for later times $V_{\rm NS} > 0.7 V_{\rm shift}$. Note that if a model has no value for a given epoch, it means that it gave only blueshifted lines at that epoch (e.g. B15-1L before 3000d).}}
\label{fig:minkicks}
\end{figure*}

\arj{For the optically thick phase, observations at viewing angles where the NS moves away from the observer can give centroid blueshifts much larger than the NS kick. At 400d the maximum factor in our grid is 14 (for model B15-1L). Thus, if the observed line is blueshifted, little could be inferred about the NS kick in this phase}.

\arj{But the results are interesting for the case of observations of a redshifted line (as in SN 1987A after 400d). 
In Fig. \ref{fig:minkicks} we plot the NS kick relative to the largest possible line redshift for each model as function of time. 
Up to 2000d, all models predict that a redshifted line (if obtainable at all) has a centroid redshift smaller than the NS kick, i.e. $V_{\rm NS} > \Delta V_{\rm redshift}$. If this is generally true, the NS in SN 1987A must have a kick velocity of at least 500 \kms, based on the set of four resolved measurements of \citet{Tueller1990} which give $\Delta V_{\rm redshift} \gtrsim 500$ \kms. This can be related to the inferred distribution of NS kicks from pulsar proper motions, X-ray binary eccentricities, and NS-SNR offsets, that have revealed a relatively broad distribution between $\sim$10-1000 \kms, with a mean value of about 400 \kms \citep[e.g.,][]{Lyne1994, Hobbs2005}. Thus, 500 \kms~would be a relatively typical kick velocity, although likely in the upper half of the distribution.}

The ``blob'' identified by \citet{Cigan2019} as possibly heated by the NS would imply a transversal 
velocity component of either 220 or 700 \kms, depending on which of two methods to identify the remnant center is used.
For the first case, if the NS 3D speed is at least 500 \kms~as inferred here, that means a line-of-sight velocity of at least 450 \kms~
and an angle between the NS motion and the direction to earth smaller than 25$^\circ$. For the second case
of a 700 \kms~transversal velocity no constraint on the angle can be put.

\arj{Most models show a rise and decline behaviour with time of the ratio between NS kick and maximum redshift (Fig. \ref{fig:minkicks}). One should note that each epoch has its own viewing angle picked out (that gives the most extreme line redshift), so this is not a behaviour for a fixed viewing angle. It is nevertheless interesting that viewing angles with larger line redshifts can be found at 200-500d than at 500-1000d. This is not an apparent property and the relatively similar morphology of the curves from different models suggests that this may a generic property of the 3D morphologies}. 

\arj{In the optically thin limit ($t \gtrsim$ 3000d), the model grid gives maximum centroid shifts for the \co~lines of $0.7-1.6$ times the NS kick. Thus, for observations in this phase one could estimate $V_{\rm NS} = (0.6-1.4) V_{\rm shift}$ (inverting the $0.7-1.6$ range), for either blue or redshifted lines (without Compton scattering there is no distinction). Therefore $V_{\rm NS} > 0.6 V_{\rm shift}$ becomes the radiative transfer-independent limit.}

\arj{In principle such curves can be derived using also \ti. As discussed in the introduction there is, however, more uncertainty in the morphology of the \ti, and these model curves would therefore be more uncertain. The maximum \ti~redshifts in the model grid at 27y span $1.1-1.8$ times the NS kick. As such, \ti~line observations could also be used to infer a NS kick of at least 0.55 times the redshift. For SN 1987A the observed \ti~redshift is $730 \pm 400$ \kms, so the lower limit to the kick becomes $330~\mbox{km s}^{-1}\times 0.55=180$ \kms~from this method, i.e. weaker than the limit from the \co~lines.  No upper limit can be placed as line shifts can be arbitrarily low at certain viewing directions}. 

\arj{Having just a single stripped-envelope model, we cannot build up a similar picture for this case. 
For \ti~in the optically thin phase, the single model gives a maximum redshift of 1.7 times the NS kick, and with the observed line shift of $1950\pm 440$ \kms, the minimum NS kick would be 890 \kms. But we do not know how suitable this model is for Cas A, e.g. would a lower He core mass give larger ejecta velocities but perhaps without increasing the NS kick. Carrying out a more systematic grid analysis will be one valuable investigation when more such models become available. Also exploring whether the \co~relation holds in a larger sample of H-rich models with a larger variation in progenitor properties and explosion energies would be of interest}.

\section{The internal gamma-ray field and deposition} 
\label{sec:gammafield}

\arj{We now proceed to study the internal gamma-ray field and deposition in the 3D models and relate this to 1D models. The internal gamma field deposition will set the luminosity and profiles of the UVOIR emission lines, which will be modelled in more detail in a forthcoming paper}. 

\subsection{Gamma-ray field}
We emit \co~decay photons in its 47 lines, using random numbers to pick packet energies to be consistent with the branching ratios. We assume local trapping of the positrons (whose kinetic energy carries 3.5\%, on average, of the decay energy). We do not treat photoelectric absorption, but this should be unimportant above $\sim$50 keV \citep{Alp2018}; below this energy the typical decay photon has lost already $\gtrsim95\%$ of its energy and what happens to the rest becomes only as a small correction. As the X-ray absorption cross sections rise steeply below 50 keV we assume these photons to be immediately absorbed; the impact on the results was less than 1\%~when opposite limits were tested.

\arj{We follow the \citet{Lucy2005} method of constructing the radiation field $J_\gamma$ from path segments traversed by the photon packets}.
\subsubsection{Optically thin limit}
\begin{figure*}
\includegraphics[width=0.49\linewidth]{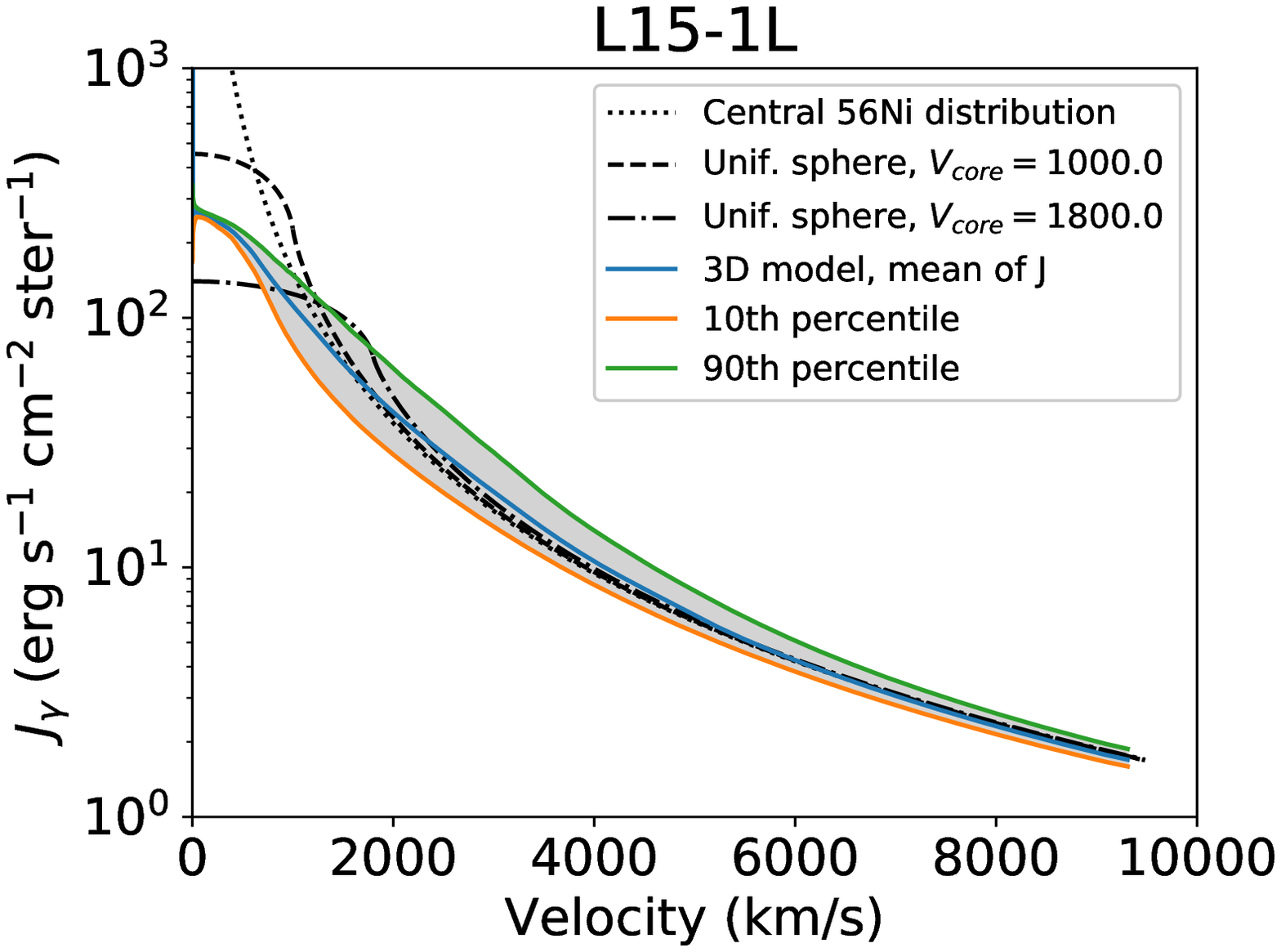}
\includegraphics[width=0.49\linewidth]{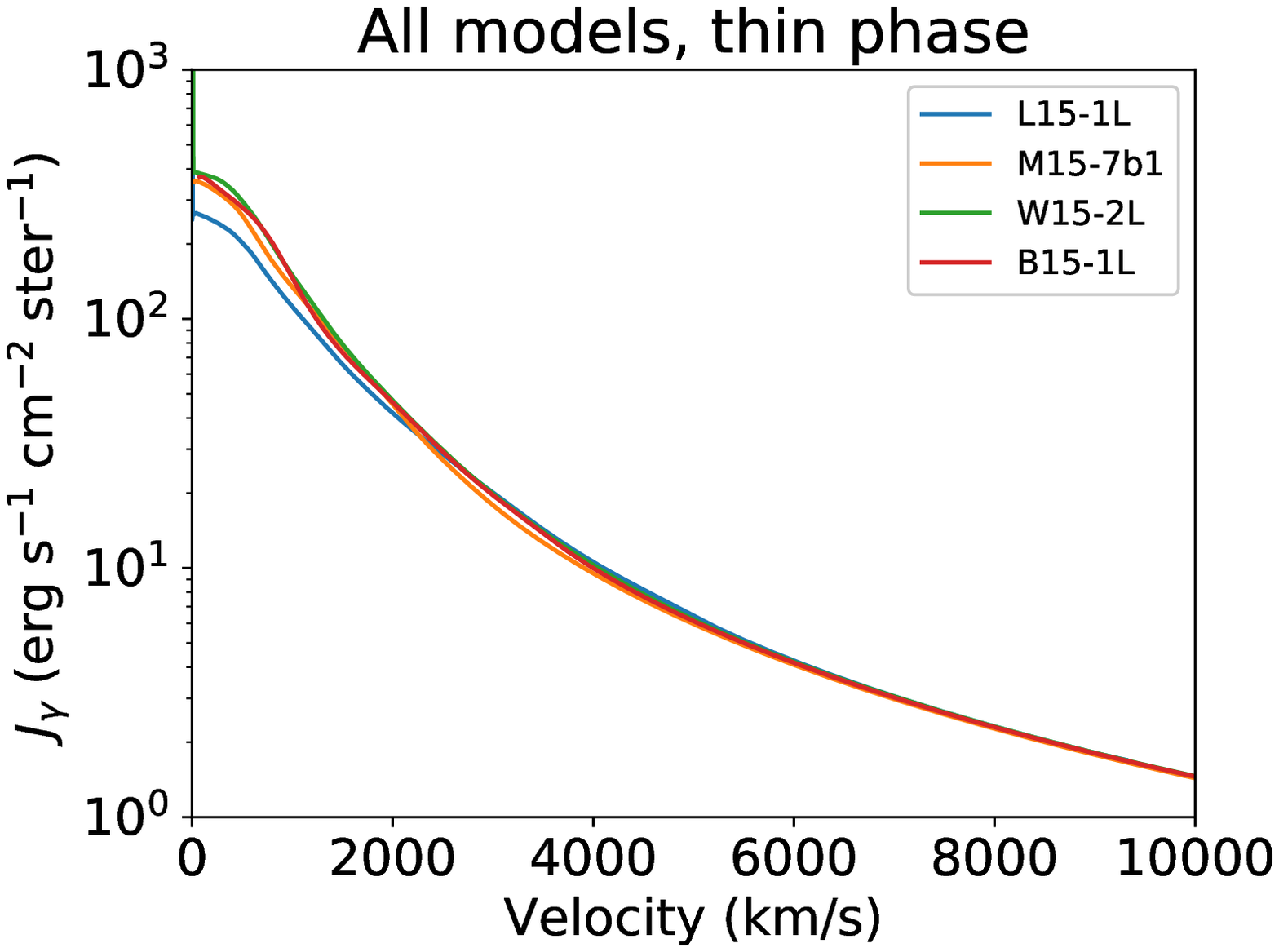}
\caption{\arj{Internal \co~gamma-ray field in optically thin limit (time-delay effects ignored). \textit{Left}: Model L15-1L, showing the mean (angle-averaged) value, and the 10th and 90th percentiles. Plotted is also a comparison to two 1D approximations. \textit{Right}: Mean value for all models.}}
\label{fig:jgamma_thin}
\end{figure*}

\arj{Fig. \ref{fig:jgamma_thin} shows properties of the bolometric gamma-ray field in the optically thin limit. In the left panel both the angle-averaged and the 10th/90th percentiles for a specific model (L15-1L) are shown.} 

\arj{For a given velocity, the gamma-ray field varies by a typical factor 2 with angle. The largest variation occurs around 2000 \kms, also in the other models, with a factor $\sim$2-3 for M15-7b1, W15-2L, and L15-1L, and a factor $\sim$1.5 for B15-1L. That B15 has the smallest variation is not surprising as it is the most symmetric model. 
Comparing the four models (right panel), the L15 model has the flattest angle-averaged distribution, whereas the other three models show small differences. Outside 2000 \kms~there are neglegible differences as most \ni~is then encompassed and its gamma rays are free-streaming. In the core region, $\lesssim$2000 \kms, the variation between models is a factor 1.5.}

\arj{Two common approximations used in 1D models, a central \ni~distribution and a uniform distribution within a spherical core region, are also shown for comparison. A central source model has a too steep gamma field, overestimating the gamma field inside $\sim$1500 \kms, and (slightly) underestimating it in the 1500-4000 \kms~range.  A uniform sphere model flattens the inner curve and brings it down to better agreement with the 3D model. This, however, comes at the expense of an overestimate at velocities around $V_{\rm core}$. } 

\begin{figure*}
\includegraphics[width=0.49\linewidth]{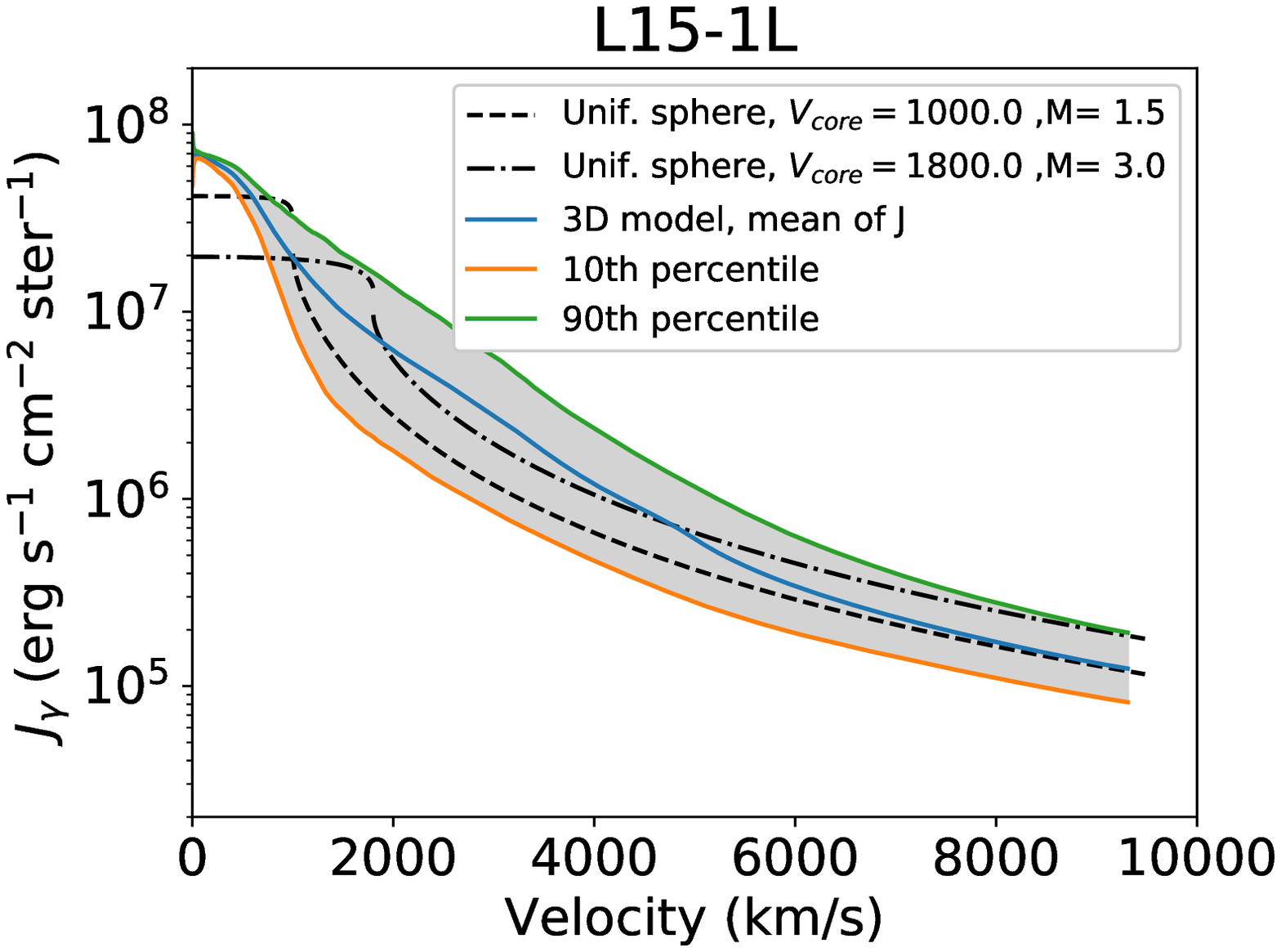}
\includegraphics[width=0.49\linewidth]{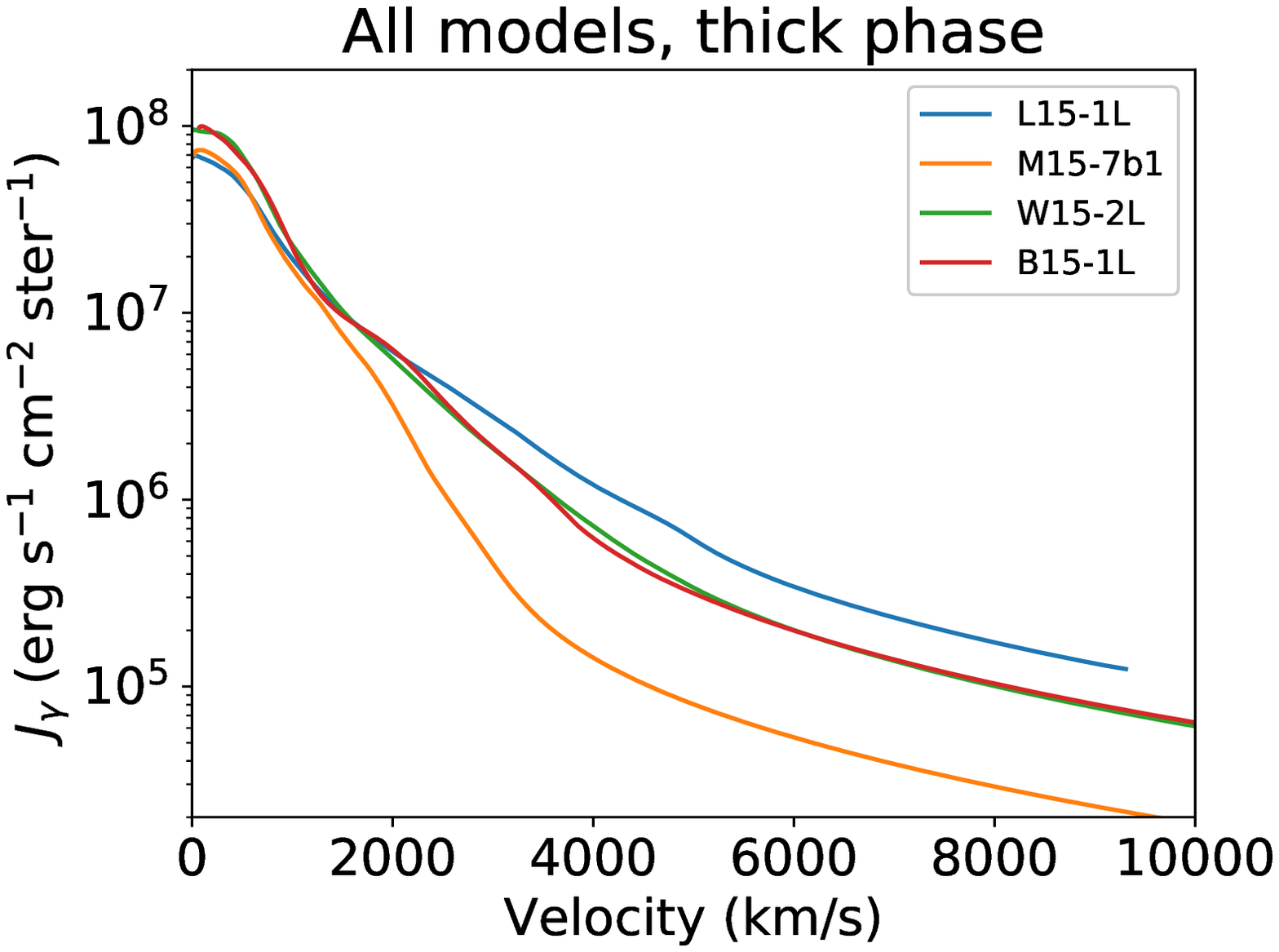}
\caption{\arj{Internal gamma-ray field in optically thick limit (300d). \textit{Left}: Model L15-1L, showing the mean (angle-averaged) value, and the 10th and 90th percentiles. Plotted is also a comparison to the case of a spherical, homogenous core (no material outside $V_{\rm core}$) for two different $(M,V_{\rm core})$ combinations, using an effective opacity of 0.04 cm$^2$g$^{-1}$. \textit{Right}: Mean value for all models.}}
\label{fig:jgamma_thick}
\end{figure*}

\subsubsection{Optically thick limit}
\arj{Fig. \ref{fig:jgamma_thick} shows properties in the optically thick limit, at 300d. Starting with the model-to-model comparisons (right panel), the \ni~morphology now severely impacts also the higher velocity regions. The gamma-field strengths at 2000 \kms~now differ by a factor $\sim$2 between models (compared to $\sim$10\% for the optically thin case), and at higher velocities by a factor $\gtrsim$10. The innermost regions, $\lesssim$ 1000 \kms, have similar degree of differences as in the thin limit. The high ejecta mass of model M15-7b1 here leads to a weak field in the outer envelope; the higher degree of \ni-mixing compared to e.g. model B15 does not offset the higher degree of trapping of the gamma rays in the central regions. Whereas the highest-velocity regions ($\gtrsim 4000$ \kms) are not important for direct, \co-powered emission, the physical conditions in them can have secondary effects on the UVOIR spectral formation due to radiative transfer processes.} 

\arj{In the left panel the angular variation for a specific model (again L15-1L) is plotted. The angular variation is here significantly larger (factor $\sim$10) compared to the thin phase (factor $\sim$2), as expected when the photons travel shorter paths and the radiation field becomes less diluted and more influenced by local morphologies. Comparison with theoretical curves for uniform spheres are also plotted. Whereas in the optically thin limit the only parameter for this is $V_{\rm core}$, in the general case with opacity one must also specify the quantity $\kappa_{\rm eff} M_{\rm core}$. The functions plotted use $\kappa_{\rm eff}=0.04$ cm$^2$g$^{-1}$ (to account for some H in the core region which raises $Y_e$) and two combinations of mass and velocity. Note that the tail of this function is for a vacuum outside $V_{\rm core}$, whereas there is significant absorption in the envelope of the models at this phase - a detailed fit outside $V_{\rm core}$ is therefore not expected and the true curves would be lower than the ones plotted outside $V_{\rm core}$. But it is of interest to see whether the core region approaches the flat $J_\gamma$ distribution obtained in uniform sphere models for $\tau \gg 1$. The 3D models do show a flattish region but only out to 500 \kms, after that $J_\gamma$ is quite steep. One may fit the innermost gamma field reasonably well by e.g. $M_{\rm core}=1.5$ \msun~and $V_{\rm core}=1000$ \kms, however then the gamma field is too weak further out in the ejecta (considering also the overestimate described above). This would be quite serious as most of the ejecta would be outside 1000 \kms~and would then receive too little powering. One can remedy this by making a larger and higher velocity core. Here $M_{\rm core}=3$ \msun~and $V_{\rm core}$=1800 \kms~produces quite a good fit throughout but at the expense of underproducing the most central regions, $\lesssim$ 1000 \kms.}

\subsection{Deposition}

\begin{figure*}
\includegraphics[width=0.43\linewidth]{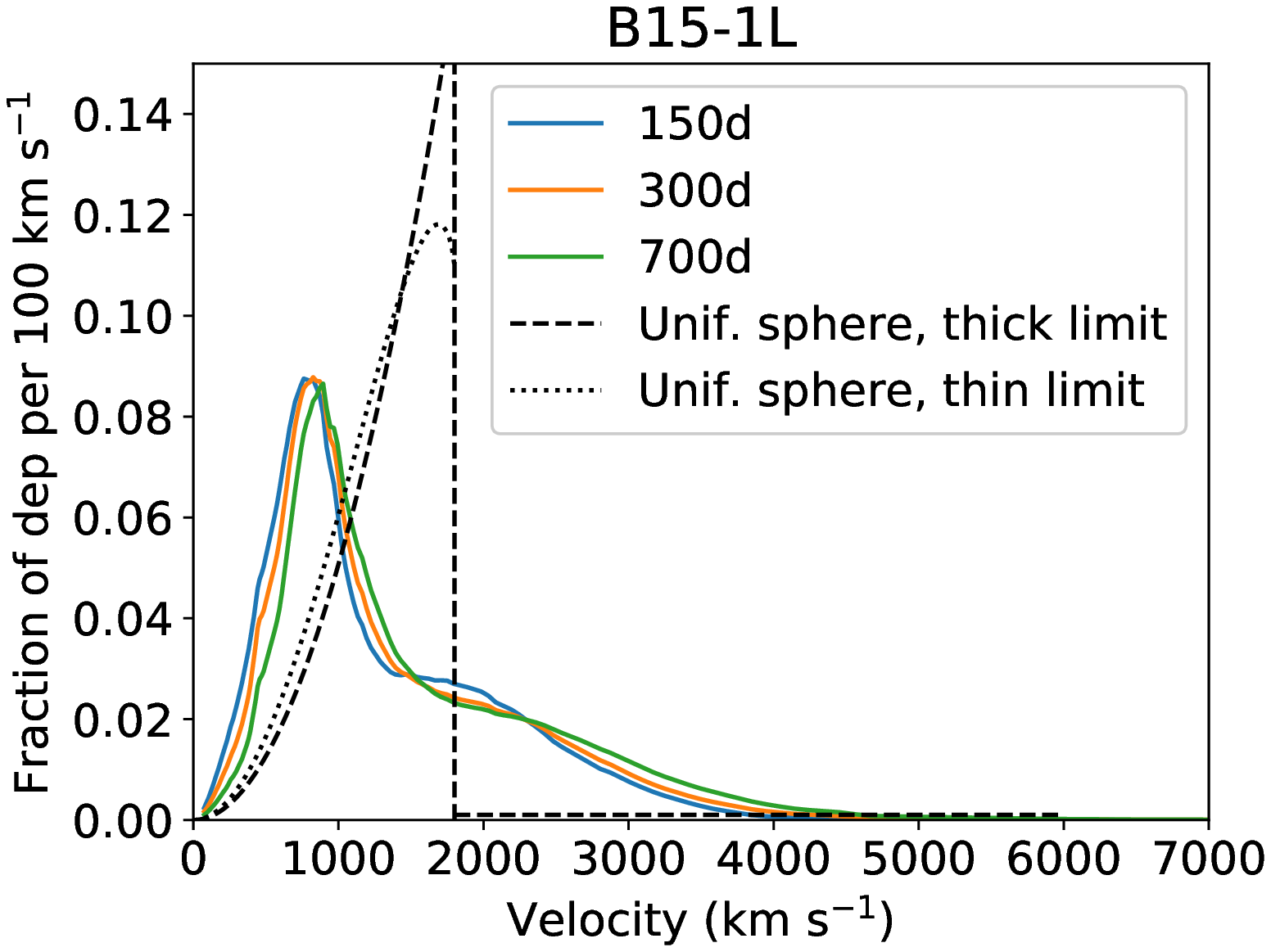}
\includegraphics[width=0.41\linewidth]{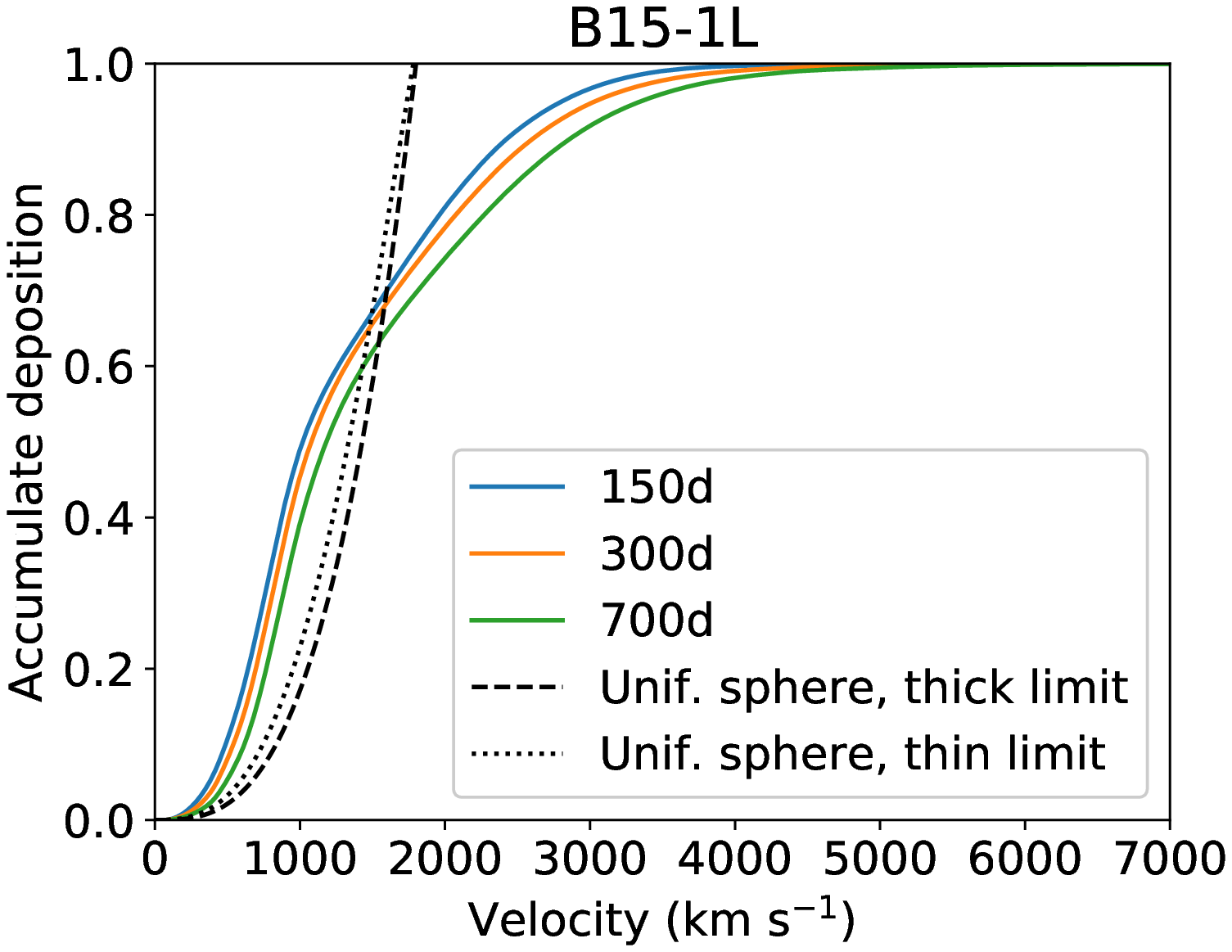}\\
\includegraphics[width=0.43\linewidth]{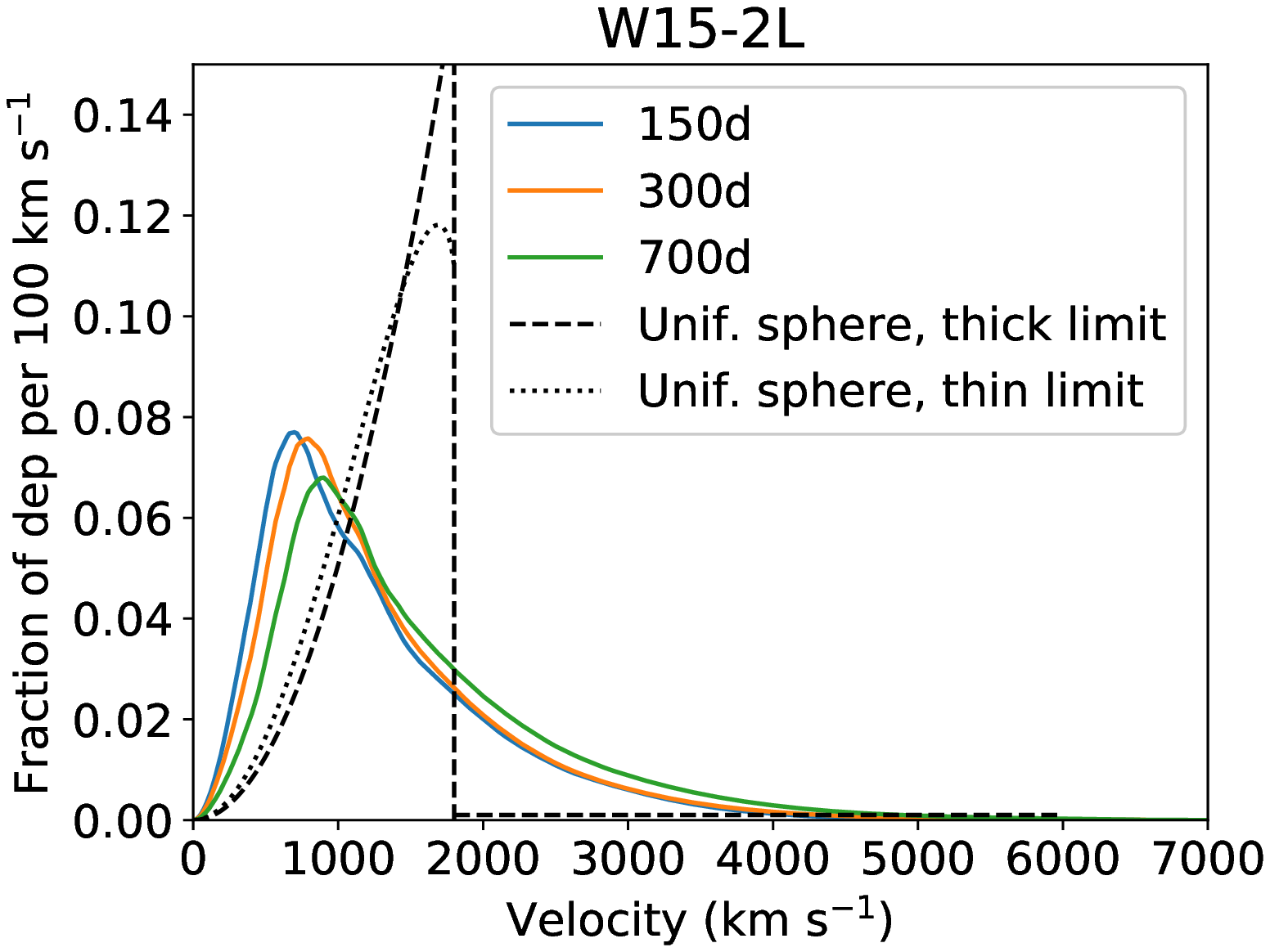}
\includegraphics[width=0.41\linewidth]{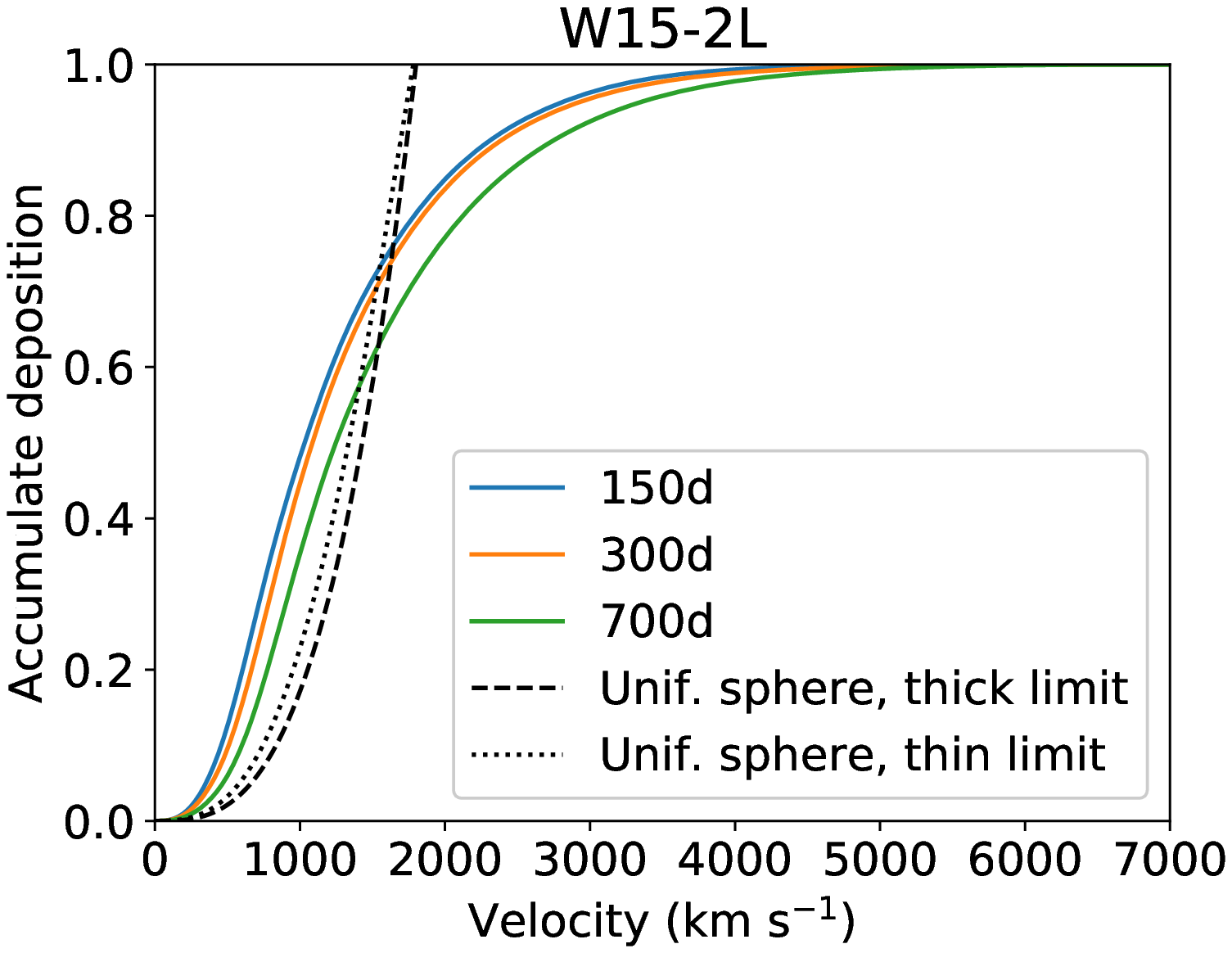}\\
\includegraphics[width=0.43\linewidth]{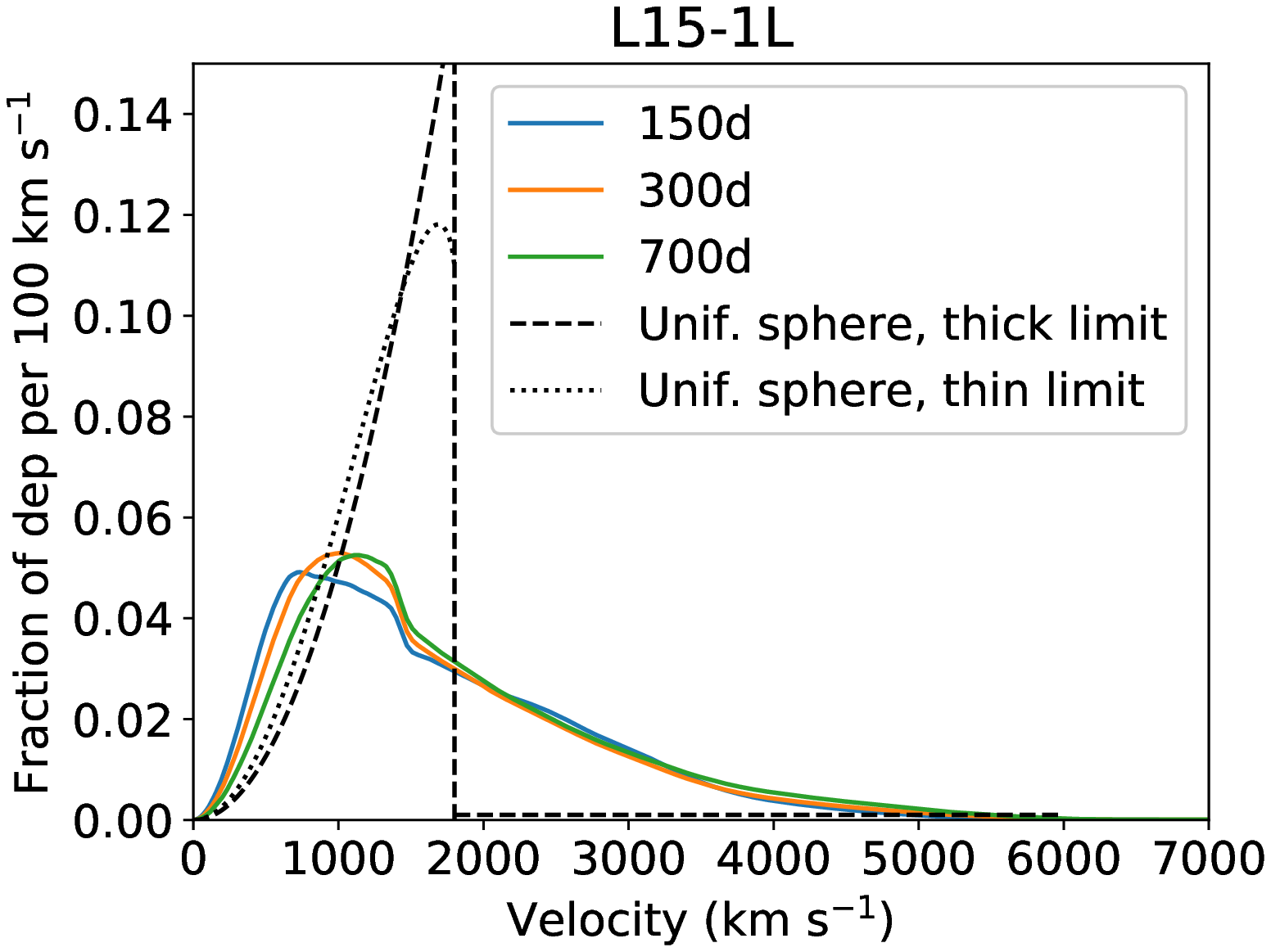}
\includegraphics[width=0.41\linewidth]{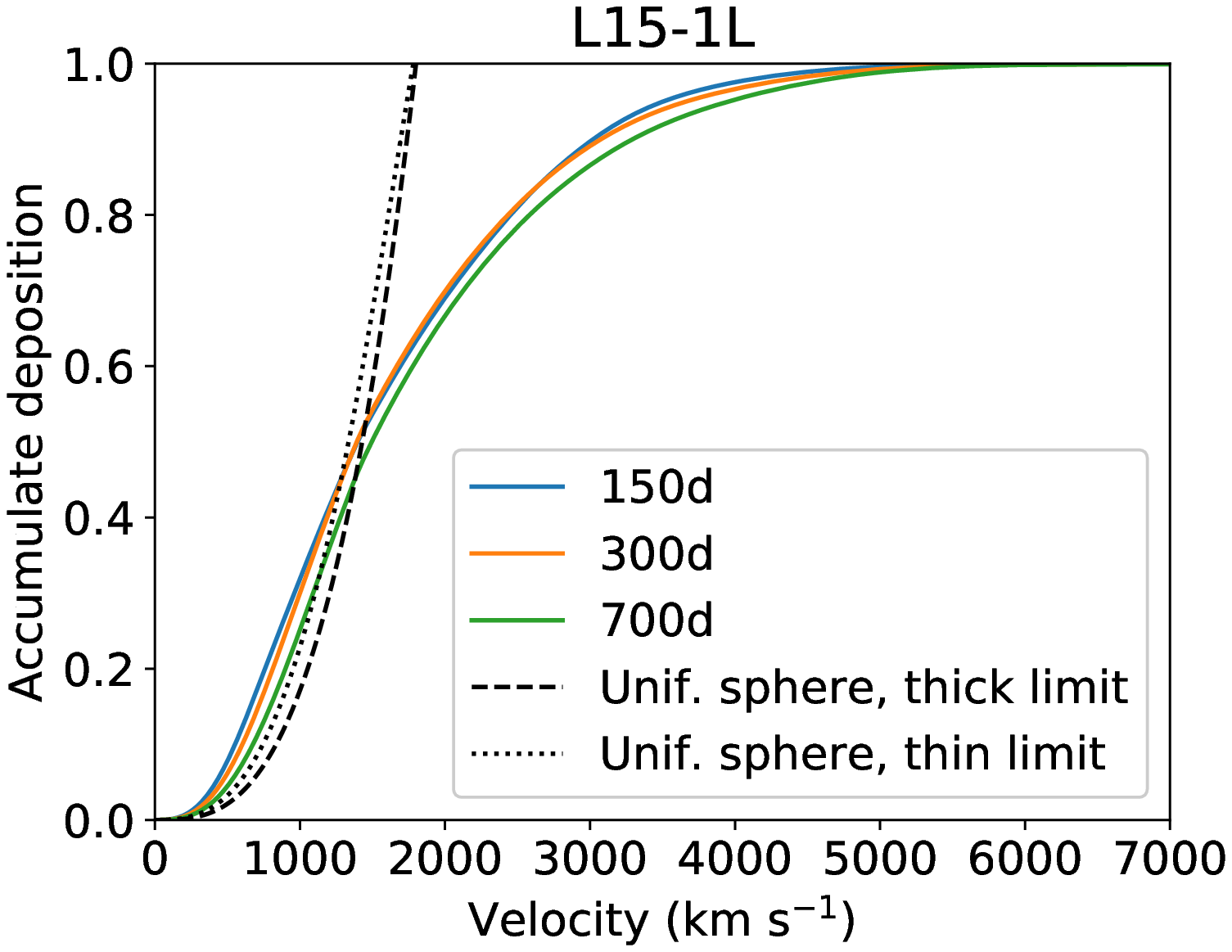}\\
\includegraphics[width=0.43\linewidth]{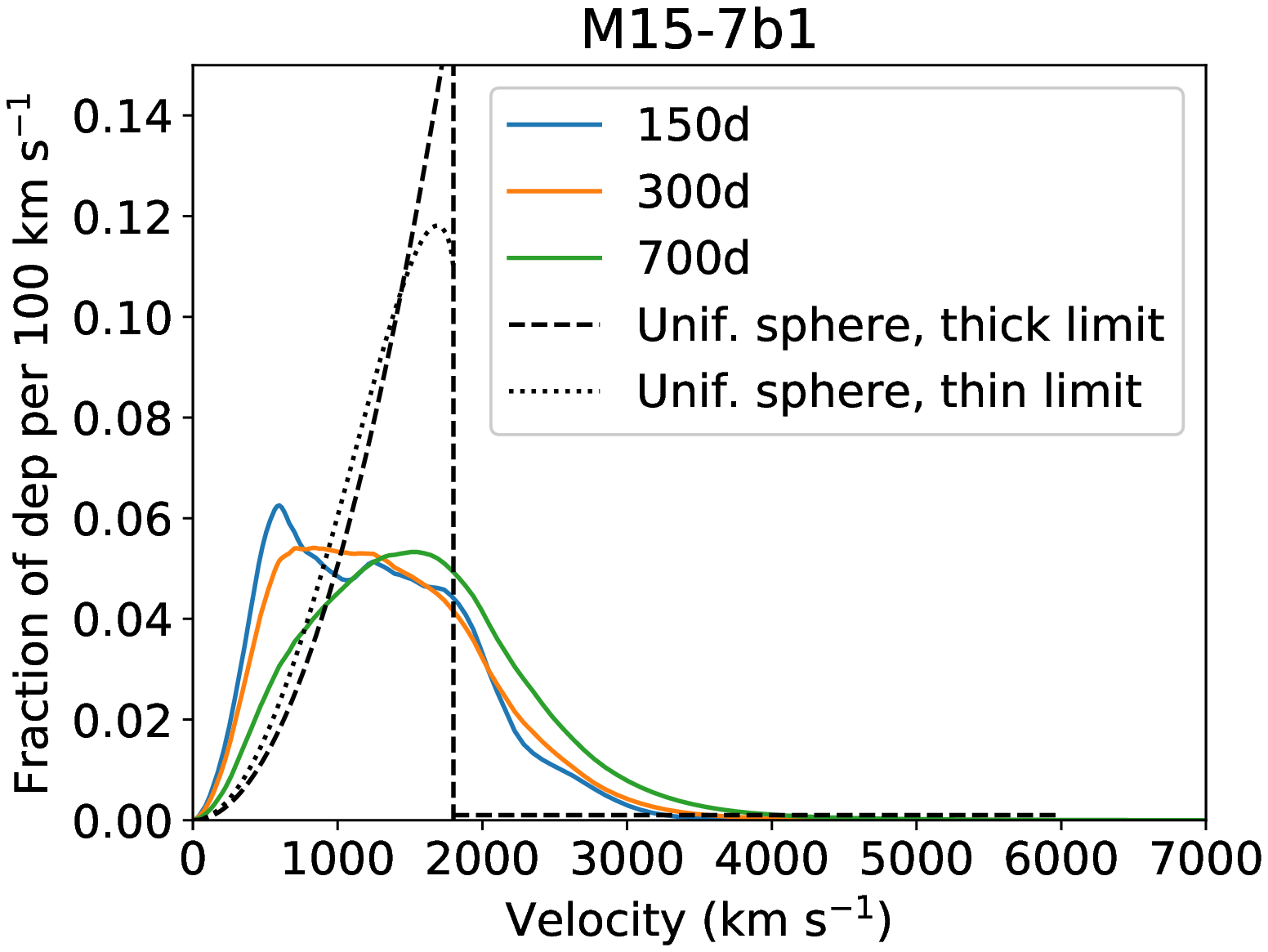}
\includegraphics[width=0.41\linewidth]{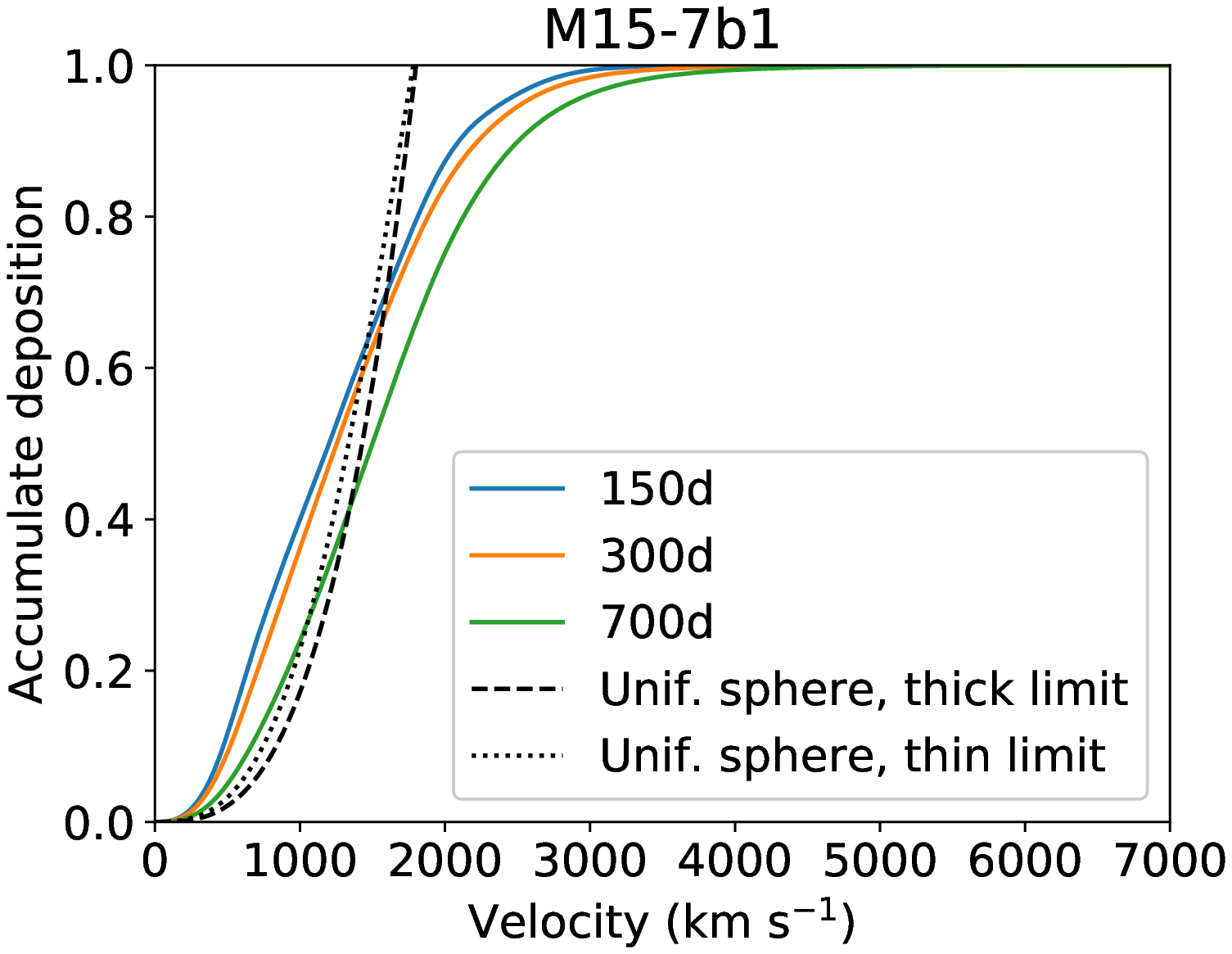}\\
\caption{\arj{Deposition fraction of gamma rays as function of velocity. \textit{Left:} Fraction of deposition per 100 \kms. \textit{Right}: Accumulative deposition. Also plotted is the function for a sphere with $V=1800$ \kms~in the optically thick (black, dashed) and thin (black, dot-dashed) limits.}}
\label{fig:deps_vsvel}
\end{figure*}

\arj{Looking at depositions, Fig. \ref{fig:deps_vsvel} shows the $\gamma$-ray deposition versus velocity, for three different times. The left panels plot the fractional deposition per 100 \kms, and the right panels plot the cumulative deposition}.

\arj{The peaks in the deposition-per-unit velocity curves range from a lowest value of 700 \kms (B15-1L, early times) to about 1500 \kms (M15-7b1, late times). If one imagines the supernova as a set of equally thick shells, the most important ones for nebular spectral formation are therefore the ones lying within $\sim$500-2000 \kms. This has the corollary that it is not overly crucial to model the innermost 500 \kms~to high accuracy; there is too little mass here to have any significant impact on the line formation. Outside 2000 \kms~the deposition is between $15-30$\% of the total, and beyond $\sim$4000 \kms~it is neglegible}. 

\arj{With time the deposition distributions move towards higher velocity material as the gamma rays can travel further distances before absorbed. This effect is however quite weak; the centroids of the deposition distributions increase by only about 20\% from 150 to 700d. The corresponding broadening of the emission lines may be searched for in spectral time-series (and the broadening rate used for model testing), although dust formation (which set in strongly after 530d in SN1987A) complicates this. After 700d, effects such as positron contribution, $^{57}$Co powering, and freeze-out will also start to affect the line formation}.

\arj{The deposition per velocity unity in a uniform sphere is $D_v \propto J_\gamma(V) V^2$, with the $V^2$ factor taking into account the area of the shells. In the optically thick case $J_\gamma$ is constant and so $D_v \propto V^2$. In the thin limit $J_\gamma = 1/2 + (1-V^2)/(4V)\ln{(V+1)/(1-V)}$ \citep{Kozma1992,Jerkstrand2017hb} so $D_v \propto 1/2V^2 + (V-V^3)/4\ln{(V+1)/(1-V)}$. Both limits are plotted in Fig. \ref{fig:deps_vsvel} for comparison with the 3D models. As expected, a uniform core model with its enforced constant density and sharp cut-off underproduces deposition in both low and high velocity regions, and over-deposits in the intermediate-velocity range. Adding an envelope tail to a uniform sphere model would bring about deposition at velocities higher than $V_{\rm core}$; this will push down the intermediate region (as the contributions are normalized) to better agreement with the 3D models, but will worsen the relative underproduction of the low-velocity region}.

\subsubsection{The total deposition}

\begin{figure}
\includegraphics[width=1\linewidth]{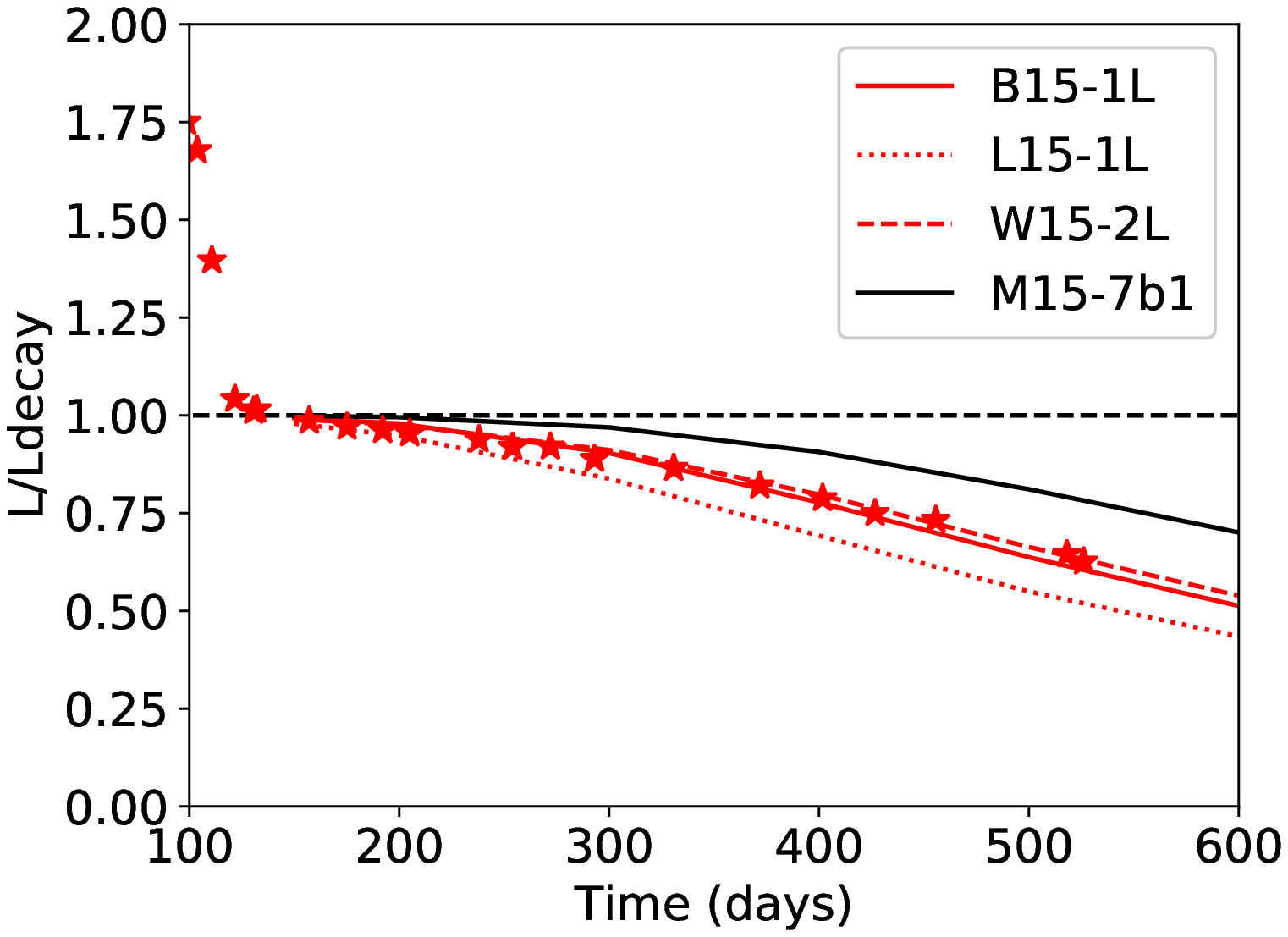}
\caption{\arj{Bolometric luminosity of SN 1987A normalized to the \co~decay luminosity (red points) \citep{Suntzeff1990,Jerkstrand2011thesis} compared to the models. The models with ejecta masses 13-14 \msun~are colored red, the one with 19 \msun~black.}}
\label{fig:depfracs}
\end{figure}

\arj{Figure \ref{fig:depfracs} shows the bolometric luminosity of SN 1987A \citep{Suntzeff1990} normalized to the decay power of 0.07 \msun~\co~\citep{Jerkstrand2011thesis}, see also \citet{McCray1993}, compared to the models. This is an important test for the models to pass, testing a basic global property independent of viewing-angle specifics. There is also little uncertainty in the data (SN 1987A was observed over all wavebands) or in the association between deposition and bolometric luminosity; neither time-dependent effects nor $^{57}$Co have any influence until later than $\sim$800d \citep{Fransson2002}, so steady-state with respect to the \co~deposition is valid over this time span}.

\arj{The B15-1L and W15-2L models (with ejecta masses of $14$ \msun) pass this test with close reproduction of the observed track. The primary property of importance for the total deposition is the characteristic column density of the ejecta, which depends on $M^2/E$ (see Section \ref{sec:optimal}). The $13-14$ \msun~ejecta models have values for this quantity of (using units of solar masses and $10^{51}$ erg for $M$ and $E$) 109 (L15), 129 (W15) and 142 (B15) and that order is maintained in the trapping (the second order effect would be the specific 3D morphology and outward mixing of \ni). Note that L15-1L has a somewhat higher explosion energy than the other models (1.7 B); this leads to faster ejecta and a somewhat too low degree of gamma trapping. The 19 \msun~model (M15-7b1) has a value $M^2/E=270$ and maintains a too high deposition compared to the data}.

\arj{These results mean that assuming a) the kinetic energy of $\sim$1.5 B is correct for SN 1987A, and b) these 3D simulations capture the morphology and mixing reasonably well, the ejecta mass is constrained to be close to 14 \msun \footnote{As the He core mass is well constrained to 4-6 \msun~from the progenitor luminosity, everything above that must be H envelope composition. Uncertainty in the division between He-core material and H-envelope material (which have different $Y_e$) therefore gives only a small uncertainty in the average opacity ($\sim$10\%), and the best-fitting $M$ then scales only with the square root of this error factor.}.  Masses as high as the one of M15-7b1 (19 \msun) struggle to avoid trapping for too long. Given the $M^2/E$ dependency, such masses would only become viable if the energy was about 3 B ($1.5 \times (19/14)^2$), which is certainly ruled out for SN 1987A.}

Granted, this result hinges on the observed UVOIR luminosity of 87A being correctly determined to better than 20-30\%. Note though that uncertainties in distance and extinction do not add to the relevant error here; the estimated \ni~mass would scale with such errors in the same way and cancel out. The relevant uncertainty comes from assumptions made in interpolating the SED between observed wavelength points. 

To our knowledge this is the first comparison of 3D model predictions to the UVOIR bolometric luminosity curve of SN 1987A. In contrast to decay line luminosities, this is a viewing-angle independent quantity and robustly tests the total gamma deposition in the ejecta. We may relate our estimate $M/E \sim 14~\mbox{M}_{\rm \odot}/1.5~\mbox{B}$ to estimates from 1D models of the diffusion phase. This is complicated by a division of $M$ into $M_{\rm env}$ and $M_{\rm core}$ in many such efforts. \citet{Woosley1988-peak} and \citet{Shigeyama1988} obtain $M/E = (10 + M_{\rm ejecta-core})/1.5$ which matches our value for $M_{\rm ejecta-core} \approx 4$ \msun.

\section{Infrared iron lines} 
\label{sec:infrared}

\subsection{Observations and method}
\arj{Observations of optical and infrared iron-group lines in SN 1987A at a few hundred days showed clear redshifts. \citet{Witteborn1989} measured redshifts (relative to LMC) of $440\pm100$ \kms for [Ni II] 6.63 $\mu$m and [Ar II] 6.983 $\mu$m at 400d. These measurements were for the main portion of the line, ignoring the red tails. \citet{Haas1990} obtained typical centroid shifts of 280 $\pm$ 140 \kms~for [Fe II] 17.94 and 25.99 $\mu$m at the same epoch; similar to \citet{Witteborn1989} these fits excluded a resolved redshifted clump at about 3900 \kms. \citet{Spyromilio1990} showed similar redshifts in [Fe II] 1.26 $\mu$m and [Ni I] 3.12 $\mu$m around day 500. The 1.26 and 17.94 $\mu$m lines have almost identical line profiles \citep{Haas1990}. These line profiles are different from the profiles that electron scattering can produce.  Whereas initial low-resolution observations interpreted excess emission on the red side as a Thomson scattering tail \citep{Witteborn1989}, higher resolution observations of other iron lines showed that this excess was dominated by a discrete \ni~clump at radial velocity 3900 \kms \citep{Haas1990}. Further, the line peaks themselves are redshifted, opposite to predictions in the electron scattering scenario \citep{Haas1990,Spyromilio1990}. Combined with the same asymmetry seen across a wide range of lines, an asymmetric element morphology appears to be the most likely explanation for the shifts. Finally, note that the observations were made before the epoch of strong dust formation in SN 1987A, and should therefore not be affected by this (and this anyway led to blueshifts, not redshifts).}

\arj{\citet{Witteborn1989} and \citet{Haas1990} report line widths of around $2500-2750$ \kms, with effects of telescope broadening removed. This is comparable to, but distinctly lower than the decay line width of around 3500 \kms~if one averages the different measurements at 400-700d. We will show that such a difference naturally emerges from the models.} 

\arj{The UVOIR emission is powered by the gamma ray deposition. A simple model for the iron lines is to estimate the emissivity as the gamma deposition times the number fraction of iron, assuming that each element emits in proportion to its abundance. These iron lines are mostly cooling lines excited by thermal collisions. As such their luminosity depends on the heating fraction, but this is relatively insensitive to conditions and should be in the 0.3-0.6 range \citep{Kozma1992}. But the cooling fraction in any given transition depends on the composition, temperature, density and ionization. As motivation for such a simple model, one may call upon previous 1D modelling results that have shown that density, ionization, and temperature are roughly similar in the inner 2000 \kms \citep[e.g.][]{Li2012}; however one should be clear that it is still a toy model. Note that we include also primordial Fe, although with the assumption to emit in proportion to mass fraction this component has no strong impact. In more realistic models this iron can emit a larger fraction of the deposited energy than the Fe number fraction due to its higher cooling efficiency compared to H and He \citep{Kozma1998-I,Jerkstrand2012}. We also do not consider the slight line profile distortions due to electron scattering; this should have a negligible impact at the epochs considered as the ionization degree $x_e  = n_e/n_{\rm nuclei} \ll 1$.}

\subsection{Results}
\begin{figure*}
\includegraphics[width=0.42\linewidth]{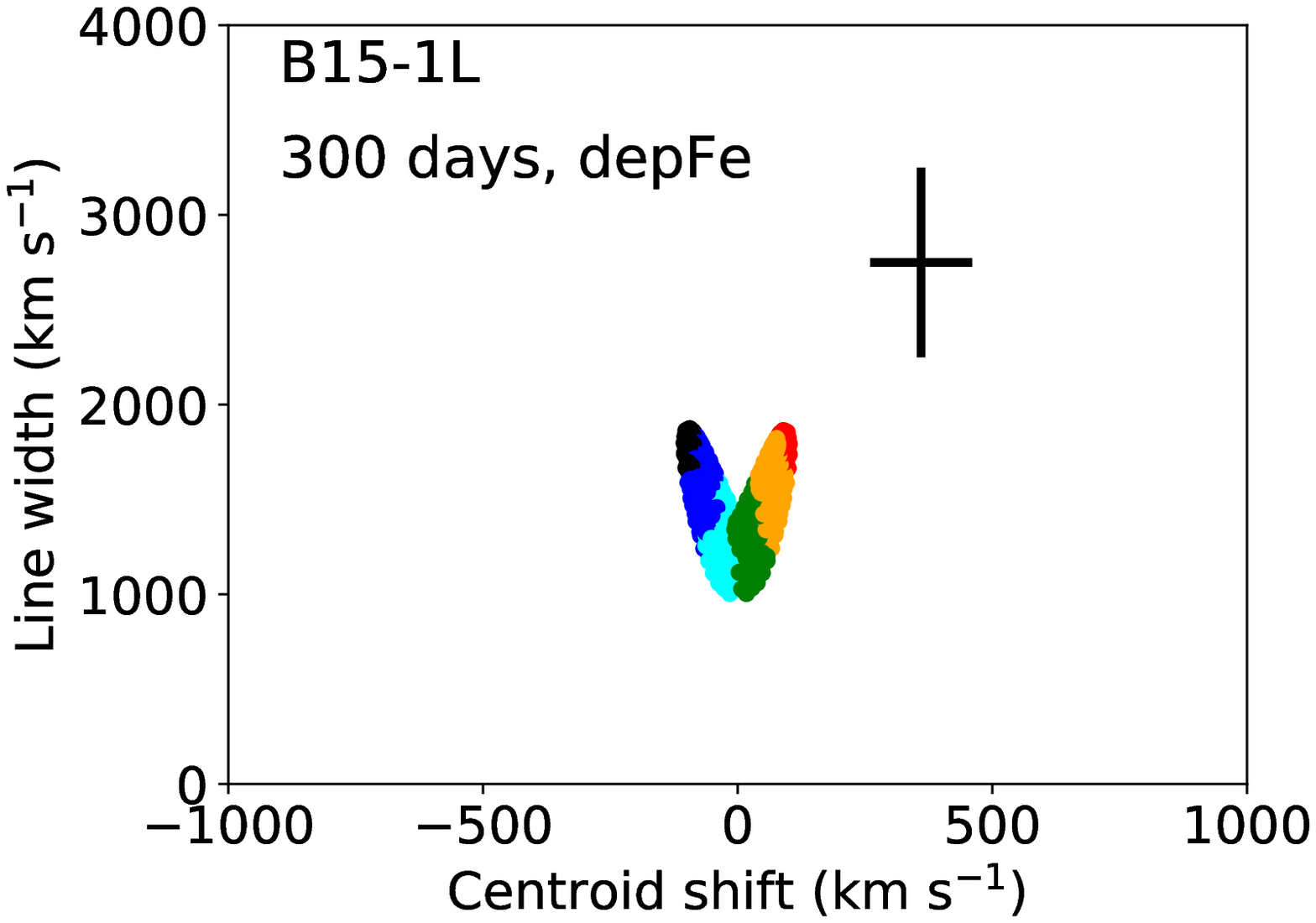}
\includegraphics[width=0.42\linewidth]{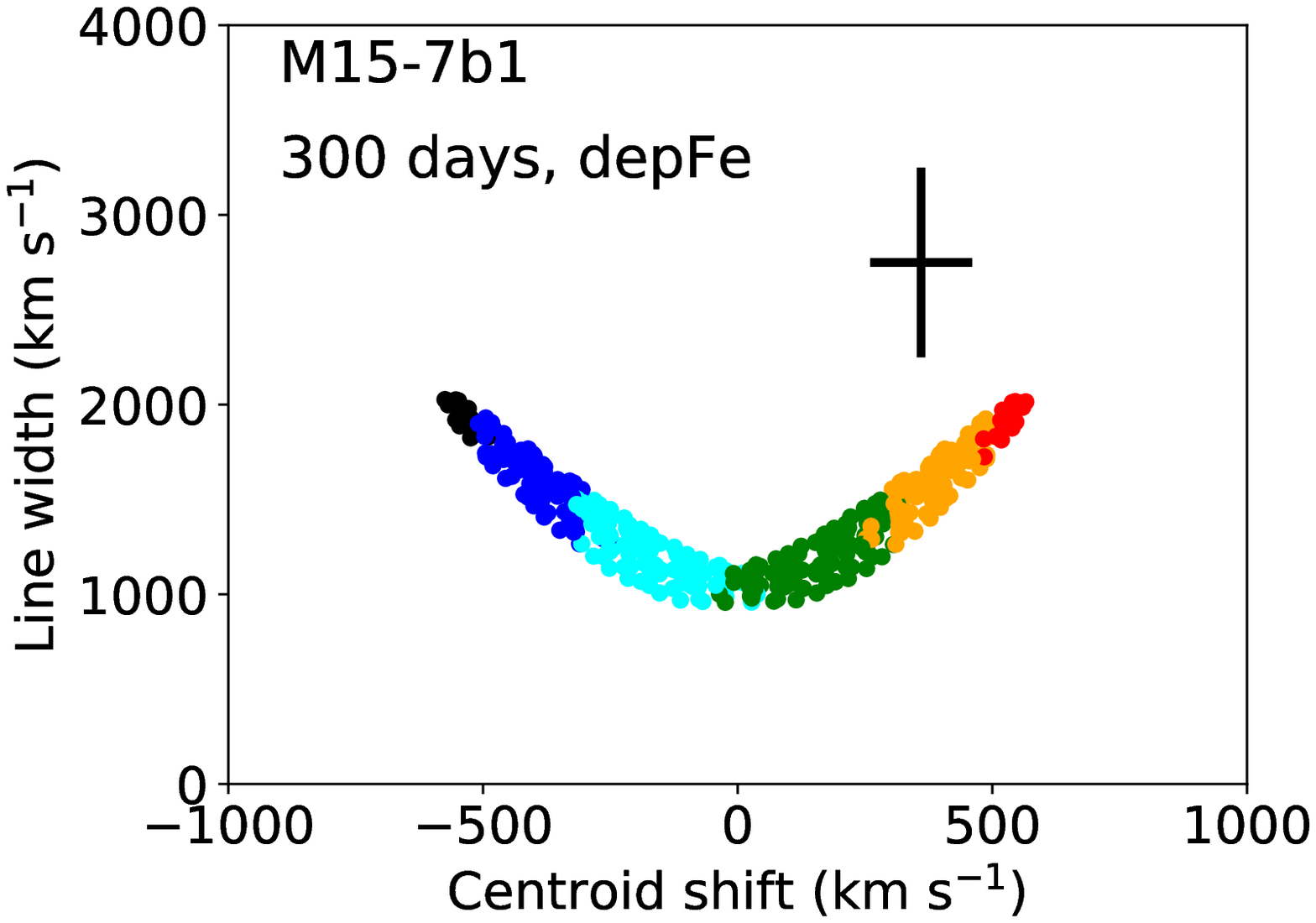}\\
\includegraphics[width=0.42\linewidth]{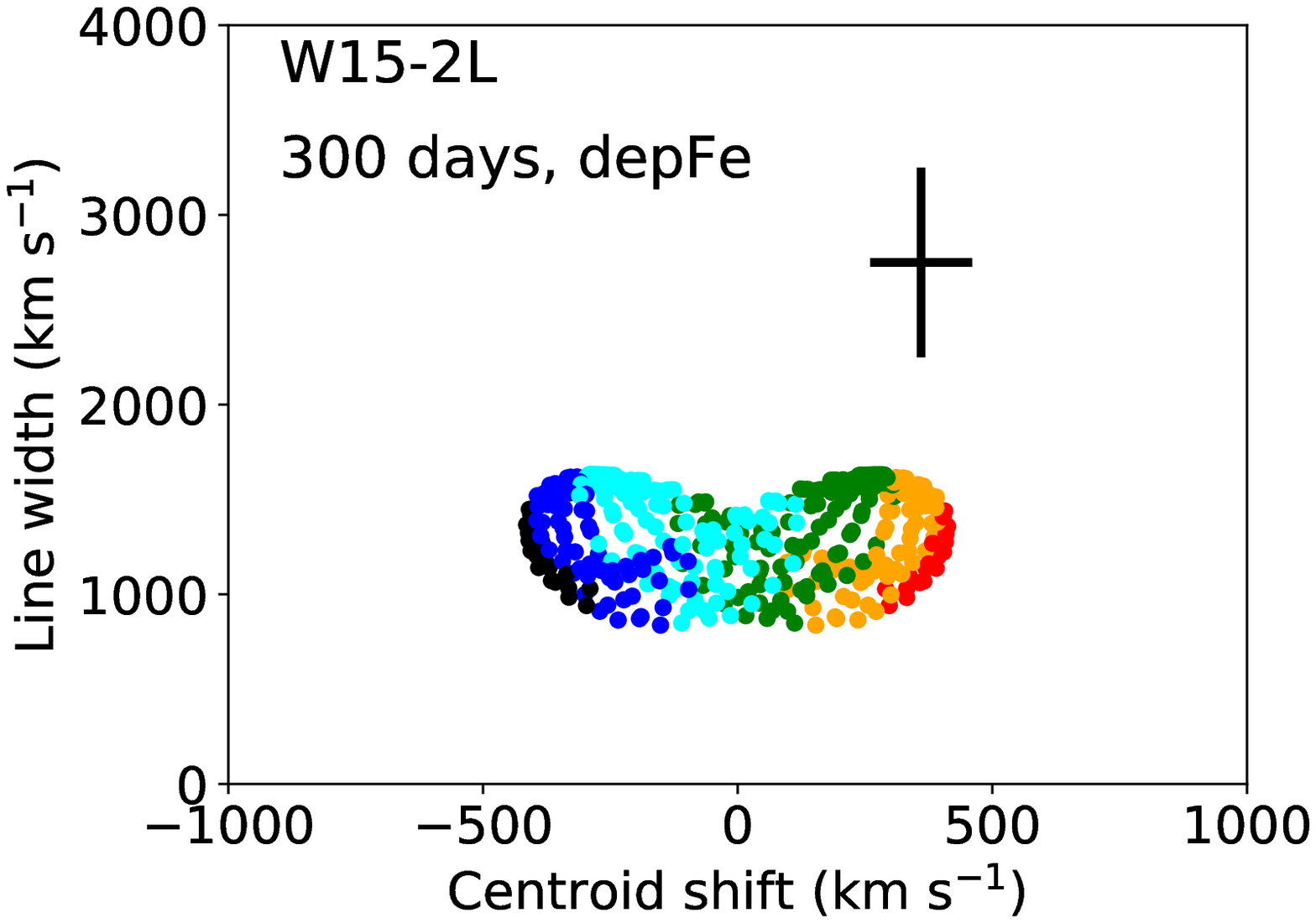}
\includegraphics[width=0.42\linewidth]{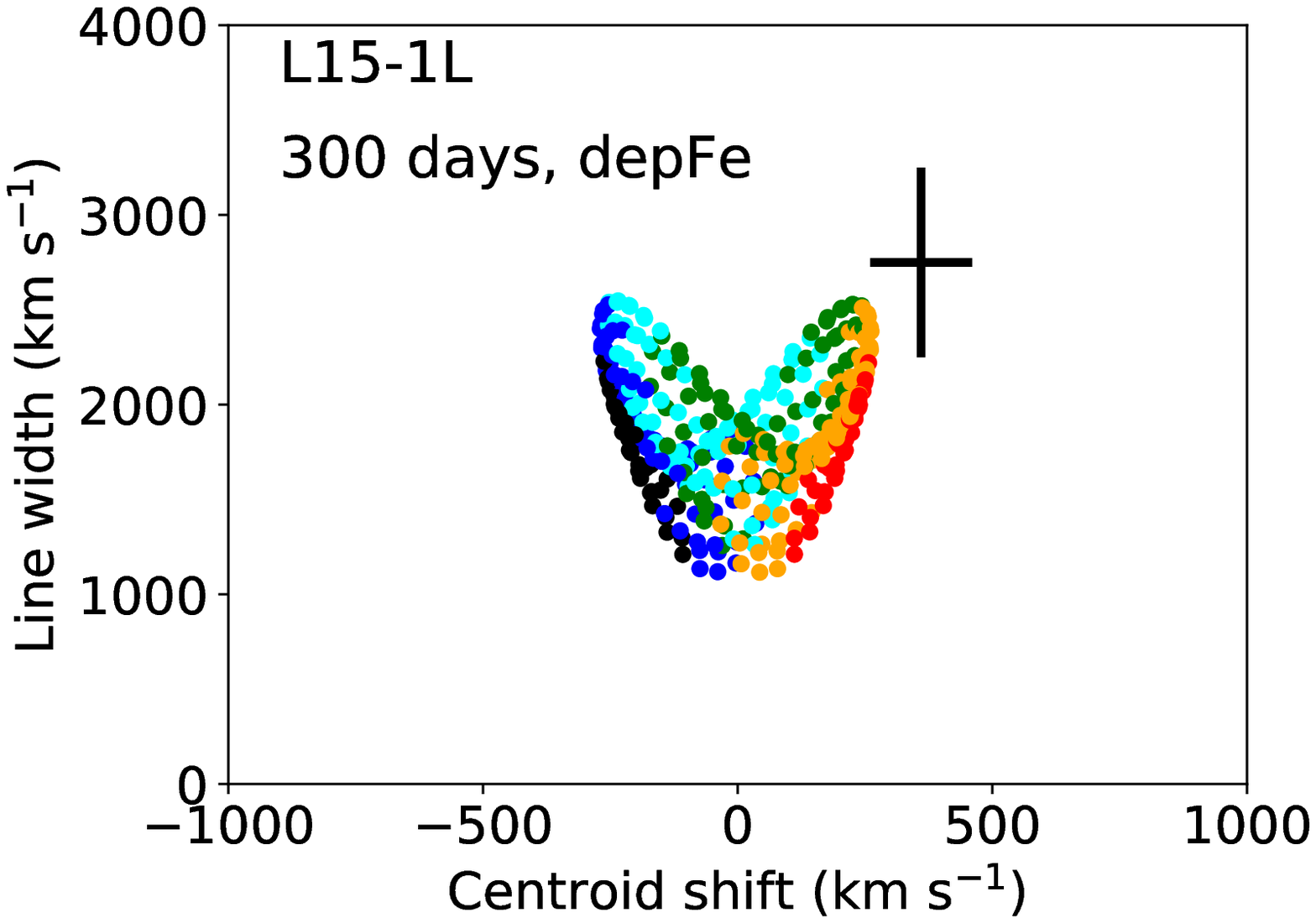}
\caption{Observed shifts (x-axis) and widths (y-axis) of Fe IR lines in SN 1987A, and the model predictions (taking the Fe emissivity to equal the gamma deposition times the Fe mass fraction) for different viewing angles. The viewing directions are color coded with the same scheme as in Fig. \ref{fig:LP_all}.}
\label{fig:IRcomps}
\end{figure*}

In Fig. \ref{fig:IRcomps} the four models are compared to the IR observations. Here, we have chosen to bundle the various measurements into a bulk estimate of a shift of 360 \kms~(the mean of the \citet{Witteborn1989} and \citet{Haas1990} measurements, with an error of $\pm$ 100 \kms (the typical scale of reported errors), and a width of 2750 \kms, with a $\pm$500 \kms~error. We note again that the reported fits excluded the most redshifted emission in the lines; thus both shift and width of the lines are in fact slightly underestimated.

Reviewing the comparisons, the picture is similar as obtained from the decay lines. Here, the data is of better quality, and radiative transfer effects on the photons is not an issue (IR lines face no significant opacity). We also probe all the iron, whereas for the case of decay lines only a minor fraction may contribute. On the other hand, the model for the emissivity is less accurate; being exactly known in the case of decay photons but just roughly approximated here for IR photons.

The comparisons show that model B15-1L has too symmetric \ni, at too low velocities. Even though B15-1L reaches high velocities for some of its \ni, the mass in this component is too small to affect this observable. Model M15-7b1 achieves the most extreme shifts of all the models, at $\pm$600 \kms. It also shows the most distinct correlation between shift and width; with stronger shifts going together with broader lines. Such a correlation is seen to some extent in all the models, although less distinct in the others (and almost absent in W15). The \ni~is, however, also here at too low bulk velocity. Model W15-2L achieves sufficient shift (450 \kms), but the lines are also here too narrow. Finally, L15-1L is the only model that achieves a sufficient line width, with values up to 2500 \kms. It marginally reaches enough redshift of the line, with maximum values close to 300 \kms. We note that this model has somewhat higher explosion energy (1.7 B) than the others (1.4-1.5 B) and this is one way to make the lines broad enough (compare e.g. to the models of \citet{Orlando2019} where $E\sim 2$ B is favoured). However, one should also note that e.g. model M15-7b1 with $E=1.4$ B has enough asymmetry of the iron, and had the ejecta mass been lower than its quite extreme 19.4 \msun, the line widths would likely also have come into agreement with observations.

\begin{figure*}
\includegraphics[width=0.42\linewidth]{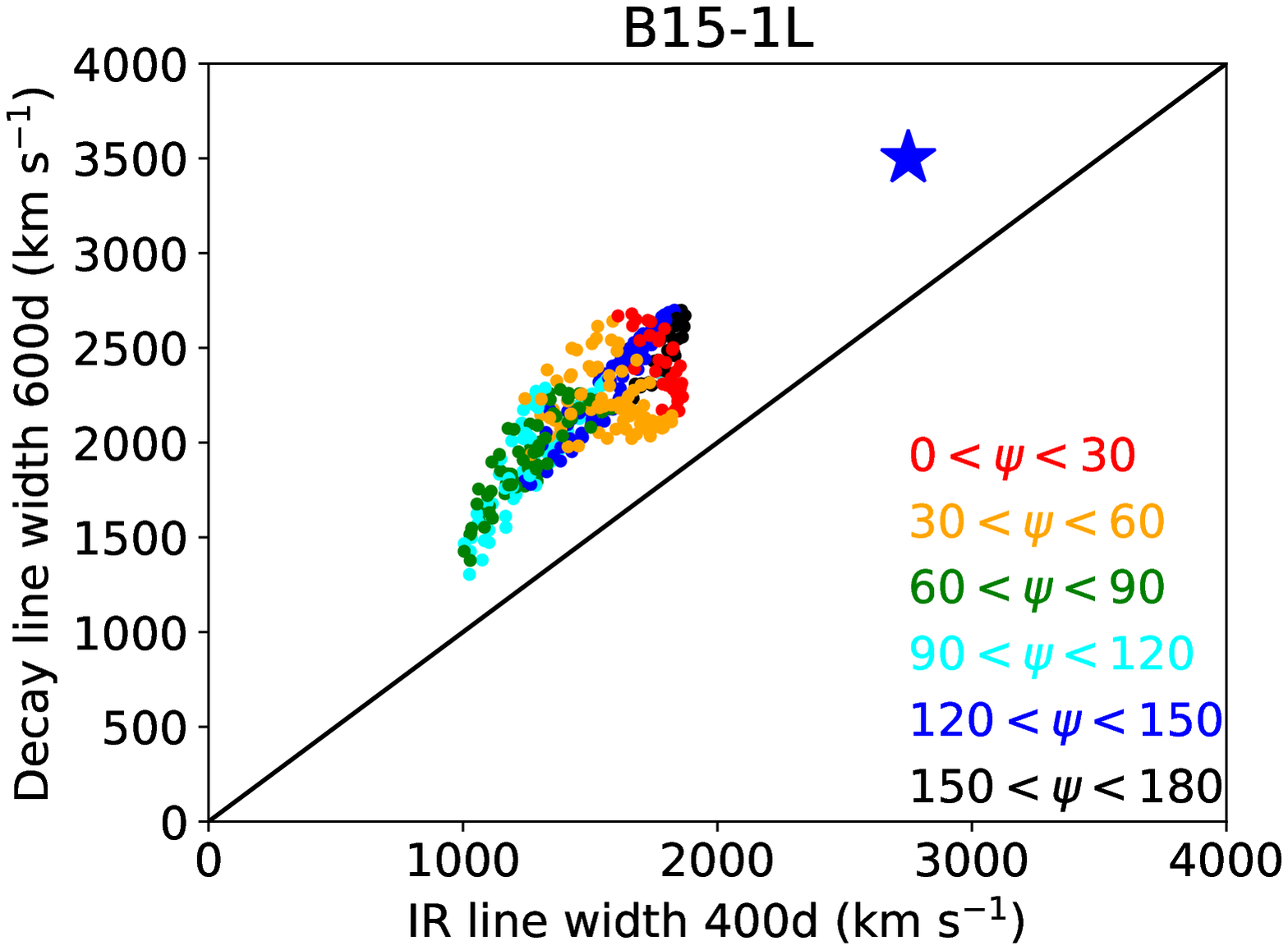}
\includegraphics[width=0.42\linewidth]{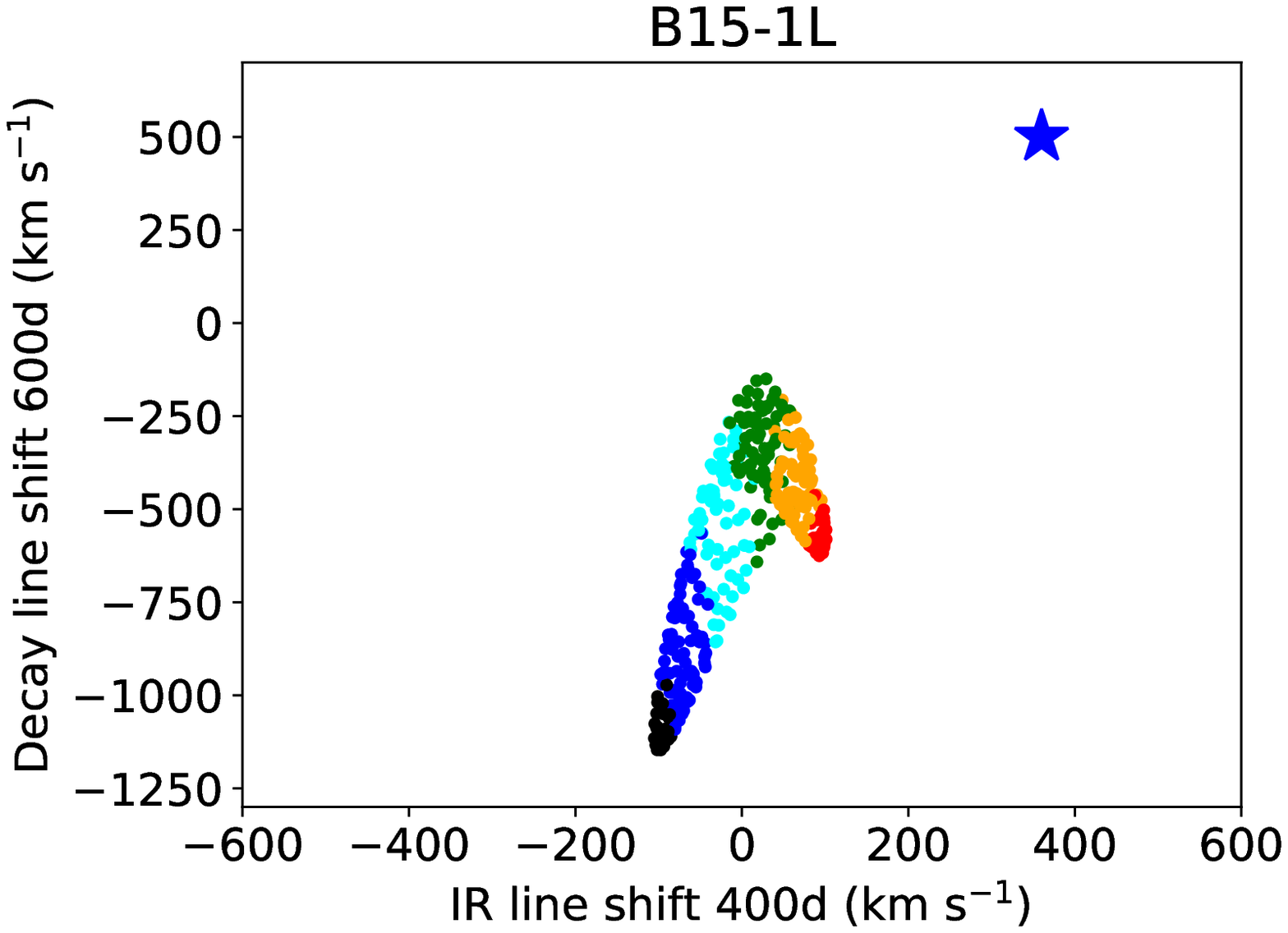}
\includegraphics[width=0.42\linewidth]{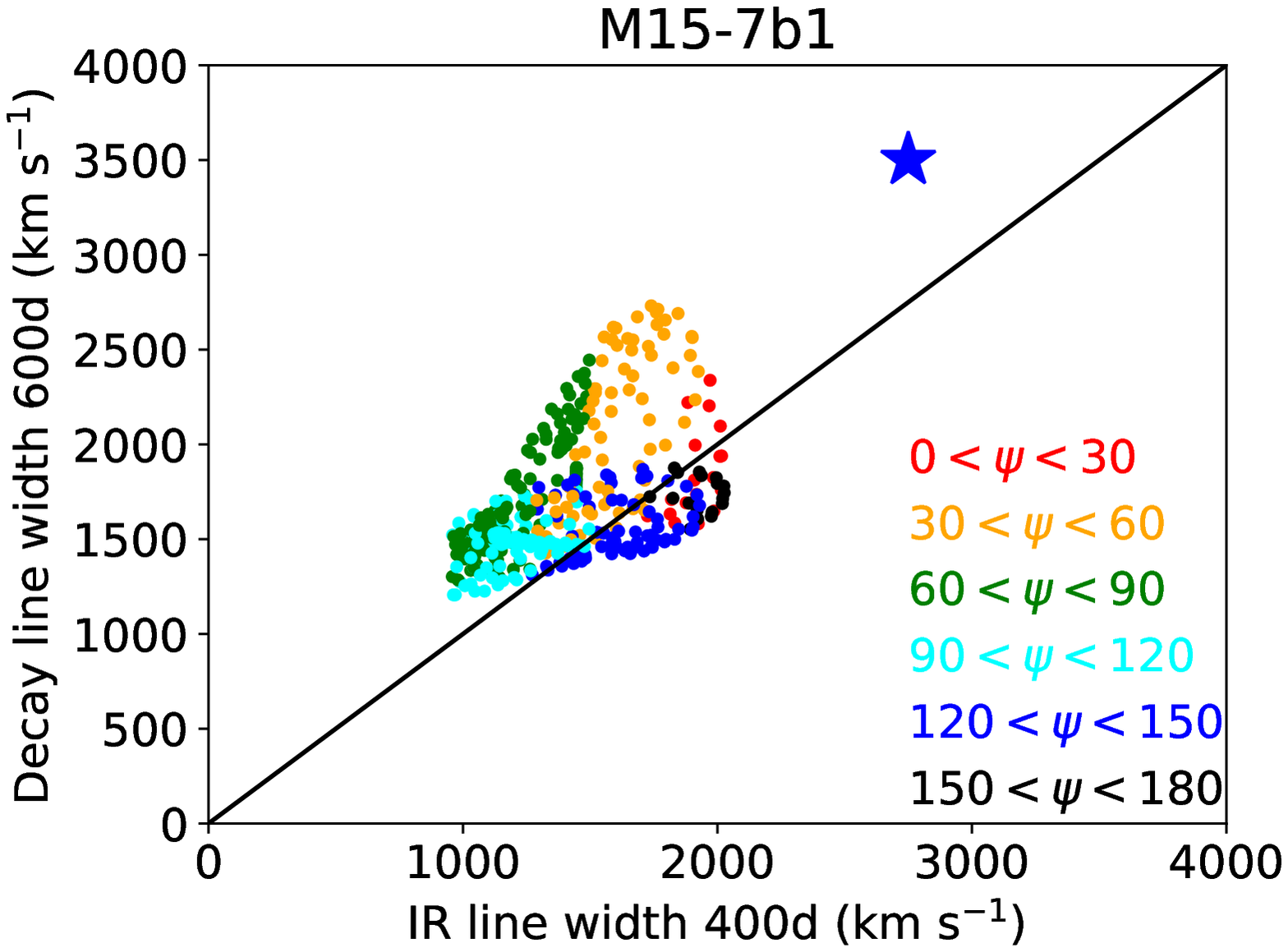}
\includegraphics[width=0.42\linewidth]{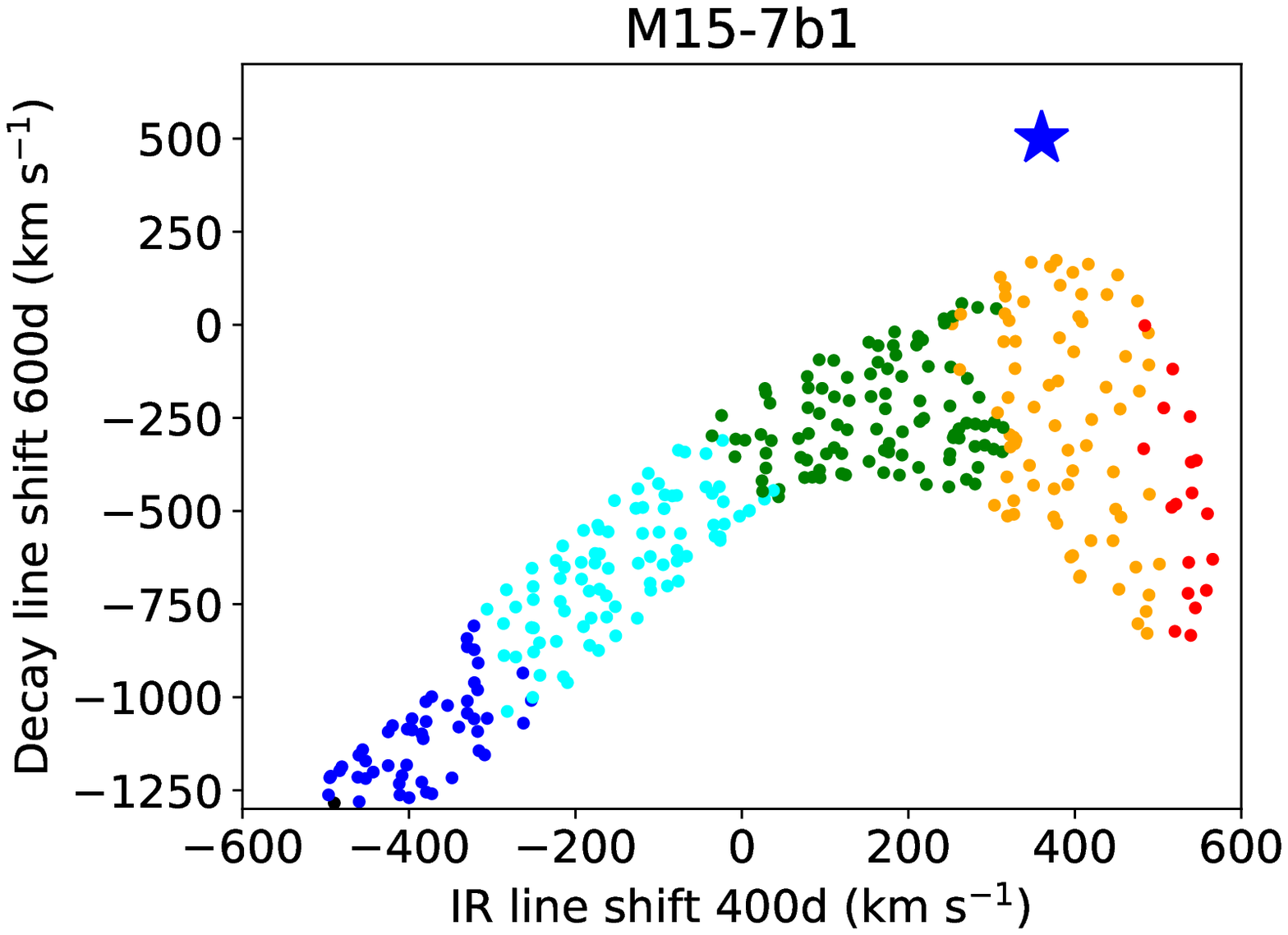}
\includegraphics[width=0.42\linewidth]{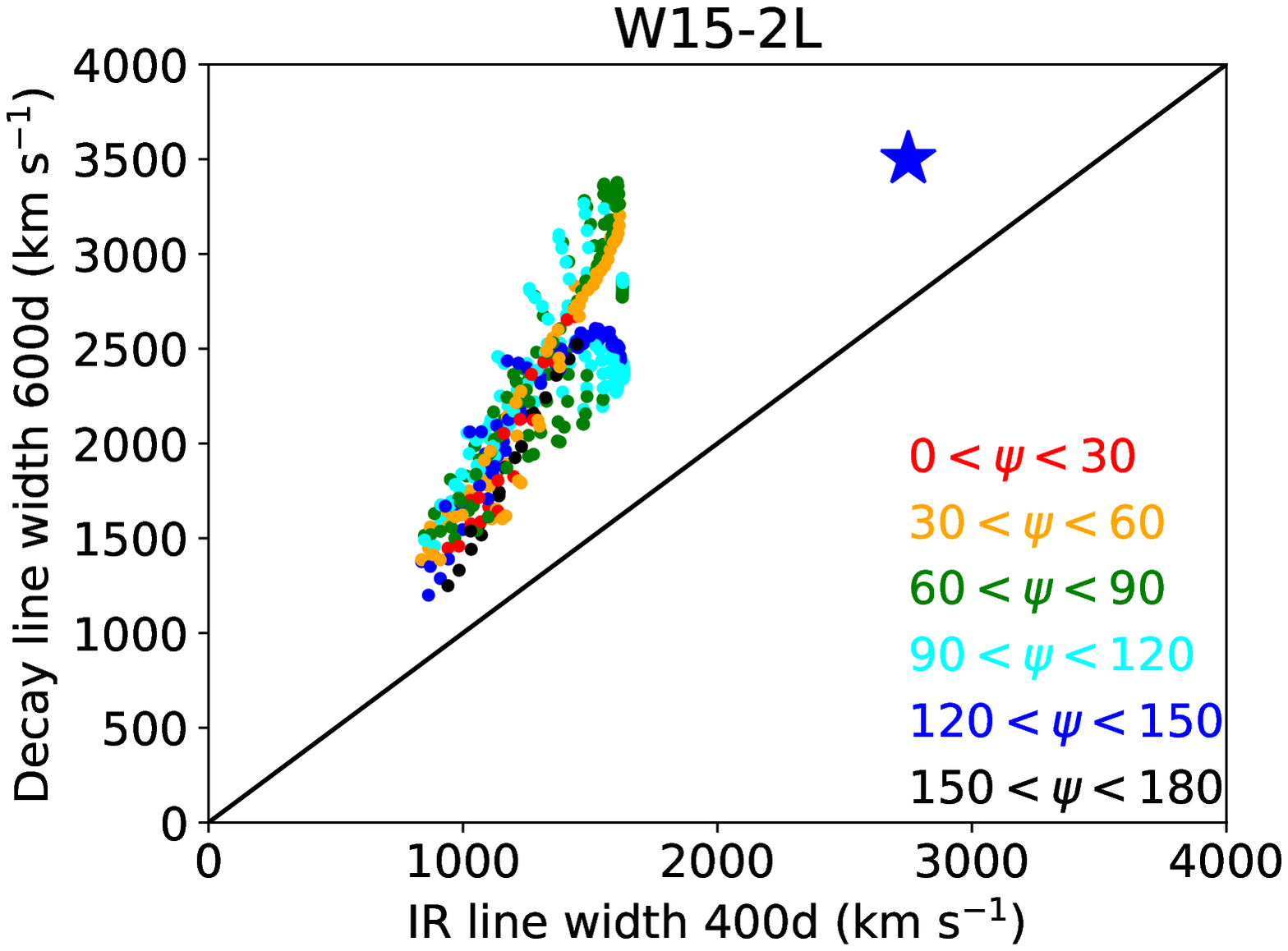}
\includegraphics[width=0.42\linewidth]{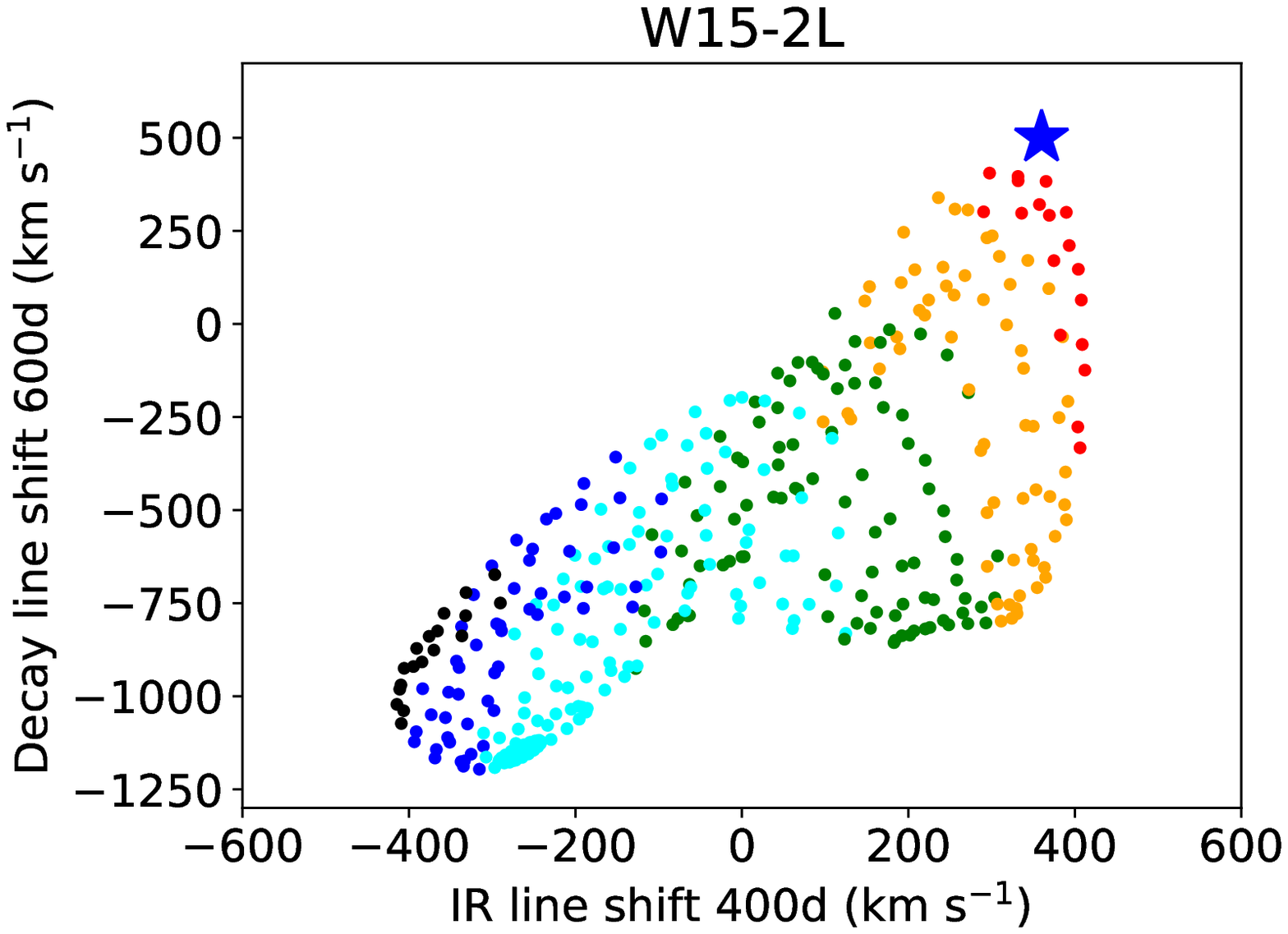}
\includegraphics[width=0.42\linewidth]{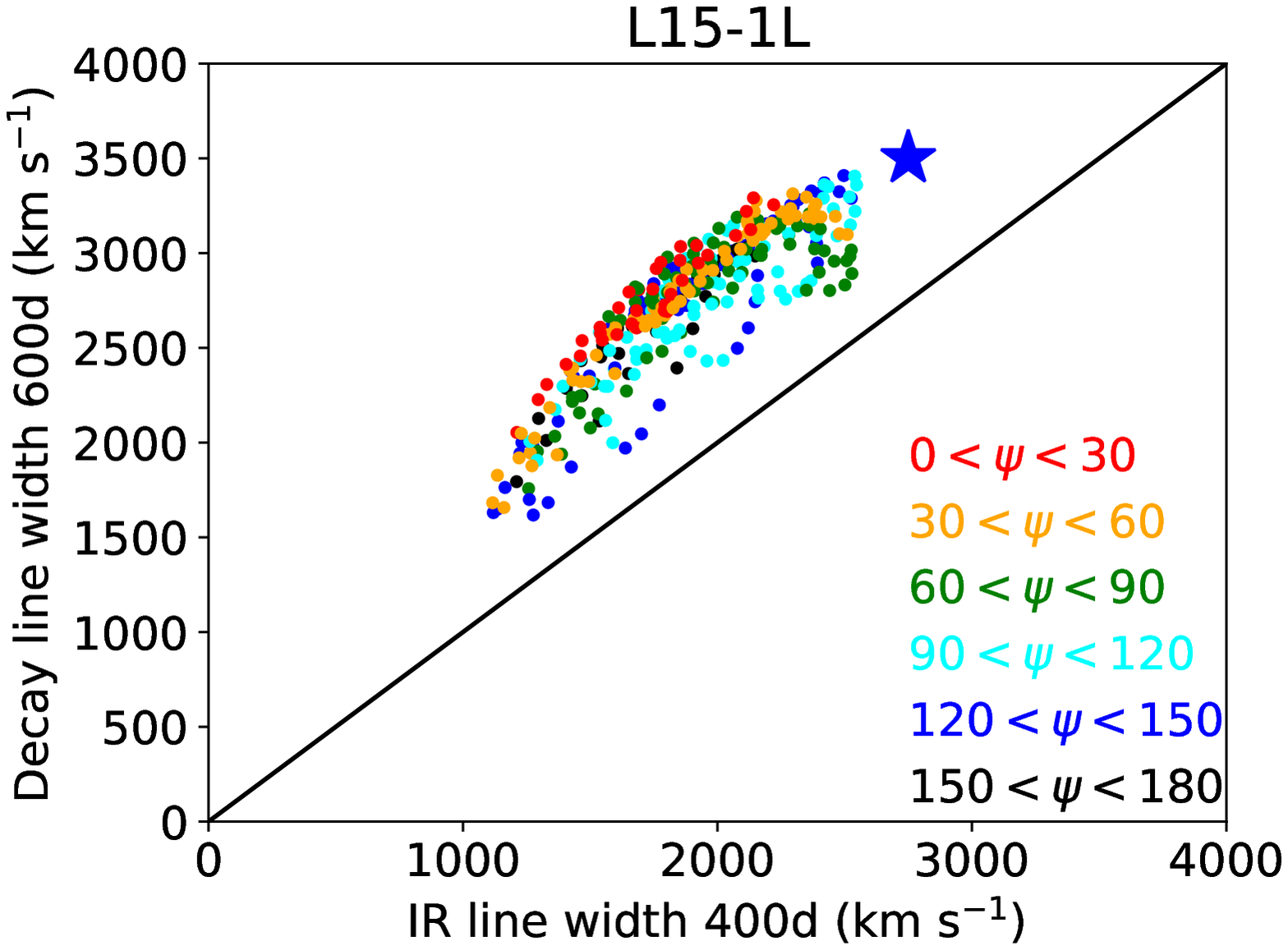}
\includegraphics[width=0.42\linewidth]{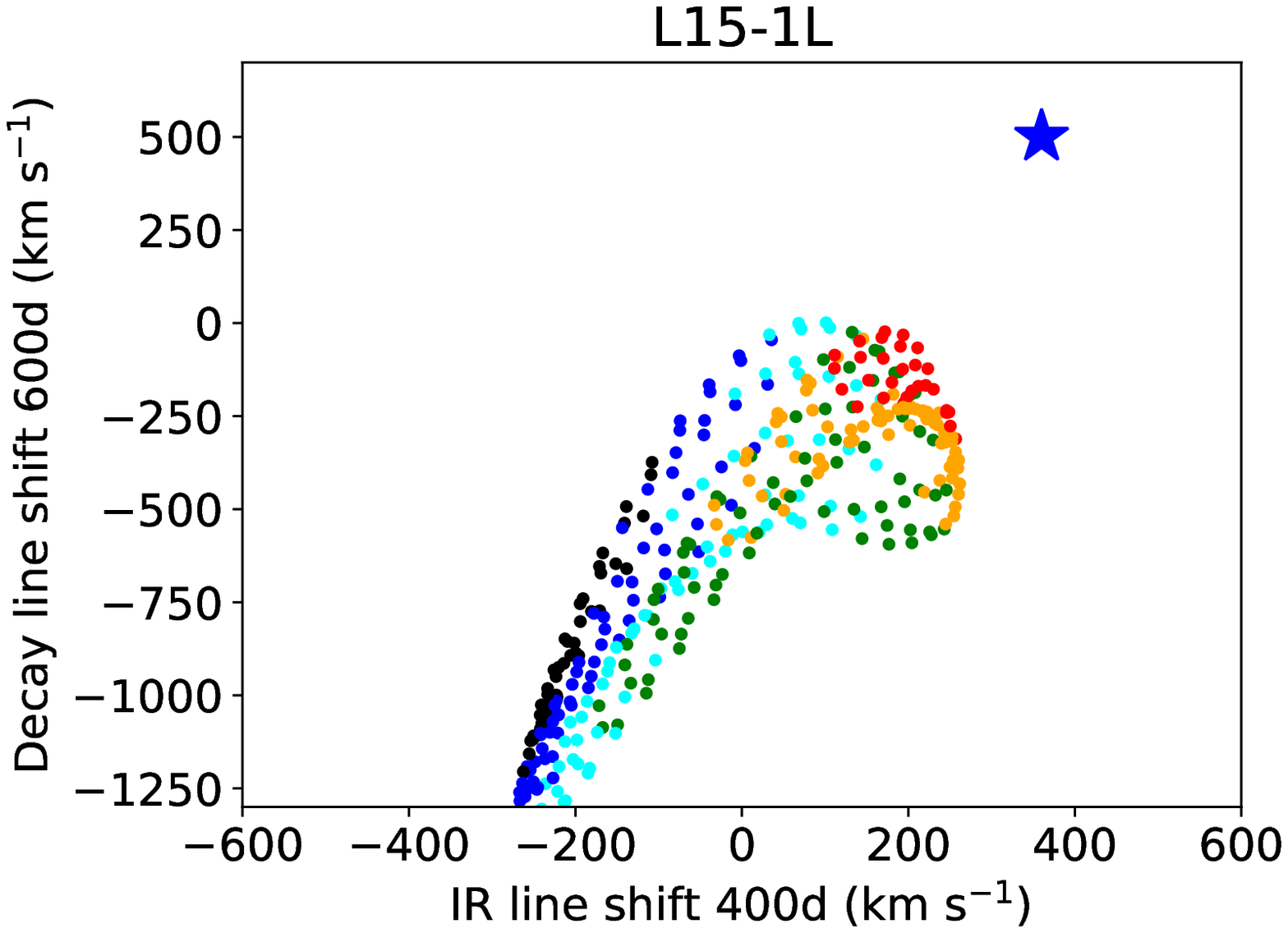}
\caption{Simultaneous comparison of properties of decay lines and IR lines, and comparison with SN 1987A (blue star). The left panel show line widths, and the right panel line shifts (redshifts with positive values), with IR lines on the x-axes and decay lines on the y-axes. The viewing directions are color coded according to the angle $\psi$ between the NS kick vector and the observer direction.}
\label{fig:gamma_and_NIR}
\end{figure*}

The summary view reinforces the interpretations from the decay line analysis section. Figure \ref{fig:gamma_and_NIR} shows the model-data comparisons with respect to both decay lines and IR lines. SN 1987A is more extreme in its observed properties of the \ni~asymmetry and bulk speed than any of the models. However, some of the models approach agreement for the right viewing angles; typically requiring a small viewing angle with the \ni~momentum vector almost parallel to our line of sight, and away from us. Note however, that there is no monotonic relation that angles with larger IR redshift also give larger decay line redshifts; there is in fact more of an anti-correlation in models B15, M15 and L15 for the highest IR line shifts caused by absorption effects for the gamma rays. One clue to understanding the 'turn-around' behaviour is to note that the most distinct turn-arounds involve cases where the line is always blueshifted, and the smallest obtainable blueshift is close to zero. If the blocking of the receding side is close to complete for all orientations, consider what happens in a toy model with a cigar-shaped ejecta. The most extreme blueshifts will occur when the cigar is oriented along the line-of-sight. This would correspond to the two bottom points in an upside-down V shape. The smallest blueshift (close to zero) would occur at the perpendicular orientation, and give the top point of the upside-down V. The 'thick' side of the cigar, when oriented towards us, would give a larger magnitude of the blueshift compared to when the short side is oriented towards us, this gives an asymmetry of the upside-down V shape. What is seen for B15-1L and L15-1L at 400d are not far from such a behaviour. For M15 and W15 emission from the receding side emerges to a larger extent for certain viewing angles.

The left panels of Fig. \ref{fig:gamma_and_NIR} show that the IR lines are distinctly narrower than the decay lines (taken at 600d) for all models, in agreement with the observed lines. An analysis of why this is so involves several factors. First, we made the comparison in the limit of optically thin decay lines. Also here are the IR lines more narrow. In a 1D picture, one replaces the $\rho_{\rm 56Ni}(v)$ distribution with $D(v) \times x_{\rm 56Ni}(v)$, where $D$ denotes the deposition. If one uses solely $D(v)$, the resulting lines are broader than the decay lines. Thus, $D$ is flatter than $\rho_{\rm 56Ni}$. But, we then multiply with a function $x_{\rm 56Ni}(v)$ that is decreasing with velocity, and does so steeply enough that the product becomes more centrally condensed. In some sense, in a 1D model one could constrain first $\rho_{\rm 56Ni}(v)$ from the decay lines, and then $x_{\rm 56ni}(v)$ from the UVOIR lines. The decay lines then show a rather weak time-dependency.

By the combined decay and IR width information, it appears that the models, perhaps with the exception of L15-1L, are on the low side in terms of bulk \ni~velocities. The mass-weighted \ni~speeds (``bulk velocity'') in the models are 1130 \kms(B15-1L), 1160 \kms(W15-2L), 1170 \kms(M15-7b1) and 1490 \kms(L15-1L). The viewing angle average values for the IR line widths are relatively close to these values. As only L15-1L formally reaches enough line widths one may draw a tentative conclusion that the bulk velocity of the \ni~should have a value of at least 1500 \kms.

\subsection{Discussion}
Due to the simplistic nature of the IR models, we do not linger on this section or draw too strong conclusions from the comparison plots. 
The key point is that the IR lines show beyond any reasonable doubt that the \ni~must preferentially reside on the receding side of the ejecta \citep[see also][]{Orlando2019,Ono2020}. As the iron from the \co~decay ``heats itself'', it is very difficult to come up with a morphology where the bulk nickel could be on the approaching side and still give redshifted IR lines. The models here solve for the gamma deposition to predict how the iron both from \ni~and primordial component is powered, and ubiquitously demonstrate this. Further, the full iron mass seems accounted for when considering optical depth effects, and neutral nickel shows the same asymmetry \citep{Spyromilio1990}, removing ionization variation as well as a dominant primordial iron component (the nickel primordial component is negligible) as possible error sources. The combined decay line and IR constraints virtually rule out any possibility that the \ni~has a bulk motion towards us. This would require simultanous suppression of the approaching side emission by two different and unrelated mechanisms; enhanced (non-local) column density for the decay lines along the right paths, and microphysical (local) anomalities to suppress the IR emissivity.

Further, the model comparisons reinforce the picture derived from the decay line analysis that this set of models achieves success only in a marginal way. Whereas the data quality of the \co~decay lines is relatively poor, there is no uncertainty from the IR (and \ti~decay lines) that the explosive ashes in SN 1987A have a high bulk velocity and a significant degree of asymmetry. It is the combined picture derived from these three independent sets of observations, and physical models operating in quite different regimes, that allow robustness to be ascribed to this conclusion. Nevertheless, the IR lines need more detailed modelling efforts before exact quantitative conclusions can be drawn from them. For example, more realistic calculation of the contribution by primordial Fe could well affect the line profiles and improve or reduce the agreement with this model set. Realistic calculations would also consider variations in temperature, ionization, and optical depth that leads to more complex links between emissivity and abundance than just a constant.

\subsubsection{The redshifted emission bump}
We finish this section by discussing the discrete emission bump seen in the highest-resolution IR data.
The high-resolution observations of [Fe II] 17.94 $\mu$m and 25.99 $\mu$m at 400d by \citet{Haas1990} show a narrow (marginally resolved, $\Delta V \lesssim 500$ 
\kms) redshifted emission bump at 3600 \kms~(in the SN frame) in both lines. The luminosity in this bump is of order 5\% of the total line luminosity. This bump was seen also in [Ni II] \citep{Colgan1989}. As there should be negligible contribution by primordial nickel, this removes any doubt that the emission seen in the Fe lines is from synthesized iron \citep[the contamination by primordial Fe can in general be a more severe issue,][]{Jerkstrand2012}.

How this emission percentage would map to a mass percentage depends on what temperature and density values are assumed. 
\citet{Haas1990} estimates the clump temperature to $2600\pm 900$ K from the ratio of the 26 and 18 $\mu$m lines, and assuming optically thin emission gets a mass of $2\e{-3}$ \msun, or 3\% of the total. Such a temperature is in reasonable agreement with model predictions for the \ni~clumps at 400d \citep[$\sim$3300 K,][] {Kozma1998-I,Jerkstrand2012}, and these lines are also not overly sensitive to errors in this range. However, these models also indicate that the optical depths in the IR iron lines are significantly above unity at 400d ($\tau_s \sim 10$). \citet{Haas1990} recognized this situation by the too low total mass inferred by the optically thin formula (0.026 \msun~vs the needed 0.07 \msun) - an optical depth of $\tau_s=3-4$ would be needed for reconciliation. If optical depth is an issue for the 3900 \kms~clump, its mass could possibly be a factor few higher than $2\e{-3}$ \msun. 

It is of relevance to consider to what extent this single clump would contribute to further bulk redshift of the IR and decay lines (the fittings done in \citet{Witteborn1989} and \citet{Haas1990} excluded it). With 5\% flux, it contributes a shift $0.05 \times 3600= 180$ \kms~to the total shift of the Fe IR lines.

For the decay lines one would expect a smaller contribution since the emission is from the far side, which is normally more obscured. But would it be possible to also have a larger effect, even to the extent that this single clump becomes responsible for most of the bulk redshift of the decay lines ($\sim$500 \kms)? The answer to this appears be a firm no. 

First, the IR Fe lines have strong redshift asymmetries also without this clump contribution ($\sim$300-400 \kms). The high-resolution observation of \citet{Haas1990} shows that there is strong continuous emission from each radial velocity range, not just from the 3600 \kms~bin. This shows that the bulk redshift emission comes from contributions by many line-of-sight velocity ranges.
Second, similarly the bins at 2000 and 2800 \kms~for the decay lines should not be affected by emission from this clump (as we know from the IR lines that it is narrowly confined to a radial velocity spread of $<$500 \kms, and the GRIS resolution is about 700 \kms). Emissions in both these bins are high, and sufficient to give a bulk redshift of the line even if the 3600 \kms~bin flux was put to zero.
Third, it would require either a lucky ``hole'' in the ejecta to exist to allow this clump to avoid absorption more than other regions (closer to the observer), or an extremely high transverse velocity to bring the clump ``outside, to the side'' of the high-opacity inner ejecta. 
There exists an extremely fine-tuned scenario where (a) The $2\e{-3}$ \msun~clump emits with no absorption (which would contribute a redshift of 500 \kms) (b) the remaining 0.068 \msun~emit with $\sim$80\% absorption but still no net blueshift. In this scenario all the bulk redshift would come from the Haas clump. But even in this scenario, point (b) would require the bulk mass to have a net redshift in order to produce no net blueshift despite heavy absorption. Then, even in this extreme scenario the \ni~distribution also without the Haas clump would be preferentially distributed on the far side.

To sum up, the IR lines suggest that a few percent of the \ni~resides in a discrete clump at receding radial velocity of 3600 \kms. This clump adds a redshift of 175 \kms~to the IR lines, however a significant redshift of these lines is present also without this component (300-400 \kms). For the decay lines, this clump contributes by an unknown extent to the emission around 3600 \kms; the exact contribution is not derivable as the particular column density to the clump is unknown. As emission from the receding side is typically blocked the most, it probably adds less than 175 \kms~redshift to the decay lines, although theoretical scenarios can be constructed where it contributes more. Several arguments suggest, however, that this is not likely. The conclusion is that the \ni~has an asymmetric redshift distribution also besides this particular clump, and that the clump is responsible only for a minor/moderate fraction of the total line shift observed.

\section{Discussion} 
\label{sec:discussion}

In \citet{Alp2019}, some of the models explored here (and a few others not explored in this paper) were investigated with respect to the X-ray/gamma-ray SEDs, and light curves for both decay lines and X-ray continuum.
Direct comparisons with SN 1987A were made for BSG models B15-1L and M15-7b2. Note that the M15-7b model used here (M15-7b1) is a different explosion simulation than the model used in \citet{Alp2019}. Figure 9 in \citet{Alp2019} shows, however, that the different M15-7b models give quite similar light curves for a fixed viewing angle, and the variation between viewing angles is larger than between models, so rough comparisons can be meaningfully made. In general, while there are certain differences both in assumptions and methods in the two papers, these should not lead to any significant differences in the model-data interpretations, so we can proceed to analyze the combined model results from both sets of calculations.

From the distinct redshifts observed for \co~decay lines, \ti~decay lines, and IR Fe lines, there is little doubt that the explosive burning ashes must be located preferentially on the receding side of the SN 1987A ejecta. The calculations here have shown that such redshifts cannot be achieved unless the \ni~momentum vector points towards the receding hemisphere, and in addition quite aligned with the line of sight. Therefore the most relevant comparisons to observations are for viewing angles between 0 and 90 degrees from the \ni~motion vector. As the fully aligned viewing direction (LOS parallel to \ni~vector) is always close to the direction of minimum flux (figure 1 in Alp et al.) the fully aligned case would correspond to the minimum flux curves plotted in Figs. 2, 3, 4 and 6 in Alp et al. However, as the angle to the motion vector is not known, one should consider also viewing angles up to 90 degrees as possibilities. This would roughly correspond to the plotted angle-average curves in Alp et al.

Starting with the decay line light curves, Fig. 4 in Alp et al. shows that models B15 and M15 both give reasonable reproduction. However, for M15 good agreement requires a viewing direction close to perpendicular to the \ni. For viewing directions more aligned with the \ni~motion, the emergent line fluxes are insufficient by a factor several. As in general such more aligned viewing directions improve the line shift agreement, there is consistency in the picture that M15-7b allows too little gamma escape. This is reinforced by figures 2 and 3 in Alp et al., where for a viewing direction close to aligned with the \ni~the emergent X-ray continuum flux is a factor $\sim$10 below observations for all epochs $\lesssim$400d. At times $\gtrsim$600d there is, however, reduced tension with the X-ray data. 

Related to this is an important point regarding the gamma decay line data. In this paper we have focused the discussion on comparisons with resolved ballon-based observations of the decay lines in SN 1987A (GRIS). The highest-cadence light curve comes, however, from unresolved satellite-based observations (SMM), and in Alp et al. much of the discussion is related to this dataset. Fig. 4 in Alp et al. shows that there is factor $\sim$2 variation between various measurements for any given epoch (although error bars are also of this order), and the GRIS observations versus SMM is no exception, with GRIS having estimated twice the flux compared to SMM at both 400d and 600d.

By looking at line shifts and widths instead of luminosities, we can at least partially circumvent this uncertainty. If we assume that any flux calibration uncertainty would not distort the line profile itself but only the overall flux levels, we can use the amplitude-independent comparisons in Fig. \ref{fig:shifts} here to test the models independent of the absolute luminosity level. M15-7b,
with its large mass, provides the largest distance between the red-blue tracks and the data, and the line widths also tend to be too low (right panel). Exceptions exist as shown by the cyan track, but the luminosity along these angles is on the other hand severely below both the GRIS and SMM level at 600d. 
Model B15, with its lower ejecta mass, provides better fits to the decay line light curves, and also the X-ray continuum light curves. As with M15, there is insufficient flux also here at early times, but the discrepancy is less severe.

Figure 6 in Alp et al. shows that the RSG models W15 and L15 have significantly earlier rises and the discrepancy with the rising light curve part is largely removed (although no explicit overplot of data was made, this can be inferred from comparing figures 4 and 6). We have shown also here that these models are in good agreement with the GRIS flux levels at 600d. As such, they appear to provide overall quite good descriptions of the gamma decay lines in SN 1987A, with generally better agreement than the BSG models B15 and M15-7b.

\subsection{BSG vs RSG models}
This leads us on to a discussion on the difference of using BSG or RSG models for the analysis. It is well established that the Rayleigh-Taylor (RT) mixing can vary quite significantly between BSG and RSG models, with RSGs providing both longer growth time ($\sim$1d vs $\sim$10min) and higher growth rate for the instability at the He/H interface (although the impact on this in the nonlinear regime is unclear), in addition to setting in later \citep{Herant1994}. There is thus, in the simplest picture, less reshaping of the ejecta morphology in BSGs and the initial explosion asymmetry may remain more intact \citep{Herant1994,Joggerst2009}. 

However, to what extent the final \ni~morphology is changed when changing only the envelope structure between BSG and RSG is a nuanced question whose answer likely varies from model to model. A full understanding of this would require running models where only the envelope structure was systematically varied, to late times at which all reverse shock, shock reflection and \co~bubble effects have played out. The final \ni~morphology depends qualitatively on three main properties of the progenitor; the CO core compactness, the He shell, and the H envelope structure \citep{Wongwat2015}. RT mixing arises from reverse shocks created not only at the He/H interface but also at the Si/O and O/He interfaces, and at these two interfaces there is no dependency on the H envelope density profile.
Further, it is not fully clear to what extent variation in the last one would impact the `bulk' properties of the \ni~that the (late) decay lines and optically thin NIR lines analyzed here mainly depends on. It is quite robustly shown in the literature that the `fastest tail' of \ni~can depend on the details of the RT mixing from the He/H reverse shock; both in velocity and in degree of fragmentation, but it is less clear how the bulk \ni~distribution is affected. One should also note that while this RT mixing changed details in the RSG models in \citet{Wongwat2015}, a general conclusion was that the fundamental morphology, arising from the explosion phase hydrodynamics, was to a large degree retained in both cases (except for one BSG not studied here).

The main observables studied here rely, to first order, on three things; the fundamental degree of asymmetry of the \ni, its bulk velocity, and the ejecta mass. Second-order effects enter for each observable, but by combining several of them (\co~decay lines, \ti~decay lines, IR lines) we can delineate a clearer picture of the requirements on these fundamental properties. None of these three properties depend on the nature of the H envelope as compact or extended.
As such the RSG models help to outline which ingredients are needed to reproduce the decay lines and IR lines. Furthermore, it is possible that these very models, if the RSG envelope was swapped for a BSG one, could retain similar outputs for the decay and IR line properties. If so, the ones where agreement with observations is satisfactory, would add their specific core structures and envelope masses to the list of stellar models that are suitable for SN 1987A with respect to these observables.

\section{Summary and conclusions} 
\label{sec:conclusions}
\arj{We have presented a new 3D photon packet transport module for the SUMO supernova spectral synthesis code, and applied this to study the transport, deposition, and emergent line profiles of gamma decay lines in 3D core-collapse models evolved to late times. We compared the model results to observations of SN 1987A and Cas A, and draw the following conclusions:}

\begin{itemize}
\item \arj{Asymmetries in the \ni~distribution in 3D CCSN models give rise to a significant variation of decay line profiles with viewing angle. Line widths vary by a factor of $\sim$3 (between $\sim$1000-3000 \kms), and line shifts by $\sim$1000 \kms~(optically thin limit) to $\sim$3000 \kms (optically thick limit). Line profiles are sometimes Gaussian-like, but sometimes have distinctly non-Gaussian shapes including multiple peaks and wings extending up to $\sim$6000 \kms}.

\item For hydrogen-rich SNe, the majority of models and viewing angles give initially blueshifted \co~decay lines (typically of order 2000 \kms~at 200d) due to Compton scattering preferentially blocking the far side of the ejecta. This Compton blueshift initially overwhelms the intrinsic asymmetry of the \co~distribution (which is of order 500 \kms). The line shifts move towards the optically thin limit on gradually redshifting tracks. Most viewing angles for which the \ni~is preferentially located on the receding side transition to actual redshifts after about two years. However, in certain models, at certain rare viewing angles, redshifted lines can be produced already at 200d. This situation may arise when favorable ``tunnels'' align with the viewer to one or a few clumps on the receding side. These tracks tend to be flat over time.

\item \arj{The observed \co~decay lines in SN 1987A are blueshifted up to 300d but develop a redshift at 400-600d. Models need two properties to allow a reproduction of this. First, a significant asymmetry in the \co~distribution, with a bulk shift (magnitude of mass-weighted velocity vector) of several hundred \kms~away from the observer. Second, a suitable ejecta mass to give sufficient Compton scattering to produce a blueshift at 300d but little enough to allow a redshift at 600d. In our model grid two models with bulk shift $\sim$500 \kms~and ejecta mass $\sim$14 \msun~achieve this, although marginally. A third model has too low \co~asymmetry (100 \kms) and a fourth has too high ejecta mass (19 \msun)}.

\item 
By adding \ti~decay lines and IR Fe lines to the analysis, degeneracies in required ejecta properties to fit SN 1987A can be resolved, and uncertainties stemming from the limited quality of the \co~decay data can be mitigated. The optically thin (but unresolved) \ti~decay lines test the degree of asymmetry required for the explosive burning ashes and show that this needs to be $\gtrsim$ 400 \kms. The resolved IR lines confirm an asymmetry of this order, and in addition test the bulk velocity (mass-weighted speed) of the \co. The bulk \co~velocities of most models in our grid are on the low side ($\sim$1100 \kms), although one model (model L15-1L with a bulk velocity of 1500 \kms) reaches the observed iron-line widths.

\item \arj{Because the neutron star and explosive nucleosynthesis ejecta momenta vectors tend to be anti-aligned \citep{Wongwat2013}, it becomes possible to constrain the NS kick from \ni/\ti/Fe line profiles. We show here that one may use observed decay line profile shifts to constrain the NS kick speed, in the case that the decay lines are redshifted. Using the H-rich model grid, we find that a constraint $V_{\rm NS} > V_{\rm redshift}$ holds in all models. Applying this to SN 1987A gives a minimum NS speed of 500 \kms. Considering that this is a lower limit, the NS star kick likely lies in the upper half of the distribution of $\sim$10-1000 \kms.}

\item \arj{Our single stripped-envelope model (a 4.5 \msun~He core with a thin H-layer exploded with 1.5 B) gives a satisfactory reproduction of the observed \ti~lines in Cas A for certain viewing angles. As with the \ti~observations in SN 1987A (and, marginally, the \co~observations), this confirms that the current generation of neutrino-driven models are capable of producing enough bulk shifts of the explosive burning ashes, although for the model set explored in this paper that success is marginal. The particular stripped-envelope model studied requires a viewing angle of around 30 degrees relative to the NS motion (towards us) to reproduce both the \ti~line shifts and the NS proper motion in Cas A.}

\item \arj{Analysis of the internal gamma-ray field shows that the angle-variation is of order 2 in the optically thin (late nebular) limit, but of order 10 in the optically thick (early nebular) phase. The models show strong variation in the extent  to which the envelope is powered at early times; the particular mixing and ejecta mass can change this by 1-2 orders of magnitudes. Looking at depositions, the deposition per unit velocity peaks in the 700-2000 \kms~range, and about 80\% of the deposition occurs inside 2000 \kms~and virtually none outside 4000 \kms. The deposition profiles move towards higher velocities with time, predicting a broadening of optical/IR lines by about 20\% from early to late nebular phase.}

\item By calculating the total gamma deposition, we provide the first prediction of the UVOIR bolometric light curve over the steady-state phase ($\sim$150-600d) using 3D models. Comparison of the model tracks to the well-determined evolution of SN 1987A shows that models with $E=1.5$ B require $M_{\rm ejecta}\sim 14$ \msun~for good reproduction. The dependency of column density, the main governing factor for the total deposition, on $M_{\rm ejecta}^2$ makes it possible to infer $M_{\rm ejecta}$ to reasonably good accuracy. These mass estimates show good agreement with those indicated from \co~decay line properties.

\end{itemize}

\section*{Acknowledgements} AJ acknowledges funding by the European Union's Framework Programme for Research and Innovation Horizon 2020 under Marie Sklodowska-Curie grant agreement No 702538. Funding by the European Research Council through grant ERC-AdG No.~341157-COCO2CASA
and by the Deutsche Forschungsgemeinschaft through grants SFB-1258 ``Neutrinos and Dark Matter in Astro- and Particle Physics (NDM)'' and EXC~2094 ``ORIGINS: From the Origin of the Universe to the First Building Blocks of Life'' is acknowledged. K.M. acknowledges support by JSPS KAKENHI Grant (18H04585, 18H05223, 17H02864). We acknowledge assistance by Markus Rampp at Max Planck Computing and Data Facility (MPCDF) for shared memory MPI algorithms and code optimization.  The simulations were performed on Hydra and Cobra of MPCDF. 
\appendix 

\section{Transport code details} 
\label{sec:transportcodedetails}

Let $\mathbf{r} = (x,y,z)$ denote the current position of the packet, and $\mathbf{r_i}= (x_{\rm i},y_{\rm i},z_{\rm i})$ the intersection point with the cell surface of interest. Let $\mathbf{\hat{n}}$ be the direction vector of the packet. Let $\mu$ be the direction cosine with respect to the radial direction. Unprimed quantities are in the stellar rest frame, primed in the comoving frame.

\subsection{Cell wall distances}

\subsubsection{Constant $r$ surfaces}
The distances to the next shell are determined by application of the law of cosines.

\subsubsection{Constant $theta$ surfaces}
The constant $\theta$ surfaces are conical sections described by 
\begin{equation}
z^2 - \mu_c^2 |\mathbf{r_i}|^2 = 0.
\label{eq:cone}
\end{equation}
where $\theta_c$ ($\mu_c = \cos{\theta_c})$ is the opening angle.
Note that also the 'ghost cone' with axis anti-parallell to $\hat{z}$ has this equation.
The trajectory is
\begin{equation}
\mathbf{r_i} = \mathbf{r} + ds \mathbf{\hat{n}}.
\label{eq:standardmove}
\end{equation}
Inserting gives
\begin{equation}
ds^2 \underbrace{\left(\hat{n}_z^2 - \mu_c^2\right)}_{c_2} + 2 ds \underbrace{\left(z \hat{n}_z - \mu_c^2 \mathbf{r} \cdot \mathbf{\hat{n}}\right)}_{c_1}
+ \underbrace{\left(z^2 - \mu_c^2 r^2\right)}_{c_0} = 0
\end{equation}
If $c_2 \ne 0$, the solution is
\begin{equation}
ds = -\frac{c_1}{c_2} \pm \sqrt{\left(\frac{c_1}{c_2}\right)^2 - \frac{c_0}{c_2}}
\label{eq:conesol}
\end{equation}

The quantity to be taken square root of is checked for positivity. If this quantity is exactly zero, the trajectory is tangent to the cone at a single point, and one can treat it as passing. 

If the square root argument is positive, there are two mathematical solutions. These may both be on the true cone, both on the ghost cone, or one on each (note that a positive $ds$ has not yet been enforced). Each of the two solutions are checked for directionality ($ds>0$), and that it hits the true cone. On the true cone the new $z$ coordinate ($z + \hat{n}_z ds$) has same sign as $\cos{\theta_c}$, whereas on the ghost cone the new $z$ coordinate has opposite sign to $\cos{\theta_c}$. 

Two valid solutions can occur when the trajectory intersects a surface ``protruding inwards''. The true solution is then the shorter distance one, i.e. the negative square root. Thus, one can first check validity (positive $ds$) of the positive square root, and store it. Then check validity of the negative square root, and update the solution with this if also valid.


Note that the cone is fully specified by $\theta_c$, so one does not need to precompute and store anything for these surfaces. To solve Eq. \ref{eq:conesol}, we need only the current position and direction, and the $\theta$ boundaries of the current cell.


\subsubsection{Constant phi surfaces}
The constant $\phi$ surfaces are planes. 
Determine the Cartesian coordinates for the four edgepoints of the surface. Then, define, for the $k$ plane, the vectors
\begin{eqnarray}
\mathbf{v_1} &=& \mathbf{r_{i+1,j+1,k}} - \mathbf{r_{i,j,k}} \\
\mathbf{v_2} &=& \mathbf{r_{i+1,j,k}} - \mathbf{r_{i,j,k}}  
\end{eqnarray}
Take the cross product and normalize to get the normal vector to the plane $\mathbf{\hat{p}}$.
Note that an intersection point $\mathbf{r_i}$ on the plane will obey $d = \mathbf{\hat{p}} \cdot \mathbf{r_i}$.
For the $k+1$ plane use instead
\begin{eqnarray}
\mathbf{v_1} &=& \mathbf{r_{i+1,j+1,k+1}} - \mathbf{r_{i,j,k+1}}\\
\mathbf{v_2} &=&  \mathbf{r_{i+1,j,k+1}} - \mathbf{r_{i,j,k+1}}
\end{eqnarray}
and then corresponding operations for $\mathbf{\hat{p}}$ and $d$.

The intersection point $(x_{\rm i},y_{\rm i},z_{\rm i}$) and the distance to this point, $ds$, are then found from solving the linear equation system
\begin{equation}
\begin{bmatrix}
1 & 0 & 0 & -\hat{n}_{\rm x} \\
0 & 1 & 0 & -\hat{n}_{\rm y} \\
0 & 0 & 1 & -\hat{n}_{\rm z} \\
\hat{p}_{\rm x} & \hat{p}_{\rm y} & \hat{p}_{\rm z} & 0
\end{bmatrix}
\begin{bmatrix}
x_{\rm i} \\ 
y_{\rm i} \\
z_{\rm i} \\
ds \\
\end{bmatrix}
=
\begin{bmatrix}
x \\ 
y \\
z \\
d \\
\end{bmatrix}.
\end{equation}
No solution means the packet is moving parallel to the plane and will not impact it. If $ds<0$, the packet is moving in the wrong direction and will not impact the plane.

The four numbers $(\hat{\mathbf{p}},d)$ are computed once and stored for all surfaces.

\subsubsection{Entry surface flagging}
If the surface through which the packet entered the cell is allowed also an as exit surface, small numeric errors could lead to the packet getting stuck at the boundary. To avoid this the entry surface is initially flagged to be disallowed as solution for which next boundary to encounter. This flagging is removed if an interaction occurs inside the cell. For two surfaces, the  ``convex out'' boundaries at the constant $r$ outer radius and constant $\theta$ conical boundary closer to the equatorial plane, it is possible for the packet to enter and exit the same boundary on a straight path. However, this occurrence is very rare, in particular for a high-resolution grid as those used here. We therefore kept the flaggings also for these cases, and was removed only after a small distance into the cell had been travelled. This treatment implies a small distortion to the flight path for photons gracing these surfaces but the effect of this should be very small, and this treatment avoids an more expensive algorithm for doing the transfer exactly. 

\subsubsection{Positional corrections}
Once the shortest distance has been calculated, the new Cartesian coordinates of the packet are calculated from $\mathbf{r_i} = \mathbf{r} + ds \hat{\mathbf{n}}$. The spherical coordinates are then calculated from these. It may sometimes happen that in these transformations small numeric errors lead to the new spherical coordinates being (slightly) outside the limits of the new cell. If that happens severe errors can occur in calculating the next movement. Therefore, at each new cell entry, this is checked and if a discrepancy has arisen, the coordinates are adjusted to be just inside (by a multiplicative factor $1 \pm 10^{-5}$) the surface in question.

\subsubsection{Next interaction point}
For continuum opacity, standard Monte Carlo procedures are followed by drawing a random optical from $\tau = -\ln{z}$ where $z$ is a random number between 0 and 1. Then
\begin{equation}
ds = \frac{\tau}{\kappa \rho}
\end{equation}

As the distances are computed in the stellar rest frame one should ideally use the direction-dependent opacity in this frame. We did not make the distinction in this application (but took $\kappa(E) = \kappa'(E')$) as doing relativistic corrections for this while ignoring live expansion effects on the density does not make much sense. It is straightforward to make this relativistically correct if desired by taking $\kappa(E) = \kappa(E')\left(1-\mu v/c\right)$ \citep{Lucy2005}.

For lines in the Sobolev treatment (not treated here but in upcoming papers), the distance to the line in velocity units is given by the relativistic Doppler formula, and from this the spatial distance follows from the known velocity gradient. 

\subsection{Code tests and comparisons}
\arj{We tested that the three transport modes gave identical results for the case of $\kappa=0$, and also that the ray-tracing and Compton scattering modes gave the same results (apart from the above-mentioned weak red tail) for a non-zero $\kappa$. We also tested the code against the independently developed gamma and X-ray transport code in \citet{Alp2019}}. 

\subsection{Parallellization}

\subsubsection{Compton scattering mode.} \arj{It is desireable for load balancing that processors have as similar spatial domains as possible, and that parallelization as far as possible is done over (similar) photons. We implemented here two different approaches. Some algorithm will determine the desired total number of packets to be emitted from any given cell, $N_i$. The average number of packets per processor is then $N_{\rm i}/N_{\rm procs}$. In mode A, we let each processor emit a number equal to $\mbox{floor} (N_{\rm i}/N_{\rm procs}) + 1\times \delta$, where $\delta=1$ if $z< N_{\rm i}/N_{\rm procs} - \mbox{floor}(N_{\rm i}/N_{\rm procs})$ and $\delta=0$ otherwise. As long as $N_{\rm procs} \gg 1$ and $N_{\rm i} \gg 1$, both of which should normally be fulfilled, a total emitted number close to $N_{\rm i}$ will be guaranteed. The emitted luminosity error will be $1/\sqrt{N_{\rm i}}$.  If $10^{10}$ photons are emitted over $10^6$ cells, the error is of order 1\% per cell}.

\arj{Another selectable mode (mode B) is that $N_{\rm floor}+1$ photons are emitted by all processors, and the energy packets are adjusted accordingly. As any cell with non-negligble amounts of radioisotope should have $N_{\rm i} \gg 1$, this should not increase the total tally from the specified one much from these cells. Cells with little radioactive mass may contribute up to $N_{\rm zones} \times N_{\rm procs}$ photons more, but since the number of cells is limited to $N_{\rm zones}\lesssim 10^7$, and typically $N_{\rm procs} \lesssim 300$, this is still moderate compared to the total tally typically of $10^{10}$.
Mode B may become expensive for very large grids, with many near-empty cells, and does not give good scaling for very large number of processors. But it removes one random noise component to the simulations that mode A entails. For our final models presented in the paper we ran in the B mode to remove this noise component.}

\subsubsection{Ray-tracing and thin modes.} \arj{For ray-tracing exactly $N_{\rm viewers}$ packets (typically a few hundred) need to be emitted in each cell, and splitting these over processors becomes cumbersome. As the calculations in this mode are fast, optimal load balancing is less important, and we parallelize over the spatial domain, with each processor obtaining a distinct set of cells to emit from}.

\subsection{Performance tests}
We made measurements of time used by the code in different parts of the execution, using model M15-7b1 as example. The code was compiled with Intel IFORT (at -O2 optimization) and run on a cluster of 2.3 GHz Intel Haswell Xeon E5-2698 processors. 

The initial reading of the explosion model takes about 20s. The loop where cell wall constants are calculated and stored takes $\lesssim 1$s. For the packet transport, we measured the amount of time spent on average per cell transport step. 
Across a wide variation in grid resolution, particle per cell number, optical depth, and CPU numbers (1-512) we found the execution time per cell transport step to consistently be in the range 400-800 ns. This appears to be a reasonable value, with of order 100 additions/multiplications and 30 trigonometric/exponential/power-law calls, for a total of order 500 FLOPS. The nature of Monte Carlo simulations make use of vectorization and pipelining difficult for the CPU, so number of flops per second is probably close to the clock frequency, giving a total estimated time of a few hundred ns for these operations. In addition a few retrievals/storages to big arrays (where Cache misses may occur) may incur latency costs of order $\lesssim$100 ns in total. With no complex interaction between the cores MPI communication costs are minor, dominated by MPI\_REDUCE of wavelength-dependent cell values at the end ($\lesssim 1$min), as is the output file writing time ($<1$s).

In this application of Compton scattering there is either no microphysical interactions in any given cell (most common), or a relatively simple scattering event. For future UVOIR simulations there will be many more sources of opacity present (continuum and lines from a large number of elements) and from experience with SUMO-1D the processing time for the in-cell interactions will likely be comparable to or exceed this geometry-dominated computation time of 400-800 ns. Therefore this performance level should be satisfactory and spherical geometry transport does not become a bottleneck in the computations.

\bibliographystyle{mn2e3}
\bibliography{bibl}

\end{document}